\documentclass{aa}  

\usepackage{ragged2e}
\usepackage{natbib}
\bibpunct{(}{)}{;}{a}{}{,}
\usepackage{graphicx}
\usepackage{adjustbox}
\usepackage{txfonts}
\usepackage[dvipsnames]{xcolor}
\usepackage{tabularx,multirow,booktabs,blindtext}

\usepackage[colorlinks=true,linkcolor=black, urlcolor=black, citecolor=black]{hyperref}
\hypersetup{citecolor=blue}
\hypersetup{linkcolor=blue}
\usepackage{graphicx}
\usepackage{subfigure} 
\usepackage{booktabs, caption} 
\usepackage{multicol}

\begin{document}

\title{The stellar activity-rotation-age relationship under the lens of asteroseismology}

\author{C. Pezzotti\inst{1}, J. Bétrisey\inst{2}, G. Buldgen\inst{1}, M. Gilfanov\inst{3,4}, I. Bikmaev\inst{5}, R. Sunyaev\inst{3,4}, E. I\c{s}{\i}k\inst{6}, E. Gosset\inst{1}, N.J. Wright\inst{7}}

\institute{STAR Institute, Université de Liège, Liège, Belgium:
\email{camilla.pezzotti@uliege.be}
\and Department of Physics and
Astronomy, Uppsala University, Box 516, SE-751 20 Uppsala, Sweden
\and Space Research Institute, Russian
Academy of Sciences, Profsoyuznaya 84/32, 117997 Moscow, Russia
\and Max Planck Institute for Astrophysics,
Karl-Schwarzschild-Str 1, Garching b. München D-85741, Germany
\and Kazan Federal University, 18
Kremlyovskaya Street, Kazan, Russia
\and Max-Planck-Institut für Sonnensystemforschung, Justus-von-Liebig-Weg
3, 37077 Göttingen, Germany
\and Astrophysics Research Center, Keele University, Keele, ST5 5BG, UK}

\date{Received ...; accepted ...}

 \abstract
   {In low-mass stars, the connection between magnetic activity, rotation period, and age provides key insights into the functioning of dynamos. Fully understanding the activity–rotation–age relationship requires stars with precise fundamental parameters, measured rotation periods, and reliable magnetic activity indicators (e.g. X-ray luminosity). Thanks to space-based photometry, asteroseismology is now the leading method for determining stellar parameters with unprecedented precision and accuracy. The best-characterized solar-like stars compose the \textit{Kepler} LEGACY sample, with highest-quality asteroseismic data for 66 stars, most of which have measured rotation periods. In the X-ray band, these stars were observed by the ROentgen Survey with an Imaging Telescope Array (eROSITA) telescope on the Russian Spektrum-Roentgen-Gamma (SRG) satellite in the course of its all-sky survey.}
   {We reviewed different components of the stellar activity–rotation–age relationship using the largest sample of solar-like stars with highly accurate fundamental parameters from asteroseismology, along with measured rotation periods and X-ray luminosities.}
   {We cross-correlated the \textit{Kepler} LEGACY sample with SRG/eROSITA source catalogue finding X-ray detections for 13 of them. We derived their fundamental parameters using the Forward and Inversion COmbination procedure and revisited widely studied activity-age and activity-rotation relationships by consistently incorporating our 13-star subsample with literature samples.}
   {By implementing revised activity–rotation-age relationships in a Star-Planet Interaction code to compute X-ray luminosity tracks and comparing the results with observations, we found improved agreement for 7 stars of our subsample. We explored the effect of the revised relationships on the mass loss of planets in the radius valley, finding a modest impact on planet size distributions.}
   {A larger and more varied sample of stars with asteroseismically-characterized parameters, rotation period and activity indicators is needed to accurately determine the multiple components of the activity-rotation-age relationship.}

\keywords{Stars: evolution -Stars: activity - Stars: rotation - Stars: solar-type - Planet-star interaction}

\titlerunning{Age-activity-rotation relationship}
\authorrunning{Pezzotti et al.}
\maketitle

\section{Introduction}

Studying the mutual dependence between magnetic activity, rotational period and age in low-mass stars (defined here as stars with initial mass $\rm M_{\star} \lesssim 1.5~M_{\odot}$, and spectral type F to M) keeps raising a growing interest in the scientific
community, as these quantities are critical observational proxies of the dynamo operating in stellar convective interiors. The rotational and magnetic properties of low-mass stars are tightly interconnected through a dynamo feedback loop \citep{Vidotto2021}, regulated by competing effects of internal differential rotation, turbulent convection, and braking at the stellar surface by magnetized winds \citep{Parker1955}. According to this feedback loop, as stars evolve, their surface rotation and magnetism globally decrease, resulting in a less powerful dynamo and angular momentum removal. 

In this context, many studies have been dedicated to observationally constraining the relationships between stellar surface rotation with age \citep[e.g.,][]{Skumanich1972, Bouvier1997, Barnes2003}, magnetic activity with age \citep[e.g.,][]{Guedel1997, PreibischFeigelson2005, Telleschi2005, Booth2017}, and magnetic activity with surface rotation \citep[e.g.,][]{Skumanich1972, Noyes1984, Wright2011,Johnstone2015,Wright2018, Magaudda2020}, trying to determine their mutual dependencies across diverse stellar evolutionary stages. Although past studies have been crucially informative on the activity-rotation-age relationship for low-mass stars up to a few billion years, the limited access to accurate ages for older stars has prevented a comprehensive investigation of this relationship for more evolved objects.

Thanks to the advent of space-based photometry missions, such as CoRoT \citep{Baglin2009}, \textit{Kepler} \citep{Borucki2010}, TESS \citep{Ricker2014}, and K2 \citep{Howell2014}, asteroseismology has established itself as a powerful tool for the derivation of fundamental stellar parameters (age, $\rm M_{\star}, R_{\star}$), with an unprecedented level of precision and accuracy. Recent investigations on the rotational period-age relationship ($\rm P_{rot}-Age$) based on asteroseismically characterized stars \citep{VanSaders2016} showed that solar-like stars more evolved than the Sun appear to rotate faster than what is predicted by typical spin-down models \citep[e.g.][]{Matt2015}, hinting at a weakening of the dynamo that would no longer be able to (fully) sustain the activity-rotation-age feedback loop. In this respect, several efforts have been dedicated to investigate the corresponding consequences on the magnetic activity properties \citep[e.g.][]{Booth2017,Metcalfe2022,Metcalfe2024}, but the dearth of robust activity indicators for evolved stars has strongly challenged the derivation of comprehensive and exhaustive results.

The stellar X-ray luminosity $\rm L_x$ is a fundamental activity indicator, unambiguously associated with the magnetic heating of the plasma in the stellar corona and has been widely used to calibrate $\rm L_x$-Age and $\rm L_x$-$\rm P_{rot}$ (or $\rm R_x$-$\rm Ro$, where $\rm R_x$ is the ratio between $\rm L_x$ and the bolometric luminosity $\rm L_{bol}$, and $\rm Ro$ is the stellar Rossby number) relationships for younger stars \citep[e.g., see recent work on more than 650 first X-ray detected Pleiades members with eROSITA, ][]{Khamitov2025}. Significantly more challenging is studying these relationships in more evolved stars that are intrinsically fainter ($\rm L_x \lesssim 10^{27}$ erg/s), requiring the use of highly sensitive telescopes and long exposure times. \citet{Booth2017} firstly studied the $\rm L_x$-Age relationship in a sample of 14 old solar-like stars with ages derived by means of more accurate techniques, among which asteroseismology was used for 6 of them. As a result, they found a steepening of the activity-age relationship for evolved stars ($\rm Age \gtrsim 1$ Gyr) compared to their younger counterparts, which appeared to be in contrast to the flattening of the $\rm P_{rot}$-Age relationship observed by \citet{VanSaders2016}. 

To develop a comprehensive picture of the activity-rotation-age relationship for evolved stars, it is crucial to expand the sample of stars that possess accurately determined fundamental parameters, measurements of the rotational period, and X-ray detections. In this study, we revisit the different components of the activity–rotation–age relationship by introducing a stellar sample whose fundamental parameters can be precisely determined through asteroseismic techniques. For this purpose, the \textit{Kepler} LEGACY sample \citep{Lund2017}, comprising 66 solar-like stars (from late K to early F type) with the highest-quality seismic data currently available until the launch of PLATO, represents the most suitable choice. In Sect.~\ref{Sec:eROSITA} we detail the target selection based on the availability of rotational periods \citep{Santos2021} and the search and analysis of X-ray detections within the
 ROentgen Survey with an Imaging Telescope Array \citep[eROSITA;][]{Predehl2021} on
board the Russian Spektrum-Roentgen-Gamma mission \citep[SRG;][]{Sunyaev2021} (SRG/eROSITA-RU) all sky surveys for stars in the \textit{Kepler} LEGACY sample. In Sect.~\ref{Sec:params}, a new derivation of the stellar fundamental parameters for our subsample of \textit{Kepler} LEGACY stars with X-ray detections is carried out. In Sect.~\ref{Sec:relationhips}, we revisit the X-ray activity vs Age, and X-ray activity vs stellar Rossby number relationships. Finally, in Sect.~\ref{Sec:models} we implement the revisited relationships in theoretical models to compute X-ray luminosity tracks and highlight any potential difference with respect to previous findings.

\section{Target selection and X-ray fluxes from eROSITA}
\label{Sec:eROSITA}

We use the X-ray luminosity of a star as an indicator of its magnetic activity. To this end, we cross-correlated the \textit{Kepler} LEGACY sample with the SRG/eROSITA source catalog obtained in the course of its all-sky survey during 2019-2022\footnote{The \textit{Kepler} field resides in the half of the sky reserved to the Russian eROSITA consortium. We invite interested readers to directly contact the members of the consortium for information about data availability.}. The \textit{Kepler} LEGACY field was observed with SRG/eROSITA during 4 surveys. We used the $0.3-2.3$ keV SRG/eROSITA source catalog derived from full-sky survey data and searched for matches within 98\% positional uncertainty of X-ray sources, which was typically in the $\sim 5-15$ arcsec range. The SRG/eROSITA matches  were filtered by the detection confidence threshold corresponding to $3\sigma$ (threshold on the likelihood value $\geq6$).  The Gaia positions of stars were corrected for the proper motion to bring them to the mean epoch of SRG/eROSITA observations. For further analysis, we only kept the stars that have measured rotation periods in \cite{Santos2021}. From the global sample of 66 stars from the \textit{Kepler} LEGACY, we kept 51 stars.

One of the eROSITA sources, SRGe J185621.8+453027, has two objects within it’s 98\% error circle, Gaia DR3 2106822131755292032 and 2106822131756839808, separated by 1.5 and 4.0 arcsec respectively. We searched for the nature and global properties of the two objects on SIMBAD \citep{SIMBAD2000}, finding that Gaia DR3 2106822131756839808 corresponds to the \textit{Kepler} LEGACY star KIC9139163, while no correspondence has been found for Gaia DR3 2106822131755292032. By querying the Gaia archive, we found that the G magnitude of Gaia DR3 2106822131755292032 ($15.70$) is significantly fainter compared to Gaia DR3 2106822131756839808 ($8.26$). In an attempt to refine the nature of Gaia DR3 2106822131755292032, we searched for photometric estimates from the G magnitude of $\rm T_{eff}$, $\rm log~g$ and $\rm [Fe/H]$, but we found no data. We computed the X-ray luminosity associated with SRGe J185621.8+453027 firstly deriving the source's distance from the Gaia DR3 parallax, which was initially corrected for the systematic offset (``parallax zero-point'', \cite{Groenewegen2021}), and then used to infer the stellar distances by means of the geometric Bayesian method of \citet{Bailer-Jones2021}. We obtained $\rm L_{x} = (9.23 \pm 1.15) \times 10^{28} \rm erg/s$, a value which appears to be compatible with the X-ray luminosities expected for solar-like star on the main sequence. With the lack of indications about the nature of the Gaia DR3 2106822131755292032 object, and the compatibility of the X-ray luminosity of SRGe J185621.8+453027 with the one typically expected for MS solar-like stars, we found reasonable to associate SRGe J185621.8+453027 to Gaia DR3 2106822131756839808 (KIC9139163).

The X-ray fluxes quoted in the eROSITA catalog were recalculated for the spectrum of thermal emission of optically thin plasma with a temperature of 150 eV assuming a zero column of neutral hydrogen in the line of sight, $\rm NH=0$, for the same 0.3--2.3 keV energy range. For a temperature of 300 (500) eV, the presented luminosities should be multiplied by a factor of 0.83 (0.81). Moderate values of column density, $\rm NH\sim 10^{20} cm^{-2}$ change these values by a few percent. We did not attempt to apply the bolometric correction of X-ray luminosities because of its potentially large uncertainty. Similarly, conversion of eROSITA fluxes to the ROSAT energy band of 0.1--2.4 keV  involves extrapolation to lower energies outside eROSITA working energy band and  may be rather uncertain,  especially for sources with softer spectra and/or  fainter sources having large uncertainties on their coronal temperature. For example, it has a moderate value of $\approx 1.25$ for the plasma temperature of kT=300 eV and  increases towards lower temperatures, reaching  the value of $\approx 2.5$ for kT=150 eV. It is worth keeping this point in mind in the following analysis and in the comparison of the revisited activity-rotation-age relationships with the theoretical X-ray tracks in Sect.~\ref{Sec:models}.

The eROSITA detected stars from \textit{Kepler} LEGACY sample, used for the following analysis are listed in Table \ref{Tab:xray}.

\begin{table*}
\centering
\caption{eROSITA detected stars from a subset of \textit{Kepler} LEGACY sample, used in this work.}
\label{Tab:xray}
\begin{tabular}{llcccc} 
 \hline\hline
 KIC ID & SRG/eROSITA ID & offset & Lkh & $\rm F_x$ & $\rm L_x$  \\ 
 & & arcsec & & $10^{-14}$ erg/(cm$^2$ s) & 
 $10^{28}$ erg/s \\
 \hline
 \noalign{\vskip 0.5ex}
  KIC2837475 & SRGe J191011.8+380457 & 1.6 & 9.6  & $1.75\pm 0.63$  & $3.10\pm1.13$  \\ 
  KIC6508366 & SRGe J190743.0+415620 & 7.5 & 7.0  & $1.19 \pm 0.48$ & $4.77 \pm 1.95$ \\
  KIC7103006 & SRGe J190655.7+424021 & 2.7 & 44.5 & $3.82 \pm 0.77$ & $11.32 \pm 2.28$\\
  KIC7206837 & SRGe J193503.7+424412 & 5.1 & 7.2  & $1.56 \pm 0.60$ & $7.14 \pm 2.76$ \\
  KIC7940546 & SRGe J185216.5+434232 & 4.2 & 204.3 & $10.53 \pm 1.14$ & $7.48 \pm 0.81$\\
  KIC8006161 & SRGe J184434.7+435001 & 2.5 & 38.2 & $3.22 \pm 0.69$ & $0.28 \pm 0.06$\\
  KIC8379927 & SRGe J194641.7+442056 & 1.4 & 195.3 & $11.42 \pm 1.29$ & $2.30 \pm 0.27$\\
  KIC9139163 & SRGe J185621.8+453027 & 4.0 & 155.0 & $7.37 \pm 0.91$ & $9.23 \pm 1.15$\\
  KIC9812850 & SRGe J184739.5+464113 & 3.6 & 40.0 & $3.20 \pm 0.64$ & $13.57 \pm 2.70$\\
  KIC10454113 & SRGe J185636.8+473927 & 4.5 & 10.0 & $1.22 \pm 0.43$ & $1.46 \pm 0.51$\\
  KIC10644253 & SRGe J184342.8+475619 & 4.3 & 14.1 & $1.24 \pm 0.39$ & $1.38 \pm 0.44$\\
  KIC11081729 & SRGe J192220.4+484143 & 1.4 & 50.32 & $3.35 \pm 0.64$ & $6.83 \pm 1.30$\\
  KIC11253226 & SRGe J194340.8+485545 & 11.98 & 7.32 & $0.89 \pm 0.37$ & $1.48 \pm 0.62$\\
 \hline
\end{tabular}
\tablefoot{KIC ID (col.~1), eROSITA source name (col.~2), separation offset with respect to Gaia DR3 coordinates in arcsec (col.~3), detection likelihood (col.~4), X-ray flux in the eROSITA $0.3-2.3$ keV energy band (col.~5), and X-ray luminosity (col.~6).}
\end{table*}

\section{\textit{Kepler} LEGACY subsample: derivation of the fundamental stellar parameters}
\label{Sec:params}

To construct our asteroseismic models, we employed the Forward and Inversion COmbination (FICO) procedure \citep{Betrisey2022,Betrisey2023_AMS_surf,Betrisey2024_AMS_quality,Betrisey2024_phd}, a sophisticated methodology that combines forward modelling with inversion techniques. This approach, rooted in pioneering works by \citet{Reese2012} and \citet{Buldgen2019f}, enables a precise and accurate determination of fundamental parameters of solar-like stars, such as mass, radius, and age. Without delving in details beyond the scope of this article, we highlight that the FICO procedure was notably designed to deal with surface effects, which are among the main limiting factors of asteroseismic characterisation. The FICO method addresses these effects through a three-step strategy. It begins by fitting individual oscillation frequencies using the \citet{Ball&Gizon2014} surface correction. The next step of the procedure consists in a mean density inversion. This inversion, based on \citet{Reese2012}, refines the estimate of the mean density in a quasi-model-independent manner. Finally, the frequency separation ratios, computed following \citet{Roxburgh&Vorontsov2003}, the inverted mean density, and the spectroscopic constraints are used for the last optimisation step. Although these ratios are almost insensitive to near-surface inaccuracies \citep{Roxburgh&Vorontsov2003,Oti2005}, they also lack information about the stellar mean density, hence the need for the prior inversion step. For a broader overview of seismic inversion techniques and their successful applications, we refer to the review of \citet{Buldgen2022c} and the references therein \citep[see e.g.][for a non-exhaustive selection]{DiMauro2004_theo,Buldgen2015a,Buldgen2015b,Buldgen2016b,Buldgen2016c,Buldgen2017c,Buldgen2019b,Buldgen2019f,Buldgen2022b,Kosovichev&Kitiashvili2020,Salmon2021,Betrisey&Buldgen2022,Betrisey2022,Betrisey2023_rot,Betrisey2023_AMS_surf,Betrisey2024_AMS_quality,Betrisey2024_MA_Sun,Betrisey2025_MA_Inv,Betrisey2025_MA_Damping}. The observational constraints are provided in Appendix~\ref{app:observational_data}.

We used the Asteroseismic Inference on a Massive Scale (AIMS) software \citep{Rendle2019} to fit individual oscillation frequencies alongside spectroscopic constraints ($\rm T_{eff}$, [Fe/H], and $\rm L$), to which an extended version of the Spelaion grid \citep{Betrisey2023_AMS_surf} was provided as input. Surface effects were corrected using the formula of \citet{Ball&Gizon2014}. The free parameters included mass, age, initial hydrogen and metal mass fractions ($X_0$ and $Z_0$), and surface correction coefficients. For stars above $\rm 1.10\ M_\odot$, overshooting was also included as free parameter. Uniform uninformative priors were considered, except for stellar age, which was uniformly constrained between 0 and 13.8 Gyr. Each AIMS run consisted in a burn-in phase followed by a solution run. The initial fit used 800 walkers, 8000 burn-in steps, and 3000 solution steps. This dense sampling ensured strong convergence, particularly for the surface correction terms.

To ensure the reliability of the final FICO step, especially near the end of the main sequence, where seismic degeneracies increase \citep[e.g.][]{White2011}, we performed an interpolation validation test. This involved regenerating a stellar model using the optimal AIMS parameters and the same evolution and pulsation codes as the original grid. A successful match between the recomputed and interpolated frequencies confirms the physical validity of the AIMS solution. Discrepancies, typically manifesting as age overestimation, indicate interpolation failure and require local re-optimisation. This issue affected a minority of our sample, namely KIC2837475, KIC6508366, and KIC11253226. For these stars, we performed an additional local minimisation using a Levenberg-Marquardt algorithm \citep[e.g.][]{Frandsen2002,Teixeira2003,Miglio&Montalban2005}, optimising the same parameters as in AIMS. Constraints included $r_{01}$ ratios, spectroscopic data, the inverted mean density, and the lowest radial-mode frequency. As indicated in Table~\ref{Tab:params}, where the fundamental stellar parameters for the 13 \textit{Kepler} LEGACY stars are listed, the agreement with the metallicity values in Table~\ref{tab:spectroscopic_constraints} is worse for KIC2837475 and KIC11253226, which is likely the sign of nonstandard processes (e.g. turbulent diffusion, undershooting, radiative accelerations; see \cite{Deal2018,Buldgen2025}, and references therein) not included in the standard physics used in our stellar models, especially considering the presence of a strong oscillatory signal in the $r_{01}$ and $r_{10}$ ratios. However, the estimated mass, radius, age, and luminosity do not seem to be significantly affected and are consistent with independent determinations of \citet{SilvaAguirre2017}. It is worth noting that our final sample is dominated by F-type stars. This reflects the composition of the global \textit{Kepler} LEGACY sample, with the majority of stars belonging to the F spectral type. These stars have thinner convective envelopes in comparison to their cooler counterparts, that may reduce the braking efficiency via magnetized winds. \citet{Kraft1967} provided a limit in $\rm T_{eff}(K) \sim 6250$ (Kraft Break) for the transition between stars with thick enough convective envelopes and efficient braking, and stars without (or with very thin) convective envelopes, and inefficient braking. In the context of the characterization of the stellar activity-rotation-age relationship, comparing the properties of our stars with the Kraft Break may give an indication about the braking efficiency of these objects. In a recent work, \citet{Beyer2024} reassessed a limit in $\rm T_{eff}(K) = 6550 \pm 200$ for the Kraft Break, by studying the rotational properties of 405 F-type stars within 33.33 pc. They found that stars above the Break have projected velocities $\rm vsin(i) = (61.5 \pm 30.6)~km/s$, while below $\rm vsin(i) = (6.7 \pm 3.6)~km/s$, with transition at $\rm vsin(i) \simeq 20 ~km/s $. By computing the rotational velocities for our sample of stars, and by checking $\rm T_{eff}$ (see Tables~\ref{Tab:params}, ~\ref{Tab:Rot_Ros}), we found that KIC2837475, KIC7103006, KIC11081729, and KIC11253226 may be in the transition region of the Kraft Break, but none of our stars clearly stands above it. In general, the actual transition into a regime with inefficient braking and disappearance of the thin convective envelope for stars does not appear to be as sharp in terms of $\rm T_{eff}$ (or equivalently of mass) as the classical Kraft Break. On the contrary, this transition would occur over a broader range ($\rm M(M_{\odot}) \simeq 1.5-2.5$), with a strong dependence on several stellar properties (e.g. age, metallicity, internal rotation at birth, angular momentum loss) \citep{Kawaler1988, Aerts2025, AmardMatt2020, AmardRoquetteMatt2020}. Given the uncertainties related with the properties of the dynamos and magnetic fields for the stars in this regime, in the following analysis we will compare the results obtained by including the whole sample of stars on the one hand, and by removing the hottest stars on the other one.

\begin{figure}
    \centering
    \includegraphics[width=\linewidth]{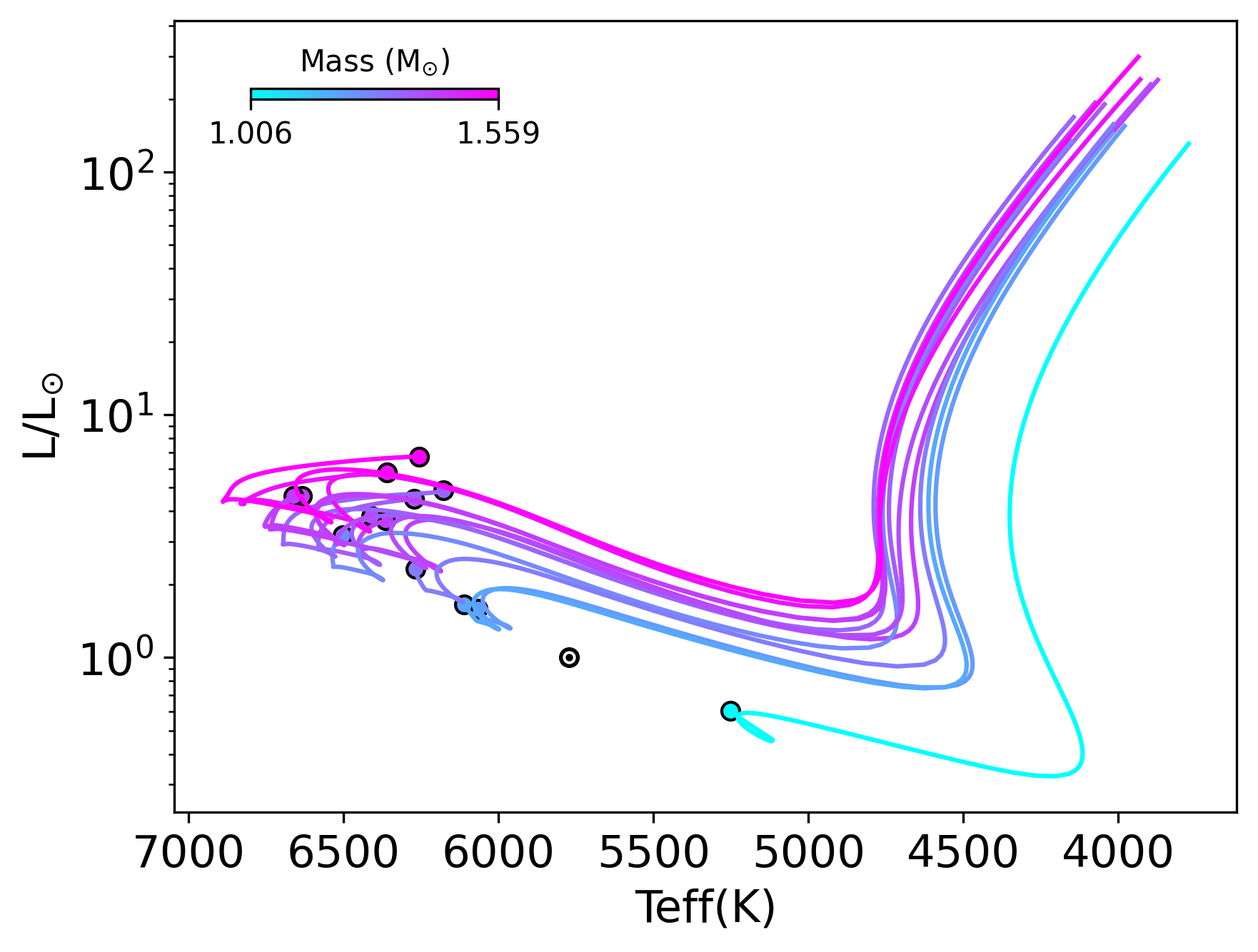}  
    \caption{Hertzsprung-Russell diagram showing the best-fit models' tracks of the 13 \textit{Kepler} LEGACY stars with X-ray detections. The location of the Sun in the $\rm L/L_{\odot}$ vs $\rm T_{eff}(K)$ plane is shown as reference.}
    \label{fig:HR}
\end{figure}

In Fig.~\ref{fig:HR} we show the distribution of the evolutionary tracks in the Hertzsprung-Russell diagram. Among the 13 stars analysed here, KIC8006161 represents the nearest solar analogue, despite its larger metallicity. In the following sections, the relative position of this star compared to the solar one is highlighted to visualize any significant impact due to higher metallicity.

\begin{table*}
\centering
\caption{Fundamental stellar parameters from asteroseismic modelling.}
\label{Tab:params}
\resizebox{\textwidth}{!}{
\begin{tabular}{lcccccccccc} 
 \hline\hline
 KIC ID & $\rm M(M_{\odot})$ & $\rm R(R_{\odot})$ & $\rm L(L_{\odot})$ & $ \rm T_{eff}(K)$ & $\rm [Fe/H]$ & $\rm Age(Gyr)$ & $\rm Age_{gyro}(Gyr)$ & $\rm X_c$ & $\rm R_{bc}/R$ & $\rm M_{ce}/M$ \\ 
 \hline
 \noalign{\vskip 0.5ex}
  KIC2837475 & $1.421 \pm 0.040$  & $1.625 \pm 0.015$ & $4.612 \pm 0.139$ & $6635 \pm 35$ & $-0.44 \pm 0.07$ & $1.74 \pm 0.37$ & $1.12 \pm 0.34$ &0.45 & 0.92 & $2.6 \times 10^{-5}$\\ 
  KIC6508366 & $1.559 \pm 0.040$  & $2.208 \pm 0.018$ & $6.727 \pm 0.189$ & $6255 \pm 41$ & $-0.05 \pm 0.08$ & $2.71 \pm 0.21$ & $0.89 \pm 0.27$ &0.21 &0.84 & $6.86 \times 10^{-4}$\\
  KIC7103006 & $1.520 \pm 0.046$  & $1.973 \pm 0.020$ & $5.750 \pm 0.173$ & $6363 \pm 43$ & $0.05  \pm 0.07$ & $2.06 \pm 0.15$ & $0.37 \pm 0.11$ &0.27 &0.87 & $2.95 \times10^{-4}$\\
  KIC7206837 & $1.394 \pm 0.045$  & $1.586 \pm 0.017$ & $3.823 \pm 0.092$ & $6409 \pm 30$ & $0.11  \pm 0.05$ & $1.93 \pm 0.18$ & $0.80 \pm 0.24$ &0.43 &0.87 & $3.80\times 10^{-4}$\\
  KIC7940546 & $1.340 \pm 0.026$  & $1.925 \pm 0.014$ & $4.958 \pm 0.061$ & $6207 \pm 27$ & $ -0.19 \pm 0.05$ & $3.55 \pm 0.12$ & $3.04 \pm 0.91$ &0.44 &0.81 & $1.767\times10^{-3}$\\ 
  KIC8006161 & $1.006 \pm 0.019$  & $0.936 \pm 0.006$ & $0.599 \pm 0.016$ & $5248 \pm 52$ & $0.34 \pm 0.05$ & $5.50 \pm 0.23$ & $3.28 \pm 0.99$ &0.43 &0.68 & $5.18 \times 10^{-2}$\\ 
  KIC8379927 & $1.194 \pm 0.027$  & $1.146 \pm 0.010$ & $1.652 \pm 0.028$ & $6111 \pm 50$ & $0.05 \pm 0.06$ & $1.55 \pm 0.08$ & $3.33 \pm 1.00$ &0.56 &0.78 & $5.91\times 10^{-3}$\\ 
  KIC9139163 & $1.423 \pm 0.039$  & $1.574 \pm 0.015$ & $3.663 \pm 0.091$ & $6364 \pm 45$ & $0.12 \pm 0.05$ & $2.01 \pm 0.15$ & $1.14 \pm 0.34$ &0.42 &0.86 & $5.60 \times 10^{-4}$\\ 
  KIC9812850 & $1.359 \pm 0.042$  & $1.796 \pm 0.019$ & $4.499 \pm 0.159$ & $6273 \pm 52$ & $-0.12 \pm 0.08$ & $3.31 \pm 0.24$ & $1.42 \pm 0.42$ &0.27 &0.83 & $1 \times 10^{-3}$\\ 
  KIC10454113 & $1.290 \pm 0.025$ & $1.286 \pm 0.010$ & $2.296 \pm 0.033$ & $6265 \pm 38$ & $0.06 \pm 0.06$ & $1.67 \pm 0.12$ & $3.68 \pm 1.10$ &0.51 &0.83 & $2.013 \times 10^{-3}$\\ 
  KIC10644253 & $1.222 \pm 0.029$ & $1.137 \pm 0.010$ & $1.566 \pm 0.032$ & $6054 \pm 54$ & $0.14 \pm 0.06$ & $1.05 \pm 0.12$ & $1.17 \pm 0.35$ &0.61 &0.78 & $6.93 \times 10^{-3}$\\ 
  KIC11081729 & $1.259 \pm 0.049$ & $1.405 \pm 0.018$ & $3.173 \pm 0.087$ & $6502 \pm 36$ & $-0.17 \pm 0.08$ & $2.03 \pm 0.30$ & - &0.45 &0.88 & $2.80 \times 10^{-4}$\\
  KIC11253226 & $1.426 \pm 0.034$ & $1.615 \pm 0.013$ & $4.628 \pm 0.091$ & $6662 \pm 24$ & $-0.49 \pm 0.10$ & $1.85 \pm 0.31$ & - &0.52 &0.93 & $1.8 \times 10^{-5}$\\ 
 \hline
\end{tabular}} 
\tablefoot{The last four columns indicate the age estimate through $\rm P_{rot}-(B-V)_0$ gyrochronology relationship \citep{Barnes2007} for which we assumed a $30\%$ uncertainty, the central abundance of hydrogen, the radius of the base of the convective envelope normalised to the total radius, and the mass contained in the convective envelope, normalised to the total mass.}
\end{table*}

\begin{table}
\caption{Rotational periods, colours and Rossby numbers for our sample of \textit{Kepler} LEGACY stars.}
\label{Tab:Rot_Ros}
\resizebox{\columnwidth}{!}{
\begin{tabular}{lccccccc} 
 \hline\hline
 KIC ID & $\rm P_{rot}(d)$ & $\rm v(km/s)$ & $\rm (B-V)_0$ &$\rm (V-Ks)_0$ & $\rm Ro_{(V-Ks)_0}$ & $\rm Ro_{T_{eff}}$ & $\rm Ro_{m}$ \\
 \hline\noalign{\vskip 0.5ex}
  KIC2837475 & $3.73 \pm 0.39$ & 22.05 & 0.432 & $0.974$ & $0.487$ &  $6.644$ & 13.52\\ 
  KIC6508366 & $3.72 \pm 0.33$ & 30.04 & 0.439 & $1.157$ & $0.437$ & $0.803$  & 1.92\\
  KIC7103006 & $4.66 \pm 0.46$ & 21.43 & 0.521 & $1.13 1$ & $0.556$ & $1.533$ & 3.97\\
  KIC7206837 & $4.05 \pm 0.26$ & 19.82 & 0.449 & $1.242$ & $0.454$ & $1.652$  & 3.28\\
  KIC7940546 & $11.03 \pm 0.82$ & 8.83 & 0.482 & $1.223$ & $1.249$ & $2.038$  & 3.61\\ 
  KIC8006161 & $31.71 \pm 3.19$& 1.49 & 0.845 & $1.834$ & $2.526$ & $1.548$  & 1.58\\ 
  KIC8379927 & $17.09 \pm 1.31$& 3.39 & 0.557 & $1.430$ & $1.718$ & $2.423$  & 3.28\\ 
  KIC9139163 & $6.18 \pm 0.62$& 12.89 & 0.473 & $1.079$ & $0.760$ & $2.043$  & 4.03\\ 
  KIC9812850 & $5.17 \pm 0.68$& 17.58 & 0.445 & $1.177$ & $0.601$ & $1.189$  & 2.34\\ 
  KIC10454113 & $14.69 \pm 1.01$& 4.43 & 0.512 & $1.362$ & $1.536$ & $3.285$ & 5.13\\ 
  KIC10644253 & $10.88 \pm 0.71$& 5.29 & 0.583 & $1.366$ & $1.135$ & $1.351$ & 1.89\\ 
  KIC11081729 & $2.86 \pm 0.29$& 24.86 & 0.399 & $0.977$ & $0.373$ & $1.947$  & 2.95\\
  KIC11253226 & $3.73 \pm 0.32$& 21.91 & 0.378 & $0.976$ & $0.487$ &  $8.407$ & 16.23\\ 
 \hline
\end{tabular}}
\tablefoot{KIC ID (col.~1), rotational period \citep{Santos2021} (col.~2), $\rm (B-V)_0$ colour (this work, col.~3), $\rm (V-Ks)_0$ colour (this work, col.~4), Rossby number computed by following \citet{Wright2018} (col.~5) and \citet{Cranmer2011} (col.~6), and from stellar models (col.~7).}
\end{table}

\section{Study of the activity-rotation-age relationship}
\label{Sec:relationhips}

After having determined both the X-ray luminosities (see Sect.~\ref{Sec:eROSITA}) and the fundamental parameters (see Sect.~\ref{Sec:params}) for our sample of 13 solar-like stars, we study the relationships linking the stellar magnetic activity with the age on one hand, and with the rotational period or dynamo efficiency, by means of the Rossby number, on the other one. This last quantity is defined as $\rm Ro = P_{rot}/\tau_{conv}$, where $\rm P_{rot}$ is the surface rotational period and $\rm \tau_{conv}$ is the convective turnover timescale, namely the timescale needed for a convective bubble to travel from the base to a certain height within the convective regions. Several methods have been proposed to determine this non-observable quantity at different heights within convective envelopes, which are usually linked to the stellar colours or effective temperatures \citep[e.g.][]{Noyes1984, Wright2011, Cranmer2011}. In this work, we computed $\rm \tau_{conv}$ with two approaches. The first one follows the empirically derived prescription of \cite{Wright2018} that is based on the stellar $\rm (V-Ks)_0$ colour and is valid for $\rm 1.1 < (V-Ks)_0 < 7$. We determined this last quantity by taking the V and Ks magnitudes from SIMBAD \citep{SIMBAD2000}, and computed $\rm (V-Ks)_0 = (V-Ks) + (-A_V + A_{Ks})$, where $\rm A_{Ks} = 0.161 \times E(B-V)$ is derived from the 3D dustmaps for Galactic reddening \citep{Green2018}, and $\rm A_V = 3.1\times E(B-V)$ \citep{Savage1979}. 
Notably, Table~\ref{Tab:Rot_Ros} shows that the $\rm (V-Ks)_0$ colours for KIC2837475, KIC9139163, KIC11081729, and KIC11253226 are slightly below the lower limit required for applying the prescription of \citet{Wright2018}. In this case, the $\rm \tau_{conv}$ is potentially overestimated, as indicated by Fig.~7 in \citet{Wright2011}. The second approach follows the parametrization of Zero Age Main Sequence (ZAMS) stellar models from \citet{Gunn1998} by \cite{Cranmer2011}, that is valid for $\rm 3300 \lesssim T_{eff}(K) \lesssim 7000$. To compute $\rm \tau_{conv}$ with this second approach, we used the effective temperature values in Table~\ref{Tab:params}. A comparison between the convective turnover timescales computed with the approaches mentioned above for each target is shown in Fig.~\ref{Fig:tau_conv_comp}. For the sake of completeness, we computed $\rm \tau_{conv}$ at the base of the convective zone derived from best-fit stellar models in Table~\ref{Tab:params}, by adopting the convention of \citet{Montesinos2001} and include their values in Fig.~\ref{Fig:tau_conv_comp}. The corresponding Rossby numbers for the three diverse approaches are listed in Table~\ref{Tab:Rot_Ros}, while in Fig.~\ref{fig:Ro_normalised_RoSun} we compare their values consistently normalised by the solar one ($\rm Ro_{\odot, (V-Ks)_0} = 2.306$, $\rm Ro_{\odot, (Teff)} = 1.95$, $\rm Ro_{\odot, m} = 2.19$). We recall that for the calibration of the activity-rotation-age relationships in the following we will use the first and second approach to compute $\rm \tau_{conv}$.

With all the key quantities determined for our sample of stars, we studied its impact on some relationships which have been intensively investigated in the literature to provide key indications on the evolution of stellar activity as a function of the age and Rossby number, which are: $\rm Log(L_x/R^2)-Log(Age)$ \citep[e.g.,][]{Booth2017}, $\rm Log(R_x)-Log(Age)$ \citep[e.g.,][]{Jackson2012}, and $\rm Log(R_x)-Log(Ro)$ \citep[e.g.,][]{Wright2011, Magaudda2020, Johnstone2021}, with $\rm R_x = L_x/L_{bol}$. In the following sections, we present the methodology and results obtained when revisiting these relationships.

\subsection{Log(L$\rm _x$/R$^2$) vs Log(Age)}
\label{Sect:Lx_Age_Booth}

\citet{Booth2017} (hereafter B17) calibrated the activity-age relationship by considering a sample of 14 stars older than about $\rm 1$ Gyr and X-ray detections from either \textit{Chandra} or \textit{XMM-Newton} observations. For 6 out of 14 stars (16 CygA, Proxima Cen, Alpha CenB, KIC7529180, KIC9955598, KIC10016239) the ages have been asteroseismically determined. With this sample of stars, they performed an orthogonal distance regression to fit the decimal logarithm of the X-ray luminosity normalized to the stellar radius versus the decimal logarithm of the age, with the units of the three quantities above given in erg/s, solar radii and years, respectively. They found a relatively steep linear dependence, of the form: 
\begin{equation}
\rm Log(L_x/R^2) = -2.80~(\pm 0.72) \times Log(Age) + 54.65~(\pm 6.98).  
\end{equation}

In our work, we revisited this relationship by joining our 13 \textit{Kepler} LEGACY stars to the 6 stars of B17 with asteroseismically derived ages. For consistency with respect to the method used to determine the stellar age, we decided to not include the rest of the B17 sample. Since the X-ray luminosities in B17 were calculated in the $\rm 0.2$-$2.0$ keV energy band, to ensure a more consistent analysis and comparison with respect to the X-ray theoretical tracks computed in Sect.~\ref{Sec:models}, we converted their X-ray fluxes into the ROSAT energy band ($0.1$-$2.4$ keV) by means of the WebPIMMS\footnote{\url{https://heasarc.gsfc.nasa.gov/cgi-bin/Tools/w3pimms/w3pimms.pl}} tool. We first rescaled the X-ray fluxes in B17 with updated distances from \textit{Gaia} EDR3. We thus provided the rescaled fluxes as input to WebPIMMS assuming an APEC model with solar abundance and uniform coronal temperature $\rm Log(T(K)) = 6.50$ \citep{Booth2017}. We also include the Sun among our sample of stars. In this case, we took the $\rm L_x$ from \citet{Wright2011} and the relative uncertainty from B17. The X-ray luminosities derived from the converted fluxes of the B17 subsample are indicated in Table~\ref{Tab:resc_flux}.\\

\begin{figure}
    \centering
    \includegraphics[width=\linewidth]{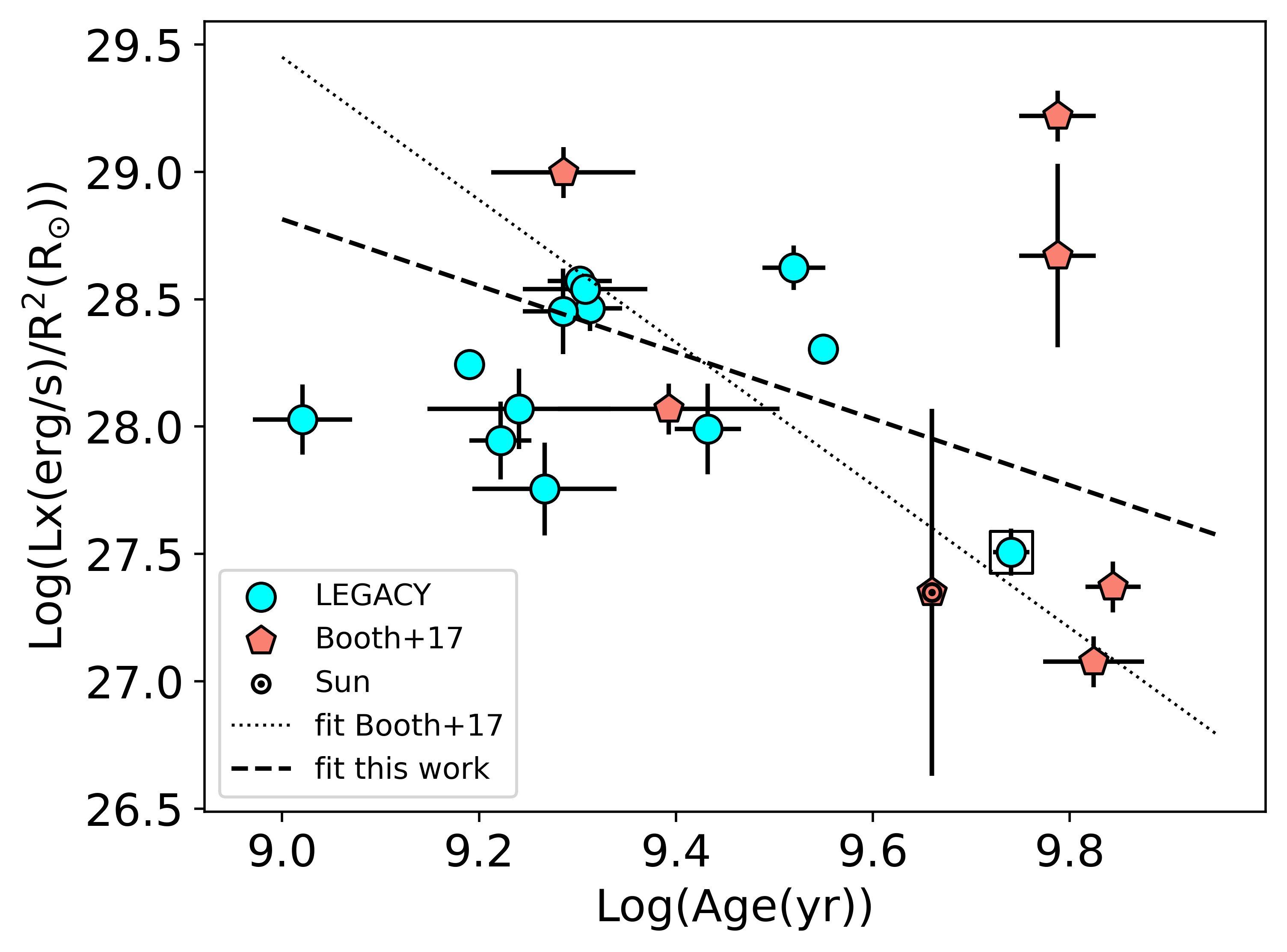}  
    \caption{Distribution of 13 stars from the \textit{Kepler} LEGACY (cyan dots) and 7 stars from B17 (orange pentagons) including the Sun (indicated by the usual symbol) in the $\rm Log(L_x/R^2)$-$\rm Log(Age)$ plane. The empty square highlights the position of KIC8006161. The dotted blue and dashed black lines show the fit obtained in B17 and in this work, respectively.}
    \label{fig:LogLxR2_LogAge}
\end{figure}

With this final sample of 20 stars (including the Sun), we performed an orthogonal distance regression on $\rm Log(L_x/R^2)$-$\rm Log(Age)$, with a weighted linear fit, by means of the SciPy ODR method \citep{SciPy2020}, finding:
\begin{equation}
\rm Log(L_x/R^2) = -1.31~(\pm 0.46)\times Log(Age) + 40.60~(\pm 4.37),  
\end{equation}
\noindent
where the uncertainties on the fitting parameters are expressed as one standard deviation. To test the goodness of the fit we computed the reduced chi-squared, $\rm \chi^2_{red} = 26.9$. This value clearly indicates a poor fit, which is mostly due to the very scattered distribution of the points in the $\rm Log(L_x/R^2)$-$\rm Log(Age)$ plane, as shown in Fig.~\ref{fig:LogLxR2_LogAge}. 

If we attempt to make a comparison with the fit found in B17, we can see that a flatter slope is obtained in our case. It is worth noticing that the average age of the stars considered in B17 for the fit is approximately 5 Gyr, which is about twice the one relative to our sample of 20 stars ($\sim 2.5$ Gyr). As indicated by \citet{VanSaders2016}, low-mass stars more evolved than a certain stage may transition into a ``weakened magnetic braking regime'' (wmb), as a result of an efficiency loss of the dynamo that sustains the stellar magnetic field. This transition may occur whenever the stellar Rossby number becomes larger than a certain threshold value $\rm Ro_{cri}$ \citep[e.g., $\rm Ro_{cri} = (\rm 0.92 \pm 0.01)~Ro_{\odot}$, where $\rm Ro_{\odot}$ is the Sun's Rossby number,][]{Metcalfe2024}. In this scenario, the stellar spin-down slows considerably, resulting in the stalling of the surface rotational period as stars age. Understanding the consequences for magnetic activity is challenging, given that evolved stars are generally less active, hindering the identification of a possible transition into a diverse regime. By comparing the Rossby number of our 13 \textit{Kepler} LEGACY stars in Table ~\ref{Tab:Rot_Ros} with respect to $\rm Ro_{cri}$ (which is 2.4 or 1.9 for $\rm \tau_{conv}$ computed as in \citet{Wright2018} or \citet{Cranmer2011}, respectively), we get that in the first case only one star would have already transitioned into the wmb regime; while in the second case, seven stars have $\rm Ro$ larger than 1.9. Similarly, if we compare the Rossby number of the additional six stars from B17 with the two values of $\rm Ro_{cri}$, in both cases we get that only two stars would have transitioned into a wmb regime. Therefore, despite the B17 stars in Table~\ref{Tab:resc_flux} having older ages on average, from a magnetic activity point of view they populate a similar parameter space with respect to the 13 \textit{Kepler} LEGACY stars, at the turn between the potential transition. In this context, while the findings in B17 suggest a steepening of the activity-age relationship in evolved low-mass stars compared to their younger counterparts \citep[e.g.][]{Jackson2012}, our results indicate a more compatible trend. This result is in agreement with the recent work of \citet{Aldarondo2025}, where the authors derived a shallower slope compared to previous findings.

It is important to note that the uncertainties on the fitting parameters obtained by means of the orthogonal distance regression in the case of a very sparse sample of objects, as the one analysed here, may be significantly underestimated. To derive more robust estimates we implemented a bootstrapping method on the basis of the SciPy ODR, and used a resampling with replacement approach for 25000 iterations. As a result, we obtained:
\begin{equation}
\rm Log(L_x/R^2) = -1.22~(\pm 1.11)\times Log(Age) + 39.85~(\pm 10.46),  
\label{Eq:boot}
\end{equation}

\noindent
where the parameters' values and relative uncertainties indicate the mean and standard deviation of the distribution of best-fitting slopes and intercepts over the total number of iterations. The larger uncertainties found with the bootstrapping method further demonstrate that it is not possible to constrain the $\rm Log(L_x/R^2)-Log(Age)$ relationship with our global sample of well-characterized 20 stars. Clearly, a statistically more significant number of targets is needed to derive a more precise relationship, and to investigate whether different magnetic activity regimes can be unambiguously identified. In Appendix~\ref{App:Lx_Age_only_cool_stars},we show the results obtained by excluding stars with $\rm T_{eff}(K) > 6250$ from the fit, finding a similar mean value for the slope of the relationship, and more degraded quality of the fit.

In the following section, we attempt to study the activity-age relationship in a broader context and to break it down by investigating its dependence on the stellar bolometric luminosity and internal structure properties (as $\rm \tau_{conv}$).

\subsection{Log(R$\rm _x$) vs Log(Age)}

To explore the dependence of the activity-age relationship with respect to mass in late-type stars, \citet{Jackson2012} (hereafter J12) selected 717 stars from 13 open clusters ($\rm \alpha$-Persei, Blanco 1, Hyades, IC2391, IC2602, NGC1039, NGC2516, NGC2547, NGC3532, NGC6475, NGC6530, Pleiades, Praesepe) and split the sample into seven $\rm (B-V)_0$ colour bins, with $\rm 0.29 \leq (B-V)_0 < 1.41$. By fitting a broken power law in the $\rm Log(R_x)$-$\rm Log(Age)$ plane, and assuming a constant slope in the saturated regime, they found a decreasing trend of the X-ray-to-bolometric luminosity ratio at saturation ($\rm Log(R_{x, sat})$) across the $\rm (B-V)_0$ range, from $10^{-3.15}$ to $\rm 10^{-4.28}$. They further identified saturation-regime turn-off ages consistent with a scatter around $\rm \sim 100$ Myr, this time without a clear trend with respect to the stellar mass.

In our work, we revisited the $\rm Log(R_x)$-$\rm Log(Age)$ relationship by employing a similar approach to J12, with additional data from our global sample of 20 stars (13 from the \textit{Kepler} LEGACY, 6 from B17 and the Sun). As a first step, we updated the cluster ages by considering more recent estimates from \citet{Bossini2019} for $\rm \alpha$-Persei, Blanco 1, IC2391, IC2602, NGC1039, NGC2516, NGC2547, NGC3532, NGC6475, NGC6530. For NGC6530, we kept the age estimate in J12 and updated the uncertainties by accounting for the work of \citet{Prisinzano2019}. Similarly, for Hyades, Pleiades and Praesepe we updated the uncertainties by accounting for the spread in age estimates in \citet{Gossage2018}. We updated as well the stellar distances, which were inferred using the geometric Bayesian method described by \citet{BailerJones2021}, which accounts for parallax measurement uncertainties and the spatial distribution of stars, using Gaia EDR3 data. Given the significant uncertainties in the distance estimates of the NGC6530 components, we decided to exclude this cluster from the fitting procedure.

Differently from J12, we adopted non-binned data to perform the fit, to consistently account for the contribution of clusters and single stars. The total range of variation for the $\rm (B-V)_0$ colour is slightly larger than the one in J12 ($\rm [0.29,1.83)$), to include Proxima Cen. We adopted the SciPy ODR method \citep{SciPy2020} to fit the following piecewise linear curve:
\begin{equation}
\rm Log(R_x) = 
\begin{cases} 
\rm Log(R_{x, sat})~ &\text{for} \quad \rm t \leq \tau_{\rm sat},   \\ 
\rm Log(R_{x,sat}) + \alpha (Log(t) - Log(\tau_{sat}))~ &\text{for} \quad \rm t > \tau_{\rm sat},
\end{cases}
\label{Eq:Rx_Age}
\end{equation}
\noindent
in which we considered $\rm Log(R_{x,sat})$, $\rm \alpha$ and $\rm \tau_{\rm sat}$ as free parameters, and $\rm t$ is the stellar age. The resulting parameters obtained from the fit on the whole sample of stars are: $\rm Log(R_{x, sat}) = -3.55 \pm 0.03$, $\rm \alpha = -1.65 \pm 0.11$; $\rm Log(\tau_{sat}) = 8.15 \pm 0.04$. The associated $\rm \chi^2_{red} = 19.74$ shows also in this case that the scatter of the points hinders the precision of the fit. In Fig.~\ref{fig:LogRx_LogAge}, we show the distribution of stars on the $\rm Log(R_x)$-$\rm Log(Age(yr))$ plane with the corresponding fit, indicated by the black dashed line. 

To investigate the dependence of Eq.~\ref{Eq:Rx_Age} on the stellar mass, or equivalently on the $\rm (B-V)_0$ colour, we split the sample in 9 bins, and carried out the same fitting as described above. In Fig.~\ref{Fig:Rx_Age_BV} we show the distribution of stars for each bin with the relative fit, while in Table~\ref{Tab:Rx_Age} the corresponding results are collected. In Fig.~\ref{fig:params_trends} we observe that $\rm Log(R_{x,sat})$ shows a non-decreasing trend as $\rm (B-V)_0$ increases, whereas $\rm Log(\tau_{\rm sat})$ and $\rm \alpha$ shows a quasi-specular behaviour for $\rm 0.290 \leq (B-V)_0 \leq 0.675$, with global maximum (minimum) around $\rm (B-V)_0 \sim 0.6$, while they both increase for $\rm (B-V)_0 > 0.675$. By comparing our results with the one in J12, we firstly highlight the different methods (ODR here and a modified \citet{Fasano1988} in J12) and number of $\rm (B-V)_0$ colour bins adopted for the fit. We decided to split the global sample in 9 instead of 7 bins in an attempt to better explore the properties of the \textit{Kepler} LEGACY stars, whose $\rm (B-V)_0$ colours especially cluster in the range 0.29-0.565. If we compare the $\rm \alpha$ indexes in absolute values, ours tend to be smaller in $\rm 0.290 \leq (B-V)_0 <0.450$, $\rm 0.790 \leq (B-V)_0 <0.935$ and $\rm (B-V)_0 \geq 1.275$, and larger in the rest of the bins. To test the impact of stars with $\rm T_{eff}(K) > 6250$, we re-performed the fitting procedure described above by excluding these stars. The results found in this case appear compatible with the ones discussed in this section for the whole sample.

In Sect.~\ref{Sec:models}, we will compare and discuss the X-ray luminosity tracks computed by following the results in J12 and the ones found in this work. 

\begin{table}
\noindent
\captionof{table}{Results of the fitting procedure of Eq.~\ref{Eq:Rx_Age}
on the global sample of stars from J12, \textit{Kepler} LEGACY and B17.}
\centering
\resizebox{\columnwidth}{!}{
\begin{tabular}{lcccc}
 \hline\hline
 Colour range & $\rm Log(R_{x,sat})$ & $\rm Log(\tau_{sat})$ & $\rm \alpha$ & $\rm \chi^2_{red}$ \\
 \hline\noalign{\vskip 0.5ex}
 $\rm 0.290 \leq (B-V)_0 < 0.450$   & $-4.33 \pm 0.18$ & $7.72 \pm 0.27$ & $-0.78 \pm 0.12$ & 10.33\\
 $\rm 0.450 \leq (B-V)_0 < 0.480$   & $-4.33 \pm 0.11$ & $8.20 \pm 0.25$ & $-0.83 \pm 0.26$ & 7.48\\
 $\rm 0.480 \leq (B-V)_0 < 0.520$   & $-4.27 \pm 0.08$ & $8.47 \pm 0.15$ & $-1.15 \pm 0.22$ & 9.25\\
 $\rm 0.520 \leq (B-V)_0 < 0.565$  & $-4.20 \pm 0.07$ & $8.47 \pm 0.08$ & $-1.78 \pm 0.29$ & 6.80\\
 $\rm 0.565 \leq (B-V)_0 < 0.675$ & $-3.78 \pm 0.06$ & $8.29 \pm 0.05$ & $-1.84 \pm 0.23$ & 7.22\\
 $\rm 0.675 \leq (B-V)_0 < 0.790$ & $-3.50 \pm 0.09$ & $8.04 \pm 0.12$ & $-1.52 \pm 0.30$ & 12.79\\
 $\rm 0.790 \leq (B-V)_0 < 0.935$ & $-3.30 \pm 0.06$ & $7.95 \pm 0.08$ & $-1.35 \pm 0.15$ & 9.59\\
 $\rm 0.935 \leq (B-V)_0 < 1.275$ & $-3.22 \pm 0.05$ & $8.05 \pm 0.09$ & $-1.13 \pm 0.20$ & 7.41\\
 $\rm 1.275 \leq (B-V)_0 < 1.830$  & $-3.15 \pm 0.06$ & $8.19 \pm 0.26$ & $-0.51 \pm 0.26$ & 6.19\\
 \hline
\end{tabular}}
\label{Tab:Rx_Age}
\end{table}

\subsection{Log(R$_{\rm x}$) vs Log(Ro)}

After having examined how our sample of stars with asteroseismically derived ages influences the relationship between stellar X-ray luminosity and age, we then analyzed the relationship between X-ray luminosity and Rossby number. In particular, we studied the impact on the $\rm Log(R_x)$-$\rm Log(Ro)$ relationship, which has been widely investigated in the literature \citep[e.g.][]{Wright2011, Wright2018, Johnstone2021}. 

As in the previous section, we considered a sample of stars for which this relationship has been thoroughly explored in past studies, which is the one of \citet{Wright2011} (hereafter W11). To this sample, we added the M-dwarf stars from \citet{Wright2018} (hereafter W18), the 13 \textit{Kepler} LEGACY stars analysed in this work, and the sample from B17 listed in Table~\ref{Tab:resc_flux}. To ensure a consistent investigation across the whole sample, we updated and recomputed some quantities in the W11 catalog: in detail, we updated the stellar distances by adopting Gaia DR3 parallaxes with the \citet{BailerJones2021} method, and rescaled the X-ray luminosities accordingly; we computed bolometric luminosities on the basis of the 2MASS Ks magnitude and $\rm (V-Ks)_0$ colours as described above; additionally, we searched for $\rm T_{eff}$, $\rm log~g$, and $\rm [Fe/H]$ estimates in Gaia DR3, specifically from the GSP-Phot module which derives parameters using low-resolution BP/RP spectra, Gaia photometry, and parallaxes \citep{Creevey2023, GaiaDR3Overview2023}. For most of the stars in the W11 catalog (94\%) we found a match in Gaia DR3. To get more reliable estimates of $\rm T_{eff}$, $\rm log~g$, and $\rm [Fe/H]$, we also searched into the APOGEE catalogue \citep[Apache Point Observatory Galactic Evolution Experiment, ][]{Majewski2017, Abdurrouf2022}, where the parameters determination is based on high-resolution spectroscopic data. Nevertheless, we found a match only for 31\% of the stars. Therefore, we decided to adopt the parameters derived from the Gaia GSP-Phot method for our analysis. We finally cleaned the revised W11 sample by filtering on Gaia astrometric and photometric quality flags, retaining sources with RUWE $<1.4$, ipd\_frac\_multi\_peak $\leq 1$,  rv\_amplitude\_robust $> 9$ and G magnitude $< 11$ \citep{Gossage2025}. The cleaned and revised W11 sample counts 411 stars.

We attempted a revision analogous to the one of W11 for the sample in W18. Nevertheless, in this case we found that only for 2 out of 16 stars with X-ray detections the Gaia GSP-Phot parameters are available. We thus preferred to keep the $\rm R_x$ estimates of W18, while we retrieved the $\rm T_{eff}$ values from \citet{Gaidos2014} for LSPM J0617+8353, \citet{Hejazi2020} for GJ 699, \citet{Hejazi2022} for LSPM J0024+2626 and LSPM J2012+0112, \citet{Kuznetsov2019} for GJ 3253, \citet{MasBuitrago2024} for GJ 3323 and GJ 1256, \citet{Kumar2023} for GJ 3417, \citet{Perdelwitz2024} for GJ 754, \citet{Antoniadis2024} for GJ 551, \citet{Jahandar2025} for GJ 1286, and \citet{Wang2022} for LSPM J0501+2237, respectively.

With this revised global sample of stars, we proceeded with the fit of a piecewise linear curve, of the form:
\begin{equation}
\rm Log(R_x) = 
\begin{cases} 
&\rm a \times Log(Ro) + b,~~\text{if}~~ \rm Ro \leq Ro_{sat},   \\ 
&\rm a \times Log(Ro) + b + c\times (Log(Ro) - Log(Ro_{sat})), \\ & ~~~~~~~~~~~~~~~~~~~~~~~~~~~~~~~~\text{if} ~~\rm Ro > Ro_{sat}, 
\label{Eq:Rx_Ro}
\end{cases}
\end{equation}
\noindent
where $\rm Ro_{sat}$ represents the value of the Rossby number separating the saturated and unsaturated regimes. 

\begin{figure}
\centering
\begin{subfigure}{
\includegraphics[width=0.45\textwidth]{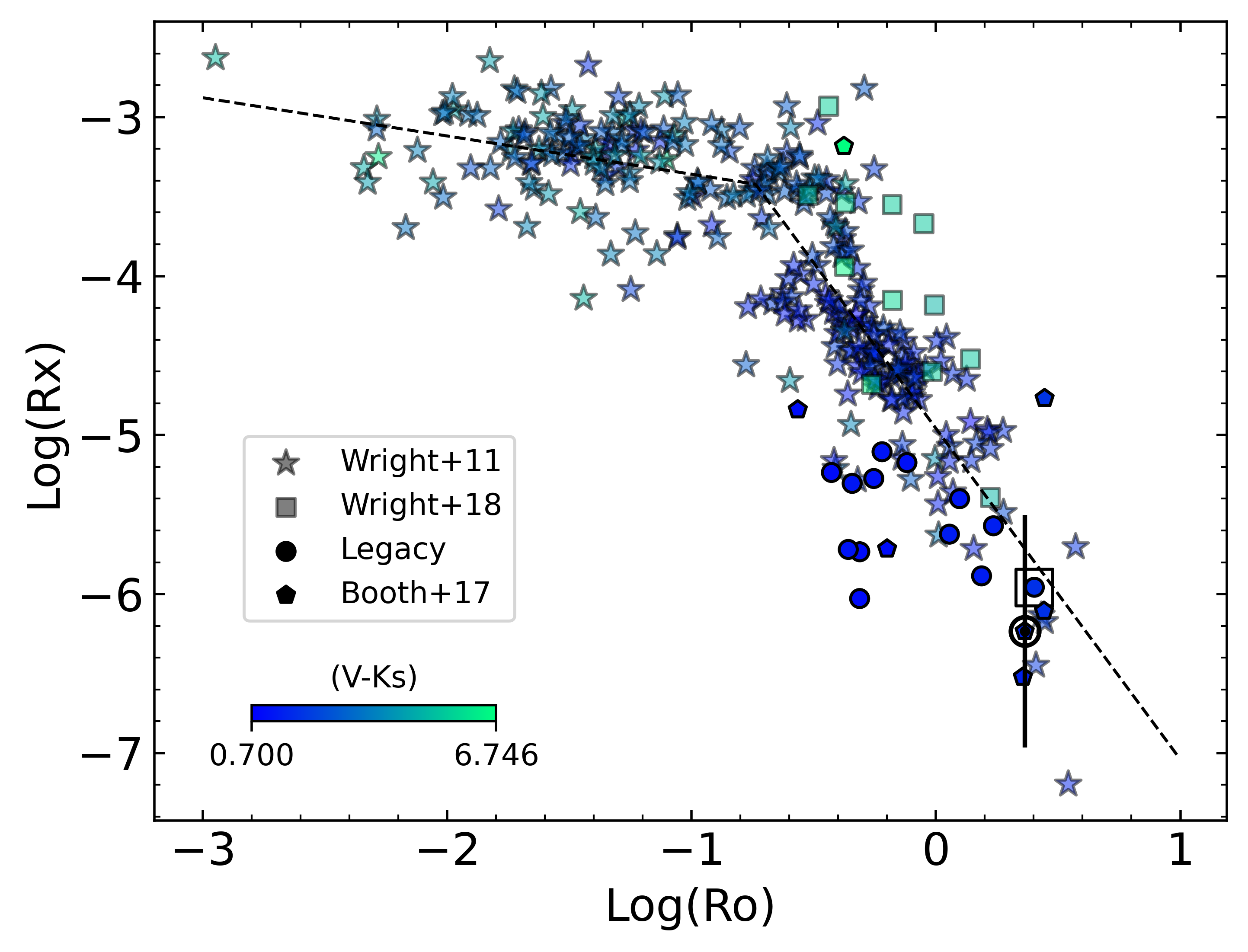}
}
\end{subfigure}
\vspace{-1.2\baselineskip} 
\par 
\begin{subfigure}{
\includegraphics[width=0.45\textwidth]{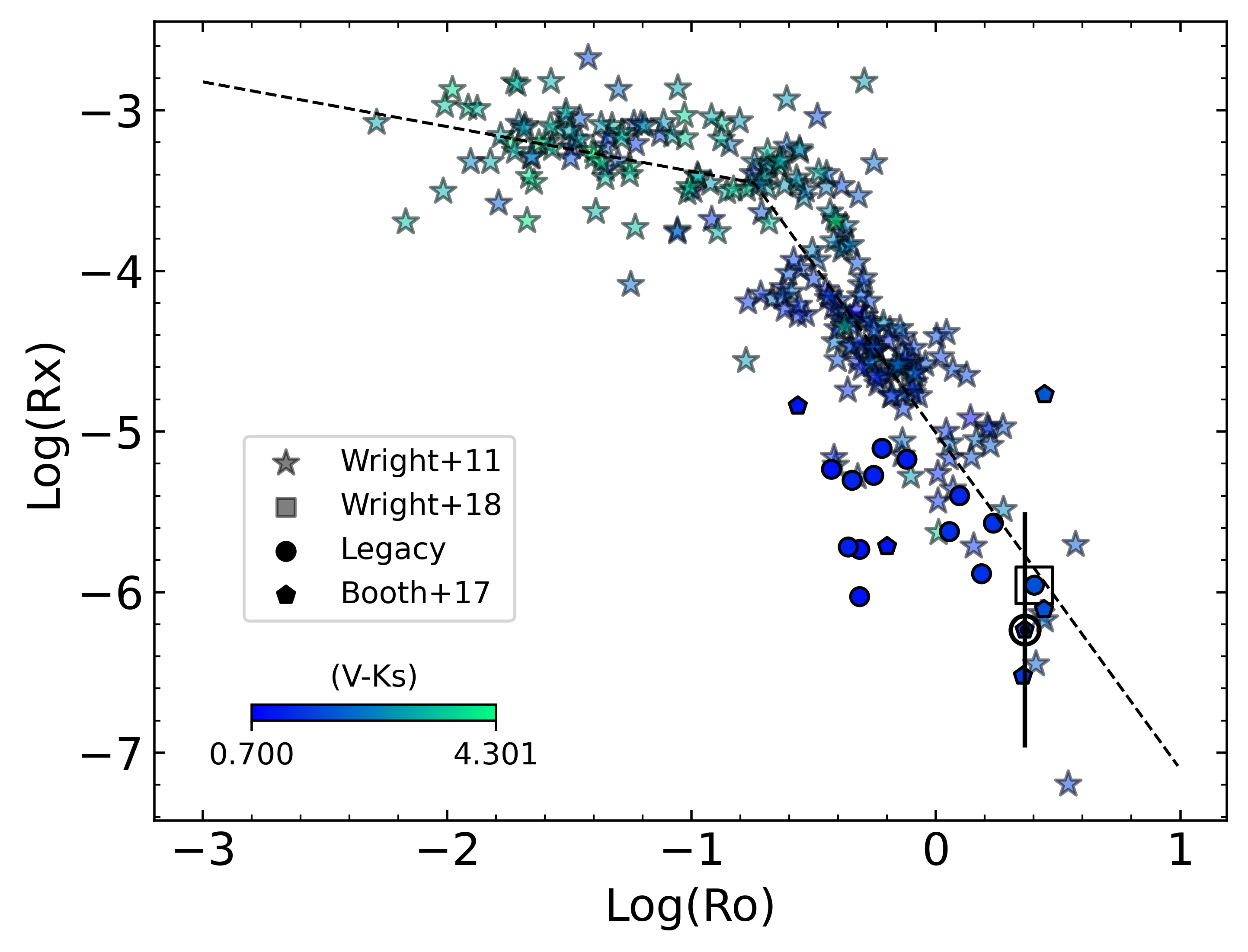}
}
\end{subfigure}
\captionsetup{skip=0.0pt}
\caption{Stars in the $\rm Log(R_x)$-$\rm Log(Ro)$ plane, with colour coded symbol according to the $\rm (V-Ks)_0$ colour. $\rm Ro$ is computed with $\rm \tau_{conv}$ as in W18. The empty square shows the position of KIC8006161. The black dashed line shows the fit to the stars. Stars with spectral type from M to F are shown in the top panel, while stars from K to F are shown in the bottom one.}
\label{Fig:Rx_Ro_VKs}
\end{figure}

\begin{figure}
\centering
\begin{subfigure}{
\includegraphics[width=0.45\textwidth]{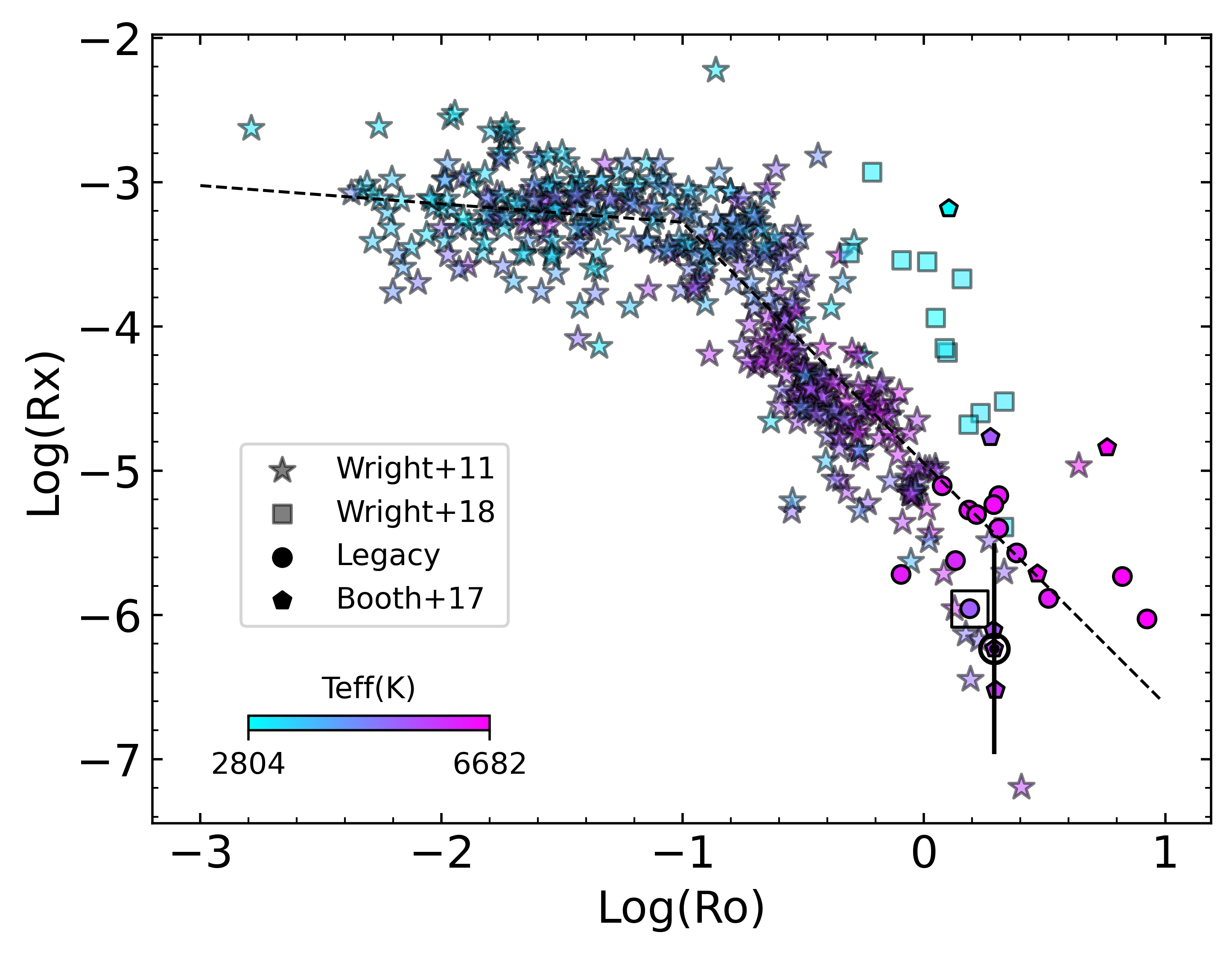}
}
\end{subfigure}
\vspace{-1.2\baselineskip} 
\par 
\begin{subfigure}{
\includegraphics[width=0.45\textwidth]{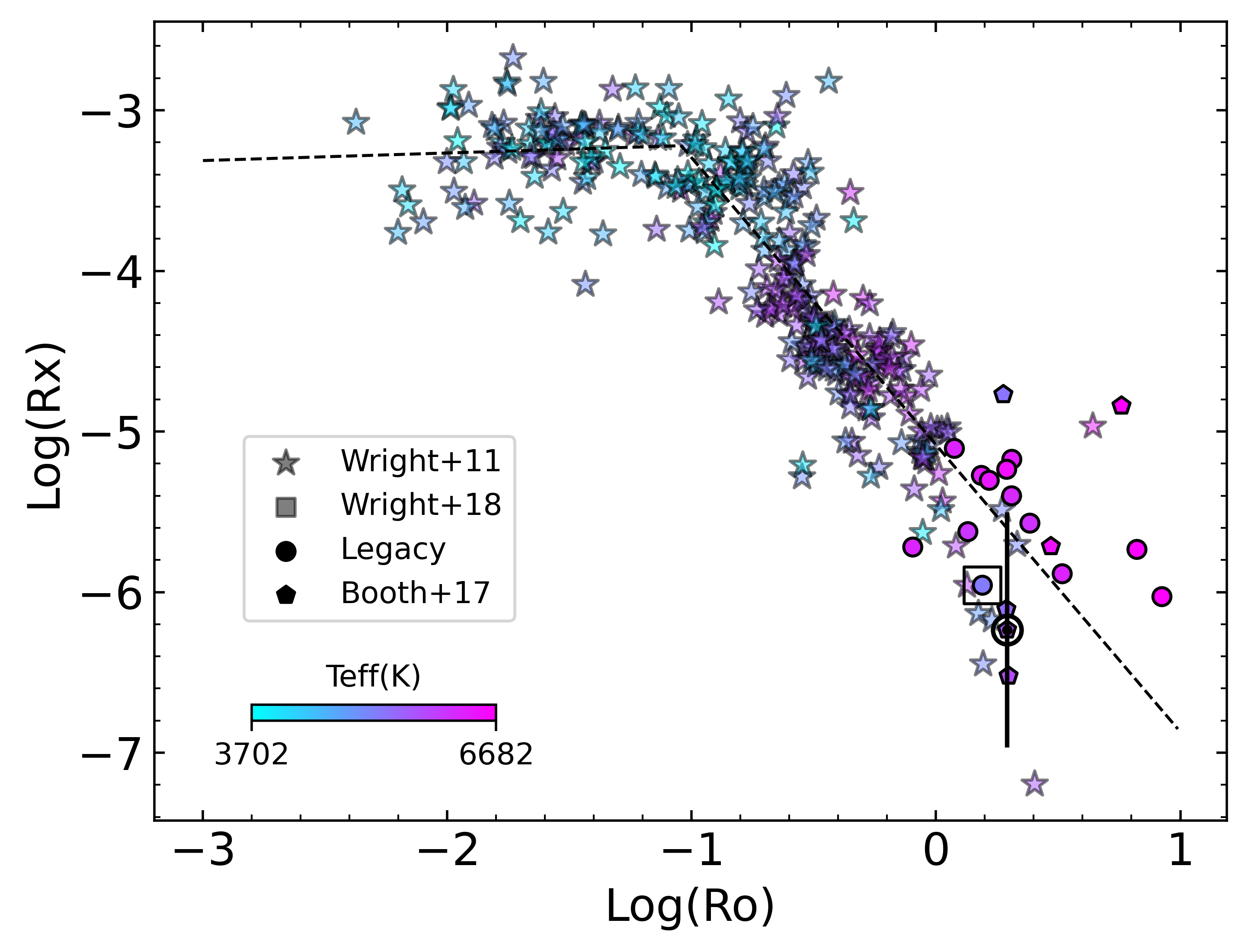}
}
\end{subfigure}
\captionsetup{skip=0.0pt}
\caption{Stars in the $\rm Log(R_x)$-$\rm Log(Ro)$ plane, with colour coded symbol according to $\rm T_{eff}$. The meaning of symbols and lines is the same as for Fig.~\ref{Fig:Rx_Ro_VKs}}
\label{Fig:Rx_Ro_Teff}
\end{figure}

To compute the Rossby number, it is necessary to get an estimate of the convective turnover timescale ($\rm \tau_{conv}$). In this work, we considered the prescription of W18, based on the $\rm (V-Ks)_0$ colour, and the one of \citet{Cranmer2011}, based on the effective temperature. Both methods provide an estimate of the convective turnover timescale at the base of the convective zone, but the correspondence between the two is not homogeneous, and using one or the other approach in the fitting procedure may lead to different results. We therefore performed two fits of Eq.~\ref{Eq:Rx_Ro}, one with $\rm Ro_{(V-Ks)_0}$ and one with $\rm Ro_{T_{eff}}$, by means of the SciPy curve\_fit method \citep{SciPy2020}, for which we considered ``a'', ``b'', ``c'', and $\rm Log(Ro_{sat})$ as free parameters. Since the majority of the stars with asteroseismically characterized ages analysed in this work have spectral types F and G, we further split our global sample of stars applying a cut in $\rm T_{eff} \geq 3700$ K, to separate M-dwarfs from FGK stars, and studied the results of the fit with respect to a relatively more restricted range of spectral types. The results relative to these fitting procedures are presented in Table~\ref{Tab:Rx_Ro}. As expected, depending on the prescription used for the computation of $\rm \tau_{conv}$, the fitted parameters assume different values, with the most significant differences found for $\rm Log(Ro_{sat})$. The cut in $\rm T_{eff}$ has a major impact on the fit procedure based on $\rm Ro_{T_{eff}}$: in particular, while the slope ``a'' in the saturated regime is negative when the whole sample is considered, it becomes positive (in average value) for the restricted sample. The slope ``b'' instead becomes steeper in the unsaturated regime. In Figs.~\ref{Fig:Rx_Ro_VKs} and \ref{Fig:Rx_Ro_Teff}, the distribution of stars in the $\rm Log(R_x)$-$\rm Log(Ro)$ plane are shown, for $\rm \tau_{conv}$ computed on the basis of $\rm (V-Ks)_0$ and $\rm T_{eff}$, respectively, together with the result of the fit indicated by the dashed black line. For each figure, both the results obtained by considering the whole, or restricted sample of stars are presented. Also for the $\rm Log(R_x)$-$\rm Log(Ro)$ we tested the impact of removing \textit{Kepler} LEGACY stars with $\rm T_{eff}(K) > 6250$ from the fitting procedure, finding results (see Appendix~\ref{App:Rx_Ro}) similar to the ones presented above.

\begin{table}
\noindent
\captionof{table}{Results of the fitting procedure on Eq.~\ref{Eq:Rx_Ro} for different prescriptions of the Rossby number on the global sample of stars (all) and on stars with spectral type from K to F (K-F) from W11, W18, \textit{Kepler} LEGACY and B17.}
\centering
\resizebox{\columnwidth}{!}{
\begin{tabular}{|l|c|c|c|c|}
 \cline{2-5}
 \multicolumn{1}{c|}{} & a & b & c & $\rm Log(Ro_{sat})$ \\
 \hline
 $\rm Ro_{(V-Ks)_0}$, all & $-0.24 \pm 0.09$  & $-3.59 \pm 0.14$ & $-1.84 \pm 0.15$ & $-0.74 \pm 0.05$ \\
 $\rm Ro_{(V-Ks)_0}$, K-F & $-0.27 \pm 0.12$  & $-3.65 \pm 0.17$ & $-1.81 \pm 0.17$ & $-0.74 \pm 0.06$ \\
 \hline
 $\rm Ro_{T_{eff}}$, all & $-0.12 \pm 0.09$  & $-3.40 \pm 0.15$ & $-1.54 \pm 0.12$ & $-1.00 \pm 0.05$ \\
 $\rm Ro_{T_{eff}}$, K-F & $0.04 \pm 0.13$  & $-3.17 \pm 0.04$ & $-1.83 \pm 0.15$ & $-1.04 \pm 0.05$ \\
 \hline
\end{tabular}}
\label{Tab:Rx_Ro}
\end{table}

We also investigated how the $\rm Log(R_x)$–$\rm Log(Ro)$ relationship depends on stellar metallicity $\rm [Fe/H]$. For the stars in W11 and B17, we adopted the $\rm [Fe/H]$ values from \textit{Gaia} DR3, while for the \textit{Kepler} LEGACY sample we used the values listed in Table~\ref{Tab:params}. Since most stars from W18 lack metallicity measurements in \textit{Gaia} DR3, we decided to not include them in this part of the analysis. We divided the overall sample into four metallicity bins spanning a range $\rm -1.5 \leq [Fe/H] \leq 0.5$, selected to ensure a statistically meaningful number of stars in each bin (N=80 for $\rm -1.5 \leq [Fe/H] < -0.5$, N=131 for $\rm -0.5 \leq [Fe/H] < -0.25$, N=141 for $\rm -0.25 \leq [Fe/H] < 0.0$, N=80 for $\rm 0.0 \leq [Fe/H] < 0.5$) while covering the full range of metallicity variations. The results of this fitting procedure are presented in Table~\ref{Tab:Rx_Ro_FeH}, while in Figs.~\ref{Fig:Rx_Ro_VKs_FeH} and \ref{Fig:Rx_Ro_Teff_FeH} we show the distribution of stars and the linear fit in the $\rm Log(R_x)$-$\rm Log(Ro)$ planes for each $\rm [Fe/H]$ bin, for the two computations of the Rossby numbers considered in this work. By looking at the trend of the fitting parameters ``a'', ``b'', ``c'', and $\rm Log(Ro_{sat})$ as a function of the metallicity bins in Figs.~\ref{fig:params_trends_metallicity_bins_VKs} and \ref{fig:params_trends_metallicity_bins_Teff}, it is possible to notice that the slope of the $\rm Log(R_x)$–$\rm Log(Ro)$ relationship in the saturated regime (``c'') is steeper in the $\rm -1.5 \leq [Fe/H] < -0.5$ bin, especially in the computation with $\rm Ro_{T_{eff}}$ for which a correspondingly smaller value of $\rm Log(Ro_{sat})$ is observed. For the remaining parameters, it is challenging to establish a clear trend with respect to the metallicity bins, given the significant uncertainties obtained from the fitting procedure.

\begin{table*}
\noindent
\captionof{table}{Results of the fitting procedure on Eq.~\ref{Eq:Rx_Ro} for different prescriptions of the Rossby number on the sample of stars from W11, \textit{Kepler} LEGACY and B17, divided in four metallicity bins.}
\centering
\begin{tabular}{|l|c|c|c|c|c|}
 \cline{2-6}
 \multicolumn{1}{c|}{} & Metallicity range & a & b & c & $\rm Log(Ro_{sat})$ \\
 \hline
 \multirow{4}{*}{$\rm Ro_{(V-Ks)_0}$} & $\rm -1.50 \leq [Fe/H] < -0.50$ & $-0.15 \pm 0.13$  & $-3.44 \pm 0.20$ & $-2.31 \pm 0.36$ & $-0.77 \pm 0.08$ \\
                    & $\rm -0.50 \leq [Fe/H] < -0.25$ & $-0.24 \pm 0.19$  & $-3.62 \pm 0.31$ & $-1.76 \pm 0.28$ & $-0.87 \pm 0.11$ \\
                    & $\rm -0.25 \leq [Fe/H] < ~0.00$  & $-0.33 \pm 0.17$  & $-3.77 \pm 0.23$ & $-1.73 \pm 0.27$ & $-0.67 \pm 0.10$ \\
                    & $\rm ~0.00 \leq [Fe/H] < ~0.50$   & $-0.27 \pm 0.24$  & $-3.63 \pm 0.31$ & $-2.03 \pm 0.35$ & $-0.64 \pm 0.11$ \\
 \hline
\multirow{4}{*}{$\rm Ro_{T_{eff}}$} & $\rm -1.50 \leq [Fe/H] < -0.50$ & $-0.39 \pm 0.09$  & $-3.87 \pm 0.14$ & $-3.12 \pm 0.62$ & $-0.51 \pm 0.07$ \\
                                     & $\rm -0.50 \leq [Fe/H] < -0.25$ & $-0.08 \pm 0.17$   & $-3.34 \pm 0.27$ & $-1.68 \pm 0.20$ & $-1.01 \pm 0.08$ \\
                                     & $\rm -0.25 \leq [Fe/H] < ~0.00$  & $-0.22 \pm 0.18$   & $-3.52 \pm 0.29$ & $-1.61 \pm 0.23$ & $-0.95 \pm 0.10$ \\
                                     & $\rm ~0.00 \leq [Fe/H] < ~0.50$   & $-0.05 \pm 0.25$   & $-3.28 \pm 0.34$ & $-1.82 \pm 0.25$ & $-0.97 \pm 0.10$ \\
 \hline
\end{tabular}
\label{Tab:Rx_Ro_FeH}
\end{table*}

In the next section, a comparison between the tracks computed by following the revisited $\rm Log(R_x)$-$\rm Log(Ro)$ found here and the ones obtained by recalibrating the prescription of \citet{Johnstone2021} as in \citet{Pezzotti2021} will be presented.

\section{Implementation in theoretical models}
\label{Sec:models}

In this section, we compare predictions by theoretical models with the measured surface rotational period (or rotation rate) and X-ray luminosity for our sample of \textit{Kepler} LEGACY stars. 

For the computation of the models, we initially derived structural evolutionary tracks for each star in the LEGACY sample by means of the Liège Stellar Evolution Code \citep[CLES][]{Scuflaire2008}, using the optimal stellar parameters derived in Sect.~\ref{Sec:params}, with the same input physics. Subsequently, we provided these structural tracks as input to a SPI (Star-Planet Interaction) code \citep{Privitera2016AII, Rao2018, Rao2021, Pezzotti2021, Pezzotti2025}, to simulate the evolution of surface rotation rate and X-ray luminosity, whose properties strongly depend on the characteristics of the input structural tracks. In the SPI code, the evolution of the stellar surface rotation rate is computed under the assumption of solid body rotation, by accounting for the braking at the stellar surface due to magnetized winds as in \citet{Matt2015, Matt2019}, with solar calibrated parameters. Given the degeneracy in the rotational history of solar-like, low-mass stars, we considered initial values of the surface rotation rate ($\rm \Omega_{in}$) ranging between $3.2$ and $18$ times the surface rotation rate of the Sun at solar age ($\Omega_{\odot}=  2.9\times 10^{-6} s^{-1}$) \citep{Eggenberger2019a}, to account for the spread observed for stars in young clusters and stellar associations \citep{Gallet2015}. 
We assumed disk-locking timescales of 2 and 6 Myr for $\rm 10 \leq \Omega_{in}/\Omega_{\odot} \leq 18$, and $\rm \Omega_{in}/\Omega_{\odot} < 10$, respectively.

\begin{figure}
    \centering
    \includegraphics[width=\linewidth]{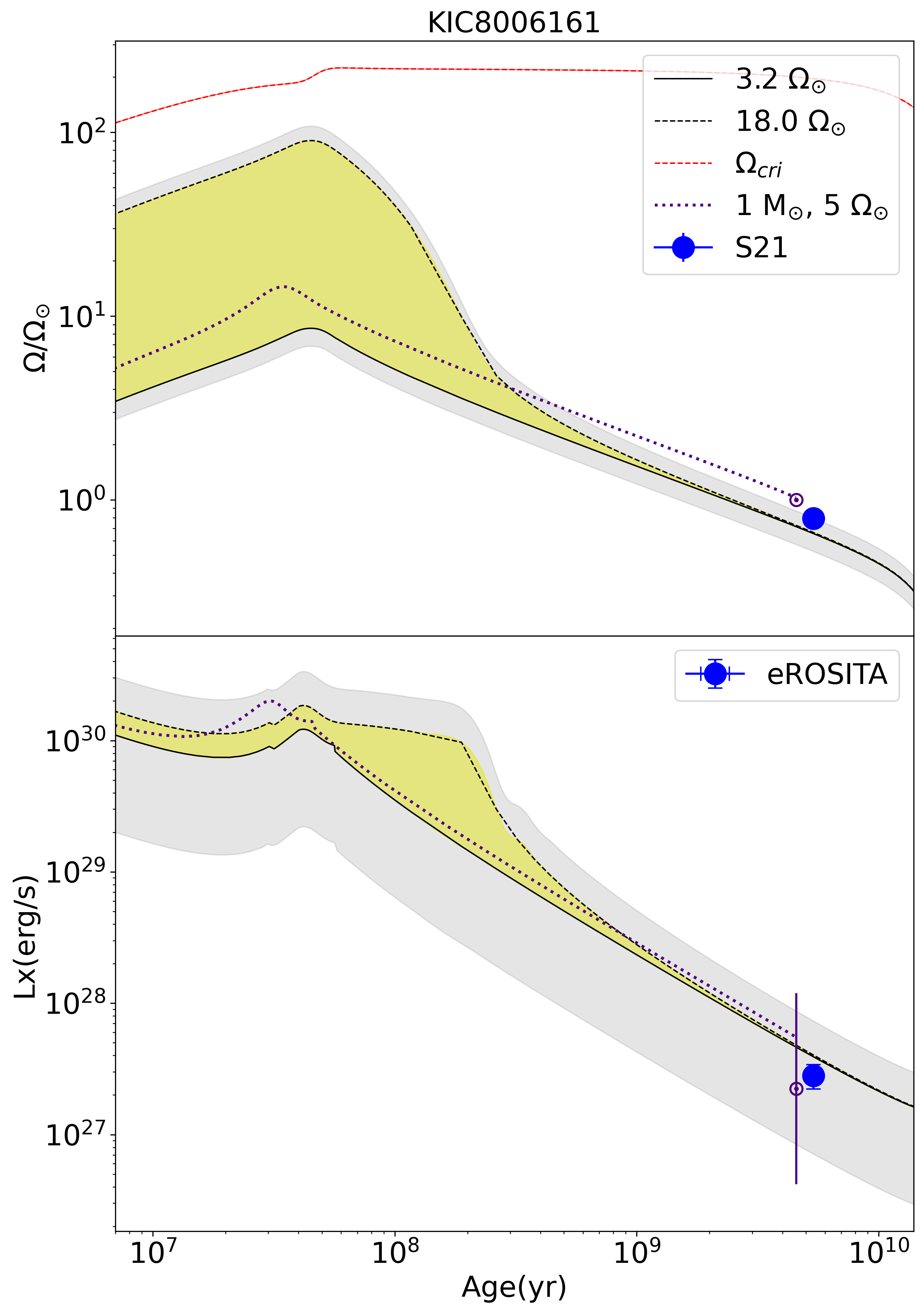}  
    \caption{Evolution of the surface rotation rate (top panel) and X-ray luminosity (bottom panel) of KIC 8006161. The yellow shaded
areas indicate the region explored by the tracks across the evolution of the star, for $3.2 \leq \Omega_{\rm in}/\Omega_{\odot} \leq 18$. The gray shaded areas show 20\% (top panel) and global one order of magnitude (bottom panel) further variation at the limit of the yellow areas, to account for deviations from more likely rotational histories and fluctuation in the X-ray luminosity due to cyclic or stochastic magnetic activity. The red-dashed line in the top panel shows the evolution of the critical rotation rate, defined as the limit at which the centrifugal acceleration at the equator equals gravity. The blue markers show observational data. For comparison, evolutionary tracks for a $\rm 1~M_{\odot}$ star are showed, with observational data from Table~\ref{Tab:resc_flux}.}
    \label{fig:Om_Lx_KIC8006161}
\end{figure}

\begin{figure}
    \centering
    \includegraphics[width=\linewidth]{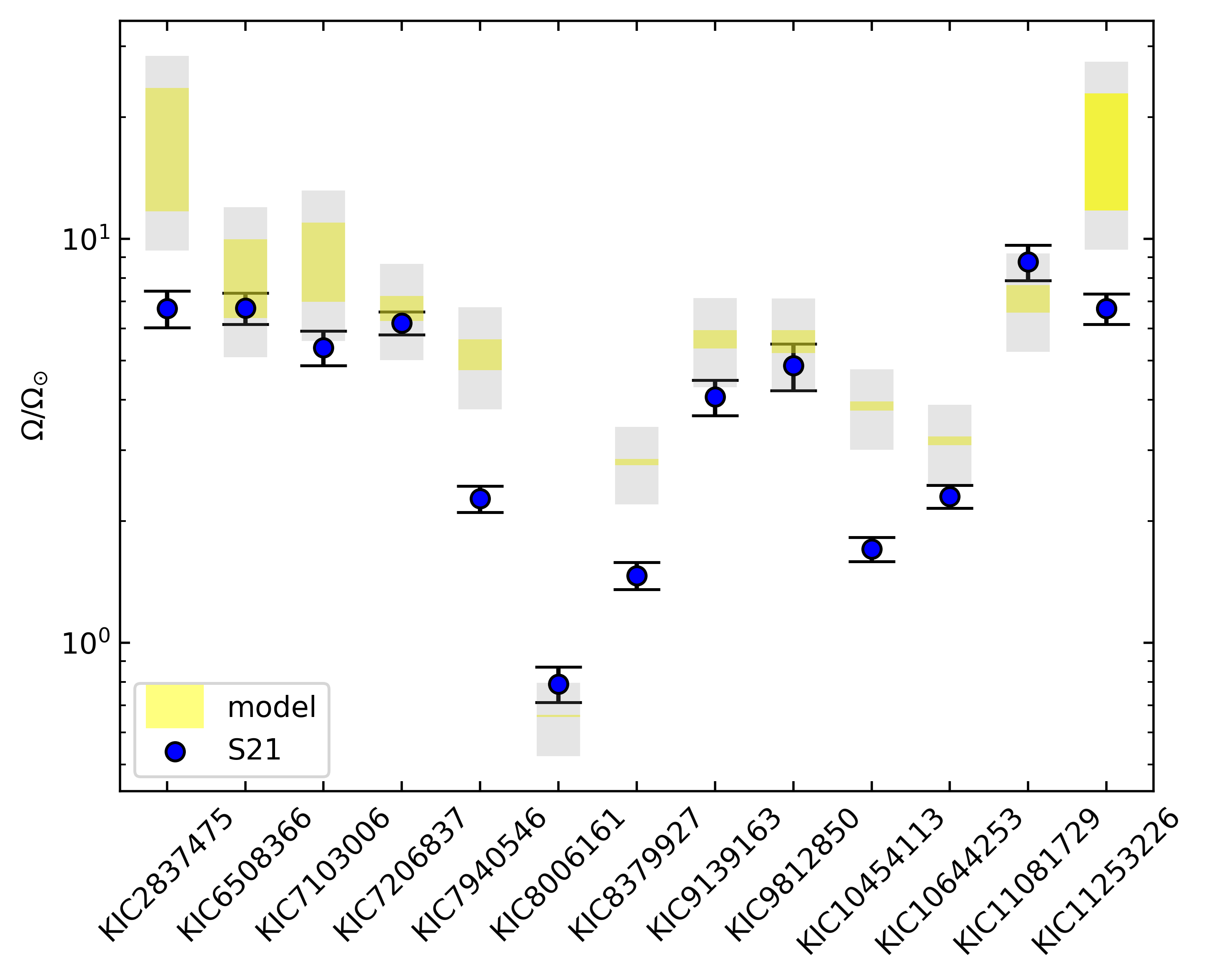}  
    \caption{Comparison between theoretical tracks and observational data for the surface rotation rates of stars in Table~\ref{Tab:params}.}
    \label{fig:Omega_J21}
\end{figure}

In the top panel of Fig.~\ref{fig:Om_Lx_KIC8006161}, an example of surface rotational evolution is shown for KIC 8006161, the nearest solar analogue of our \textit{Kepler} LEGACY sample, where the yellow region indicates the area included between the 3.2 and 18~$\rm \Omega_{\odot}$ tracks, the gray area shows a further 20\% variation with respect to the slow and fast rotator tracks, and finally the blue marker indicates the observational data. In this figure, it can be seen that the measured rotation rate \citep[][S21 in the following]{Santos2021} is slightly higher than the one predicted by the $\rm \Omega_{in} = 18 \Omega_{\odot}$ track, but it is compatible with the gray area. In Fig.~\ref{fig:Omega_J21}, we collect the results for all the stars, showing a comparison between the tracks and the measured rotation rate at the stellar age. In this figure, we see that 7 stars tend to rotate more slowly than what is predicted by models, while for the remaining 6 the tracks are compatible with observational data. It is worth noticing that for the computation of the rotational tracks we have not considered the transition into a weakened magnetic braking regime for Rossby numbers larger than $\rm Ro_{cri}$, where $\rm \tau_{conv}$ is computed as in \citet{Cranmer2011}. Despite a few stars having Rossby numbers larger than the critical value ($\rm Ro_{cri}$), according to our models, the assumption of a weakening of the magnetic braking is not necessary to reconcile theoretical models with observational data. On the contrary, the inclusion of this mechanism would increase the discrepancy with measured rotational periods.

For the computation of the X-ray luminosity tracks, the SPI code includes by default the prescription of \citet{Jackson2012} and \citet{Johnstone2021}, which we recalibrated in \citet{Pezzotti2021} according to our computation of $\rm Ro$. To these two prescriptions, we added their revisited versions (see Eq.~\ref{Eq:Rx_Age}, Table~\ref{Tab:Rx_Age} for $\rm Log(R_x)$-$\rm Log(Age)$, and Eq.~\ref{Eq:Rx_Ro}, Table~\ref{Tab:Rx_Ro} for $\rm Log(R_x)$-$\rm Log(Ro)$), with the aim of consistently comparing X-ray luminosity tracks and test the compatibility with observational data. In the bottom panel of Fig.~\ref{fig:Om_Lx_KIC8006161}, an example of the tracks computed with the recalibrated prescription of \citet{Johnstone2021} is presented and compared with the e-ROSITA X-ray luminosity for KIC 8006161. The yellow area in this figure corresponds to the region explored by the tracks, relatively to different rotational histories. At the extreme of the yellow region, we assumed a further variation of one order of magnitude, that stars may experience because of cyclic (as for the Sun) or stochastic activity.

\begin{figure}
\centering
\begin{subfigure}{
\includegraphics[width=\linewidth]{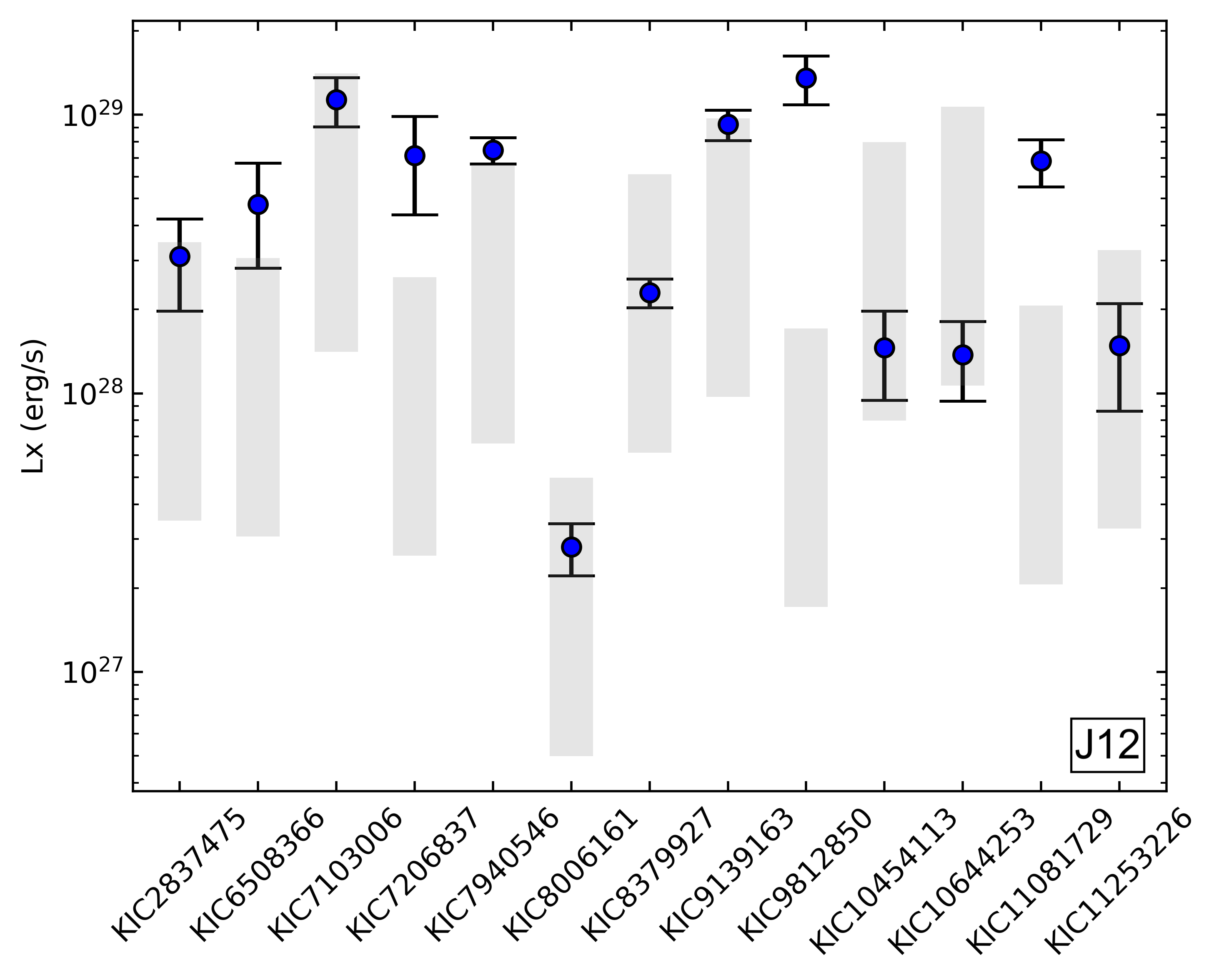}
}
\end{subfigure}
\vspace{-5.1\baselineskip} 
\par 
\begin{subfigure}{
\includegraphics[width=\linewidth]{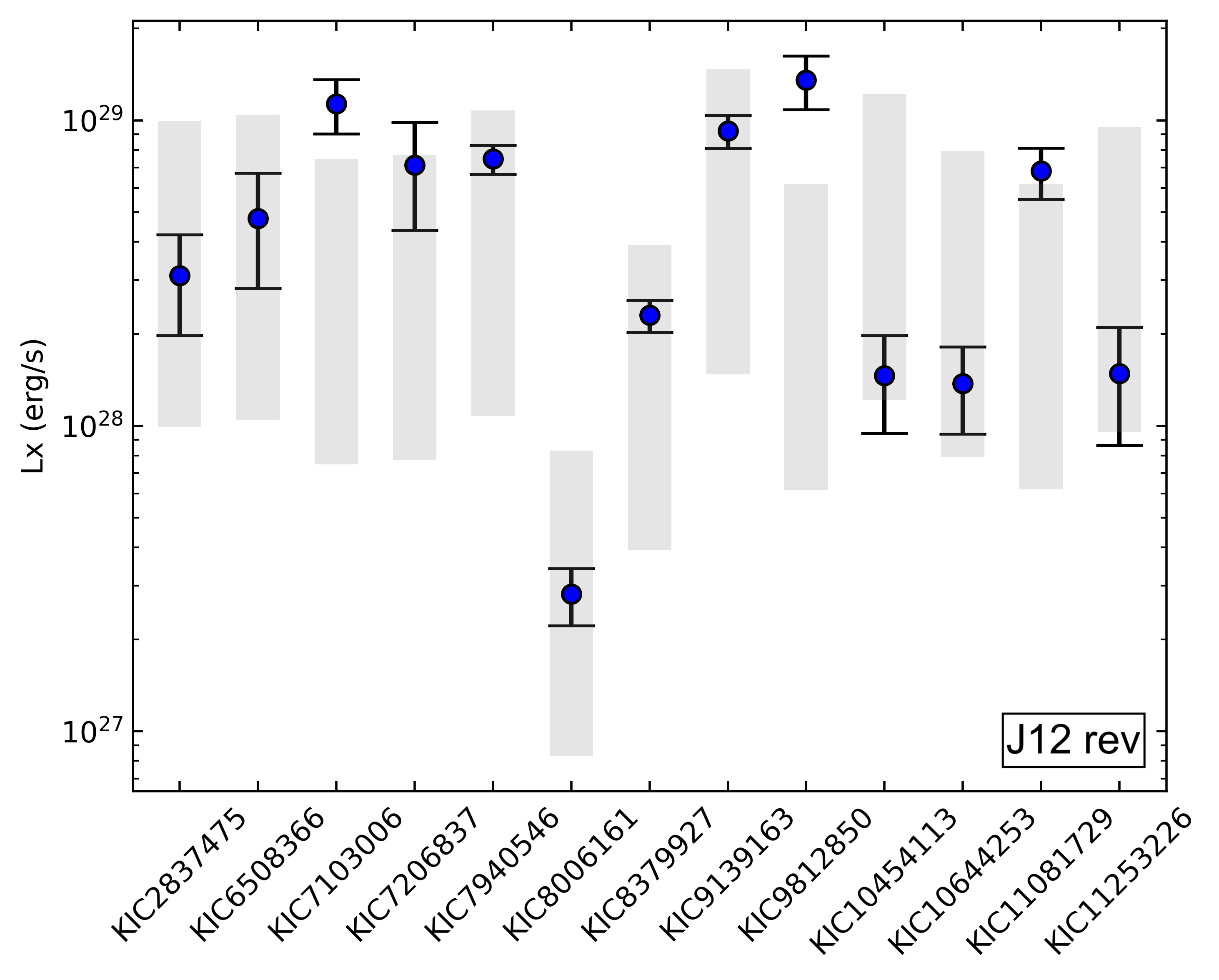}
}
\end{subfigure}
\captionsetup{skip=0.0pt}
\caption{Comparison between theoretical tracks computed with the original prescription of J12 (top panel) and its revisited version (this work, bottom panel) and observational data for stars in Table~\ref{Tab:params}.}
\label{Fig:Lx_J12}
\end{figure}

\begin{figure}
\centering
\begin{subfigure}{
\includegraphics[width=\linewidth]{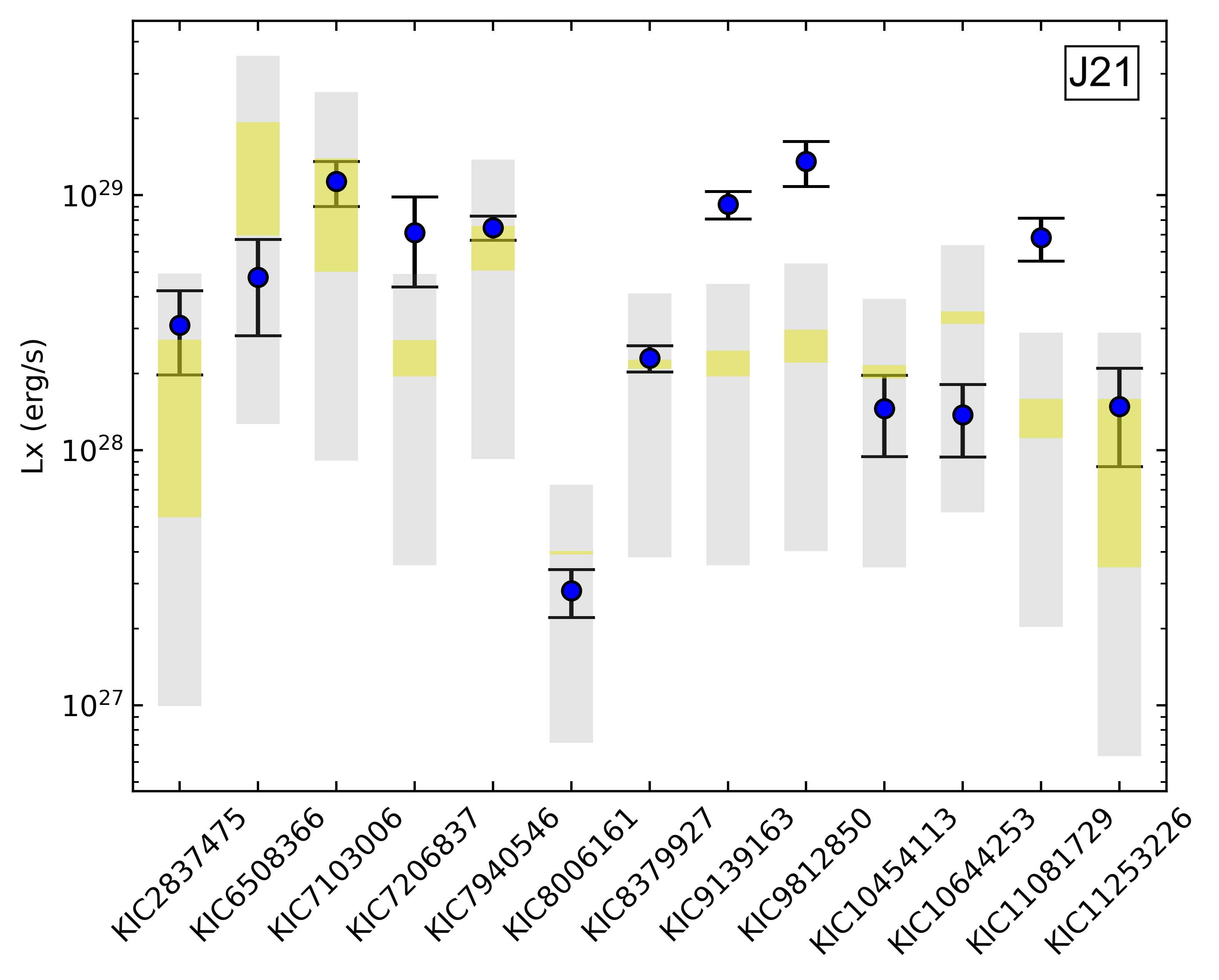}
}
\end{subfigure}
\vspace{-5.1\baselineskip} 
\par 
\begin{subfigure}{
\includegraphics[width=\linewidth]{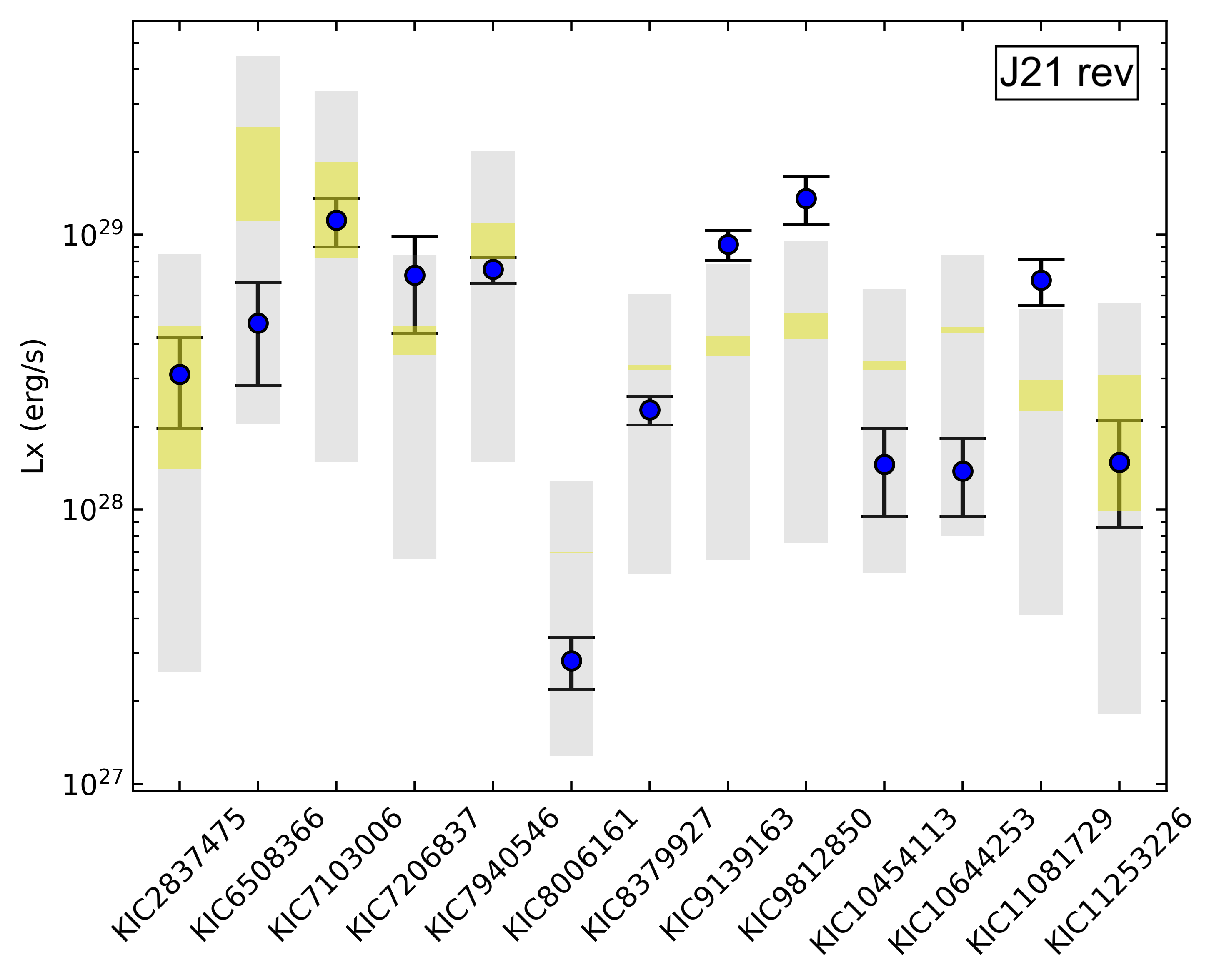}
}
\end{subfigure}
\captionsetup{skip=0.0pt}
\caption{Comparison between theoretical tracks computed with the recalibrated prescription of J21 (top panel) and its revisited version (this work, bottom panel) and observational data stars in Table~\ref{Tab:params}.}
\label{Fig:Lx_J21}
\end{figure}

In Fig.~\ref{Fig:Lx_J12}, we present the comparison between observational data and X-ray luminosity tracks computed by following the original prescription in J12 (top panel) and the revisited version derived in this work (bottom panel). Thanks to the revisited relationship, a higher compatibility between tracks and observational data is retrieved in the bottom panel for 7 stars (KIC 2837475, KIC 6508366, KIC 7206837, KIC 7940546, KIC 9139163, KIC 9812850, KIC 11081729), while for KIC 7103006 a lower compatibility is obtained. Similarly, in Fig.~\ref{Fig:Lx_J21} we present the comparison between observational data and X-ray luminosity tracks computed by using the recalibrated prescription of \citet{Johnstone2021} (top panel), and the revisited version found in this work (bottom panel). Also in this case, we observe that a higher compatibility between tracks and observational data is retrieved in the bottom panel for 4 stars (KIC 7206837, KIC 9139163, KIC 9812850, KIC 11081729).

The improvement in the overlap between theoretical tracks and observational data given by the revisited $\rm Log(R_x)$-$\rm Log(Age)$ and $\rm Log(R_x)$-$\rm Log(Ro)$ relationships is not totally surprising, since we directly included data for the \textit{Kepler} LEGACY stars in the fitting procedures. Despite the relatively small size of our sample of well-characterized stars, that tends to populate the more evolved region of the unsaturated regime generally characterized by scarcity of data, it provides an important contribution in shaping its properties.

\subsection{Impact on planetary atmospheric mass loss}

We investigated and compared the impact of implementing in our SPI code the revisited relationships in Eq.~\ref{Eq:Rx_Age} and \ref{Eq:Rx_Ro} on the computation of the planetary mass loss, by computing a set of 1520 models for each considered prescription: ``J12'', \citet{Jackson2012}, ``J12 rev'', revision of this work, ``J21'', \citet{Pezzotti2021}, and ``J21 rev'', revision of this work. We selected a set of initial inputs aimed at exploring the properties of the ``radius valley'' \citep{Fulton2017}, a bimodal size distribution of planets peaked at $\rm \sim 1.3$ and $\rm \sim 2.4~R_{\oplus}$ revealed by the \textit{Kepler} survey, whose nature can be related to atmospheric mass loss, together with other mechanisms \citep[e.g.][]{Venturini2020, Burn2024, Nielsen2015}. Our set of initial inputs is: orbital distance $\rm 0.1 \leq a_{in}(AU) \leq 1.05$, in steps of $\rm 0.05~AU$; planetary core mass $\rm 2 \leq M_{core}(M_{\oplus}) \leq 10$, in steps of $\rm 0.1~M_{\oplus}$; percentage of primordial atmosphere fixed at 5\% of the total planetary mass; $\rm 1~M_{\odot}$ parent star, with age range from the PMS to 4.57 Gyr, $\rm Z = Z_{\odot}$, $\rm (B-V)_0 = 0.656$, and initial surface rotation rate fixed at $\rm \Omega_{in} = 5~\Omega_{\odot}$.

For the estimation of the mass loss, we used the analytical formulae of \citet{Kubyshkina2021}. The stellar EUV flux that is given as input to these formulae is computed by rescaling the X-ray luminosity as in \citet{SanzForcada2011} for ``J12'' and ``J12 rev'', and the X-ray flux as in \citet{Johnstone2021} for ``J21'' and ``J21 rev'' relationships. The evolution of the planetary radius is computed by means of the parametrized formula of \citet{Lopez2014}, for planets with rocky cores and H-He rich gaseous envelopes.

The results of the simulations are collected in Fig.~\ref{Fig:planets_radius_bin} where we compare the size distribution of planets obtained when using ``J12'' and ``J12 rev'', and ``J21'' and ``J21 rev'', for radius bins from $1.1$ to $\rm 3.1~R_{\oplus}$, in steps of 0.1. From these figures we deduce that using the revised prescriptions does not significantly impact the location of the valley, which may shift by one bin between the revised and non-revised prescription. Nevertheless, a maximum difference of 5 and 15 planets is found for the $\rm [3.0, 3.1)$ bin, indicating that using the revised prescriptions yields a difference, though modest with respect to the simulations analyzed here.

\section{Summary and conclusions}
\label{Sec:Conclusion}

We revisited different components of the activity-rotation-age relationship by introducing the largest sample of solar-like stars with highest-quality asteroseismic data, surface rotational periods, and X-ray detections.  

We studied the $\rm Log(L_x/R^2)$-$\rm Log(Age)$ relationship finding that the fitting procedure is hindered by the very scattered distributions of stars and the low statistics. By comparing the average value of the slope with the results in \citet{Booth2017}, we found indications of a flatter linear dependence.

We recalibrated the $\rm Log(R_x)$-$\rm Log(Age)$ relationship by joining our sample of well-characterized stars to the one in \citet{Jackson2012}, and recomputed the slopes in the saturated and unsaturated regimes for the global and split samples of stars in nine $\rm (B-V)_0$ colour bins.  

We recalibrated the $\rm Log(R_x)$-$\rm Log(Ro)$ relationship by considering the whole spectral range on the one hand, and by applying a cut on spectral type from K to F on the other one. The major impact on the fitting parameters due to the cut is recovered for Rossby numbers, or convective turnover timescales, computed on the basis of $\rm T_{eff}$. We further analysed this relationship by splitting the sample into $\rm [Fe/H]$ bins, finding that relatively more metal-poor stars tend to follow a steeper slope in the unsaturated regime.

We finally implemented the revisited relationships for $\rm Log(R_x)$-$\rm Log(Age)$ and $\rm Log(R_x)$-$\rm Log(Ro)$ in our SPI code to compute rotational and X-ray luminosity tracks for the 13 \textit{Kepler} LEGACY stars analysed in this work. By comparing rotational tracks computed without weakened magnetic braking with measured rotational periods, we notice that our stars on average tend to rotate more slowly than what is predicted by the models. This result shows that the hypothesis of a weakened magnetic braking taking over for $\rm Ro \geq Ro_{cri}$ is not strictly necessary in our case. By comparing the X-ray tracks computed with the relationships found in this work and the ones in the literature with luminosities from eROSITA, we found that the revisited relationships provide improved agreement for a good fraction of the sample. To test the impact of the revisited relationships on the mass loss from planetary atmospheres, we simulated the evaporation of planets in the radius valley, accounting for a $\rm 1~M_{\odot}$ parent star, with properties analogous to the ones of our Sun. By comparing the size distribution of planets obtained with the revisited and non-revisited prescriptions, we observe the major difference for the bin with the largest radius. This difference, though, appears modest if compared to the total number of planets involved in the simulation.

Larger samples of stars with well-characterized parameters, rotation period, and activity indicators are needed to accurately determine the multiple components of the activity-rotation-age relationship. In this context, asteroseismology will play a crucial role in releasing precise characterizations for thousands of stars with the advent of PLATO.

\begin{acknowledgements}
We thank the anonymous referee for useful comments that helped to improve the quality of this manuscript. This work used data of eROSITA telescope onboard SRG observatory. The SRG observatory was built by Roskosmos in the interests of the Russian Academy of Sciences represented by its Space Research Institute (IKI) in the framework of the Russian Federal Space Program, with the participation of the Deutsches Zentrum für Luft- und Raumfahrt (DLR). The SRG/eROSITA X-ray telescope was built by a consortium of German Institutes led by MPE, and supported by DLR.  The SRG spacecraft was designed, built, launched and is operated by the Lavochkin Association and its subcontractors. The science data are downlinked via the Deep Space Network Antennae in Bear Lakes, Ussurijsk, and Baykonur, funded by Roskosmos. The eROSITA data used in this work were processed using the eSASS software system developed by the German eROSITA consortium and proprietary data reduction and analysis software developed by the Russian eROSITA Consortium.

CP thanks the Belgian Federal Science Policy Office (BELSPO) for the financial support in the framework of the PRODEX Program of the European Space Agency (ESA) under contract number 4000141194.

JB acknowledges funding from the SNF Postdoc.Mobility grant no. P500PT{\_}222217 (Impact of magnetic activity on the characterization of FGKM main-sequence host-stars).

MG, IB acknowledge support by the
RSF grant N 23-12-00292.

GB is funded by the Fonds National de la Recherche Scientifique (FNRS).

\end{acknowledgements}


\newpage

\begin{appendix}


\section{Observational constraints for the asteroseismic characterisations}
\label{app:observational_data}

We used the acoustic oscillation frequencies from \citet{Lund2017} as seismic constraints. The spectroscopic constraints were taken from \citet{Furlan2018}, except for KIC8379927 for which we used the data from \citet{Pinsonneault2012}. The absolute stellar luminosity was derived using
\begin{equation}
\rm \log\left(\frac{L}{L_\odot}\right) = -0.4\left(m_{\lambda} + BC_{\lambda} -5\log d + 5 - A_{\lambda} -M_{\mathrm{bol},\odot}\right) \, ,
\label{eq:luminosity}
\end{equation}
where $\lambda$ is the wavelength band. Specifically, we used the 2MASS $\rm Ks$-band, except for KIC10454113 for which we used the 2MASS $\rm H$-band, given the very large uncertainty in the $\rm Ks$-band. In this equation, $\rm m_{\lambda}$ is the apparent magnitude, $\rm BC_{\lambda}$ is the bolometric correction computed following \citet{Casagrande2014,Casagrande2018}, $\rm d$ is the distance in pc from \textit{Gaia} EDR3 \citep{Gaia2021}, $\rm A_{\lambda}$ is the extinction computed with the dust map from \citet{Green2018}, and $\rm M_{\mathrm{bol},\odot} = 4.75$ is the absolute solar bolometric magnitude. The distance was estimated using two approaches, either by inverting the parallax corrected according to \citet{Lindegren2021} or by using the distance from \citet{Bailer-Jones2021}. Both methods led to similar distances and we adopted the distance from \citet{Bailer-Jones2021} for our final luminosities. The spectroscopic constraints and luminosities are summarised in Table~\ref{tab:spectroscopic_constraints}.

\begin{table}
\caption{Observational constraints for the asteroseismic characterisations.}
\centering
\resizebox{\columnwidth}{!}{
\begin{tabular}{lccccc}
\hline \hline
KIC ID & $\rm T_{\mathrm{eff}}$ (K) & $\rm \log g$ & [Fe/H] & $\rm L/L_\odot$ & Reference \\
\hline
KIC2837475 & $6488 \pm 100$ & $4.29 \pm 0.20$ & $-0.07 \pm 0.10$ & $4.39 \pm 0.23$ & 1 \\ 
KIC6508366 & $6249 \pm 100$ & $4.14 \pm 0.20$ & $-0.06 \pm 0.10$ & $6.53 \pm 0.34$ & 1 \\ 
KIC7103006 & $6362 \pm 100$ & $4.16 \pm 0.20$ & $~0.05 \pm 0.10$ & $5.55 \pm 0.28$ & 1 \\ 
KIC7206837 & $6325 \pm 100$ & $4.29 \pm 0.20$ & $~0.12 \pm 0.10$ & $3.75 \pm 0.18$ & 1 \\ 
KIC7940546 & $6126 \pm 100$ & $3.77 \pm 0.20$ & $-0.27 \pm 0.10$ & $4.91 \pm 0.25$ & 1 \\ 
KIC8006161 & $5422 \pm 100$ & $4.47 \pm 0.20$ & $~0.32 \pm 0.10$ & $0.67 \pm 0.04$ & 1 \\ 
KIC8379927 & $6067 \pm 147$ & $4.24 \pm 0.20$ & $-0.10 \pm 0.10$ & $2.24 \pm 0.18$ & 2 \\ 
KIC9139163 & $6350 \pm 100$ & $4.29 \pm 0.20$ & $~0.09 \pm 0.10$ & $3.60 \pm 0.17$ & 1 \\ 
KIC9812850 & $6314 \pm 100$ & $4.13 \pm 0.20$ & $-0.18 \pm 0.10$ & $4.51 \pm 0.22$ & 1 \\ 
KIC10454113 & $6136 \pm 100$ & $4.31 \pm 0.20$ & $-0.07 \pm 0.10$ & $2.93 \pm 0.15$ & 1 \\ 
KIC10644253 & $6020 \pm 100$ & $4.36 \pm 0.20$ & $~0.09 \pm 0.10$ & $1.51 \pm 0.08$ & 1 \\ 
KIC11081729 & $6416 \pm 100$ & $4.24 \pm 0.20$ & $-0.13 \pm 0.10$ & $3.07 \pm 0.15$ & 1 \\ 
KIC11253226 & $6474 \pm 100$ & $4.21 \pm 0.20$ & $-0.19 \pm 0.10$ & $4.13 \pm 0.20$ & 1 \\ 
\hline
\noalign{\vskip 0.5ex}
\end{tabular}}
{\par\raggedright\small\justify\textbf{Notes.} Seismic data from \citet{Lund2017}. Spectroscopic data from (1) \citet{Furlan2018} and (2) \citet{Pinsonneault2012}. The luminosity was derived using Eq.~\eqref{eq:luminosity}. \par}
\label{tab:spectroscopic_constraints}
\end{table}


\section{Comparison of different prescription for the computation of $\rm \tau_{conv}$}
\label{app:tau_conv_comp}

\begin{figure}
    \centering
    \includegraphics[width=0.9\linewidth]{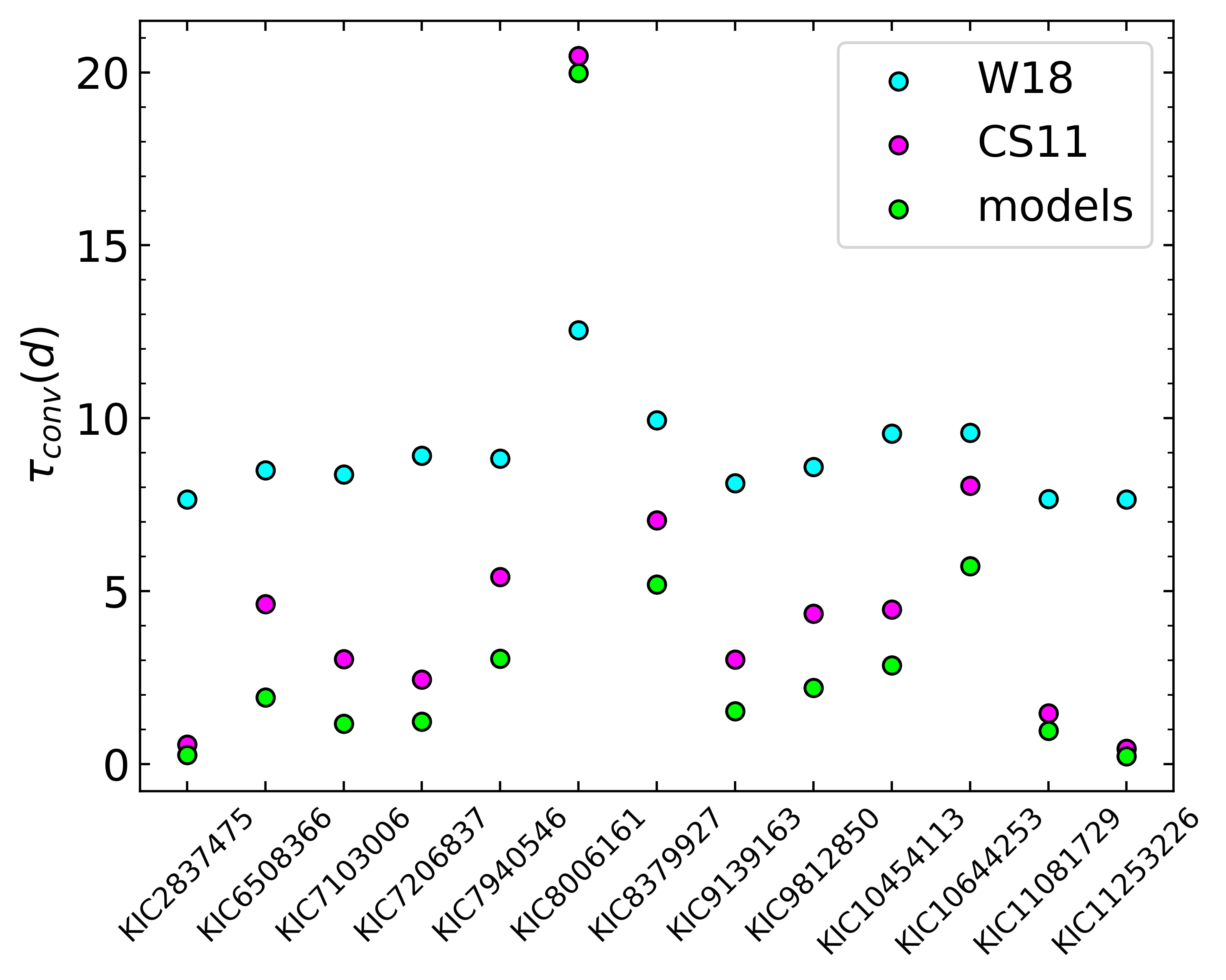} 
    \caption{Comparison among the convective turnover timescales computed for the 13 stars of the \textit{Kepler} LEGACY sample with the prescription from \citet{Wright2018} (cyan dots), \citet{Cranmer2011} (magenta dots) and from the asteroseismic models (lime dots) with the \citet{Montesinos2001} convention.}
    \label{Fig:tau_conv_comp}
\end{figure}

\begin{figure}
    \centering
    \includegraphics[width=0.87\linewidth]{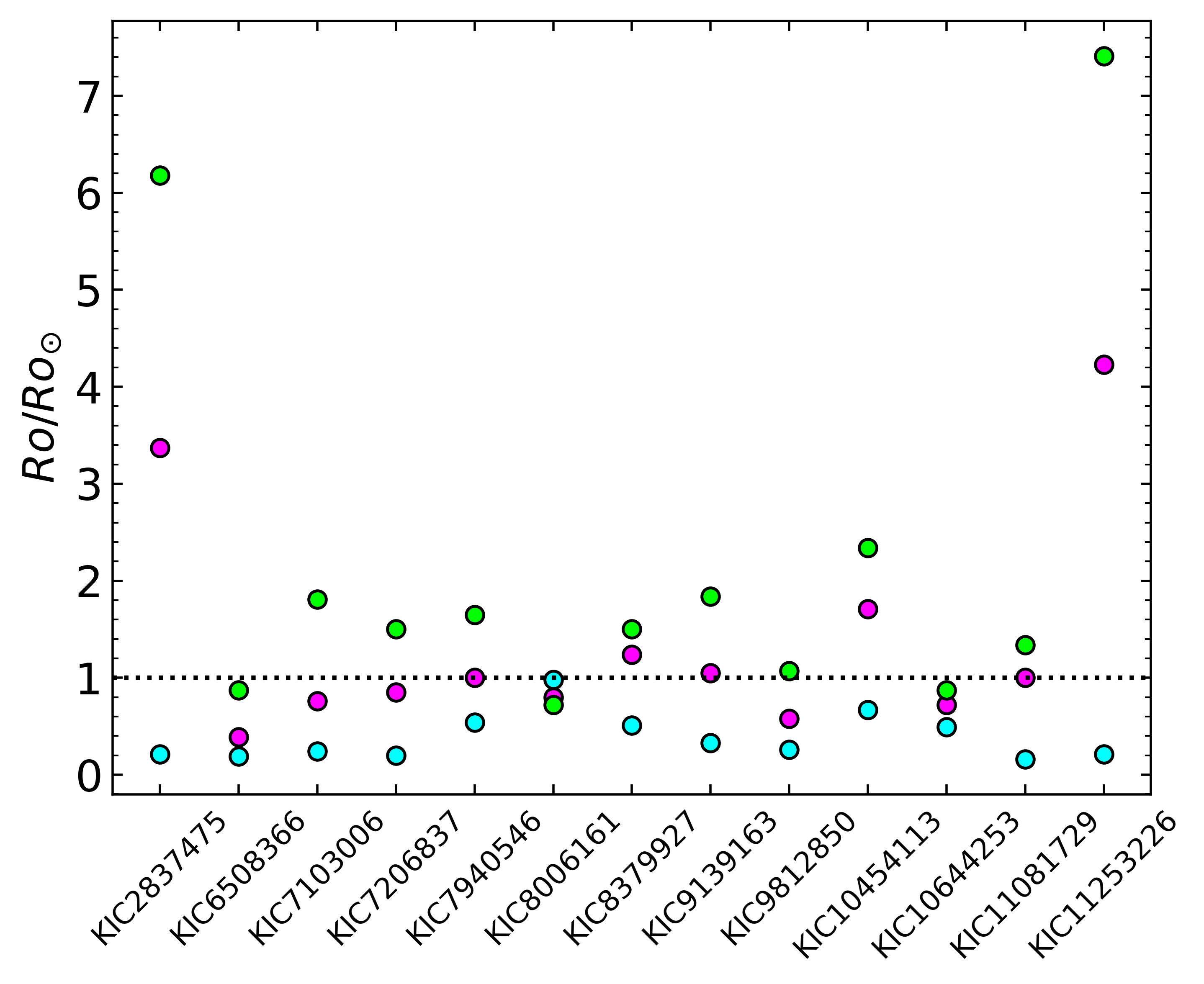} 
    \caption{Comparison among normalised Rossby numbers with respect the Rossby number of the Sun for the 13 stars of the \textit{Kepler} LEGACY sample, with convective turnover timescale computed as in \citet{Wright2018} (cyan dots), \citet{Cranmer2011} (magenta dots), and from the asteroseismic models (lime dots) with the \citet{Montesinos2001} convention.}
    \label{fig:Ro_normalised_RoSun}
\end{figure}

In Fig.~\ref{Fig:tau_conv_comp}, we show a comparison between the convective turnover timescales computed with the prescriptions from \citet{Wright2018}, \citet{Cranmer2011}, and from best-fit stellar models using the convention of \citet{Montesinos2001} for the 13 \textit{Kepler} LEGACY stars in Table~\ref{Tab:params}. With respect to the first approach, we notice that the $\rm (V-Ks)_0$ colour of KIC2837475 and KIC11253226 lies below the validity limit for the \citet{Wright2018} prescription, causing an overestimation of $\rm \tau_{conv}$. In Fig.~\ref{fig:Ro_normalised_RoSun}, we show the normalised Rossby number with respect to the solar one for $\rm \tau_{conv}$ computed by means of the three approaches mentioned above. For $\rm \tau_{conv}$ computed as in \citet{Wright2018}, $\rm Ro/R_{\odot} < 1$ for all stars, except for KIC8006161. For $\rm \tau_{conv}$ computed as in \citet{Cranmer2011}, six out of thirteen stars have $\rm Ro/R_{\odot} < 1$. Finally, for $\rm \tau_{conv}$ computed from our best-fit stellar models, only three stars have $\rm Ro/R_{\odot} < 1$. It is worth recalling that for the computation of the theoretical surface rotation rate and X-ray luminosity tracks with our SPI in Sect.~\ref{Sec:models}, we used the prescription of \citet{Cranmer2011} to compute $\rm \tau_{conv}$.

\section{Subsample of stars from \citet{Booth2017}}

In Table~\ref{Tab:resc_flux} we indicate the quantities of interest for the present study relatively to the subsample of stars with asteroseismically determined ages extracted from \citet{Booth2017}.

\begin{table*}
\noindent
\centering
\captionof{table}{Fundamental quantities for the subsample of 7 stars extracted from \citet{Booth2017}, including the Sun.}
\begin{tabular}{lccccccc}
 \hline\hline
 Target & $\rm \log(L_{x_{\mathrm{c}}}(erg/s))$  & $\rm P_{rot}(d)$ & $\rm (B-V)_0$ &$\rm (V-Ks)_0$ & $\rm Ro_{{(V-Ks)_0}}$ & $\rm Ro_{{T_{eff}}}$ & Age(Gyr) \\
 \hline\noalign{\vskip 0.5ex}
 KIC7529180         & $29.32 \pm 0.10$ & $2.12 \pm 0.22$  & 0.408 & 1.008 & 0.272 & 5.745 & $\rm 1.93 \pm 0.33$\\
 KIC9955598         & $27.26 \pm 0.10$ & $34.81 \pm 5.69$ & 0.709 & 1.839 & 2.767 & 1.941 & $\rm 6.98 \pm 0.45$\\
 KIC10016239        & $28.27 \pm 0.10$ & $4.89 \pm 0.49$  & 0.516 & 0.998 & 0.631 & 2.953 & $\rm 2.47 \pm 0.64$\\
 Alpha Centauri B   & $28.54 \pm 0.36$ & $36.9 \pm 5.00$  & 0.88 & 1.930 & 2.783 & 1.887 & $\rm 6.13 \pm 0.55$\\
 Proxima Centauri   & $27.68 \pm 0.10$ & $89.09 \pm 4.00$ & 1.82 & 6.746 & 0.420 & 1.268 & $\rm 6.13 \pm 0.55$\\
 16 Cyg A           & $27.25 \pm 0.10$ & $23.8 \pm 1.65$  & 0.648 & 1.524 & 2.268 & 1.982 & $\rm 6.67 \pm 0.79$\\
 \hline
  Sun           & $27.35 \pm 0.72$ & $25.00 \pm 0.03$  & 0.656 & 1.58 & 2.306 & 1.95 & $\rm 4.57 \pm 0.02$\\
 \hline
\end{tabular}
\label{Tab:resc_flux}
\tablefoot{Target name (col.~1), X-ray luminosities derived from the converted fluxes from the $\rm 0.2$-$2.0$ keV to $\rm 0.1$-$2.4$ keV energy band (col.~2), rotational period (col.~3), $\rm (B-V)_0$ (col.~4) and $\rm (V-Ks)_0$ colour (col.~5, SIMBAD database, \citet{SIMBAD2000}, this work), Rossby number computed with $\rm \tau_{conv}$ as in \citet{Wright2018} (col.~6) and \citet{Cranmer2011} (col.~7), and asteroseismically derived age \citep[col.~8,][and references therein]{Booth2017} of a subsample of stars from \citet{Booth2017}.}
\end{table*}


\section{Log(L$\rm _x$/R$^2$) vs Log(Age) for stars with $\rm T_{eff}(K) \leq 6250$}
\label{App:Lx_Age_only_cool_stars}

We performed the fit of the Log(L$_x$/R$^2$) vs Log(Age) relationship by excluding stars with $\rm T_{eff}(K) > 6250$, that could be above the Kraft break. We found the resulting fit:

\begin{equation}
\rm Log(L_x/R^2) = -1.27~(\pm 0.64)\times Log(Age) + 40.21~(\pm 6.07),  
\end{equation}
\label{LxR2-Age_Teff_6250}

\noindent
with reduced $\rm \chi_{red}^2 = 44.59$. We further applied a bootstrap method as described in Sect.~\ref{Sect:Lx_Age_Booth}, finding:

\begin{equation}
\rm Log(L_x/R^2) = -1.54~(\pm 1.62)\times Log(Age) + 41.98~(\pm 15.58).  
\label{Eq:boot}
\end{equation}
\label{Bootstrap_Teff_620}

\begin{figure}
    \centering
    \includegraphics[width=0.8\linewidth]{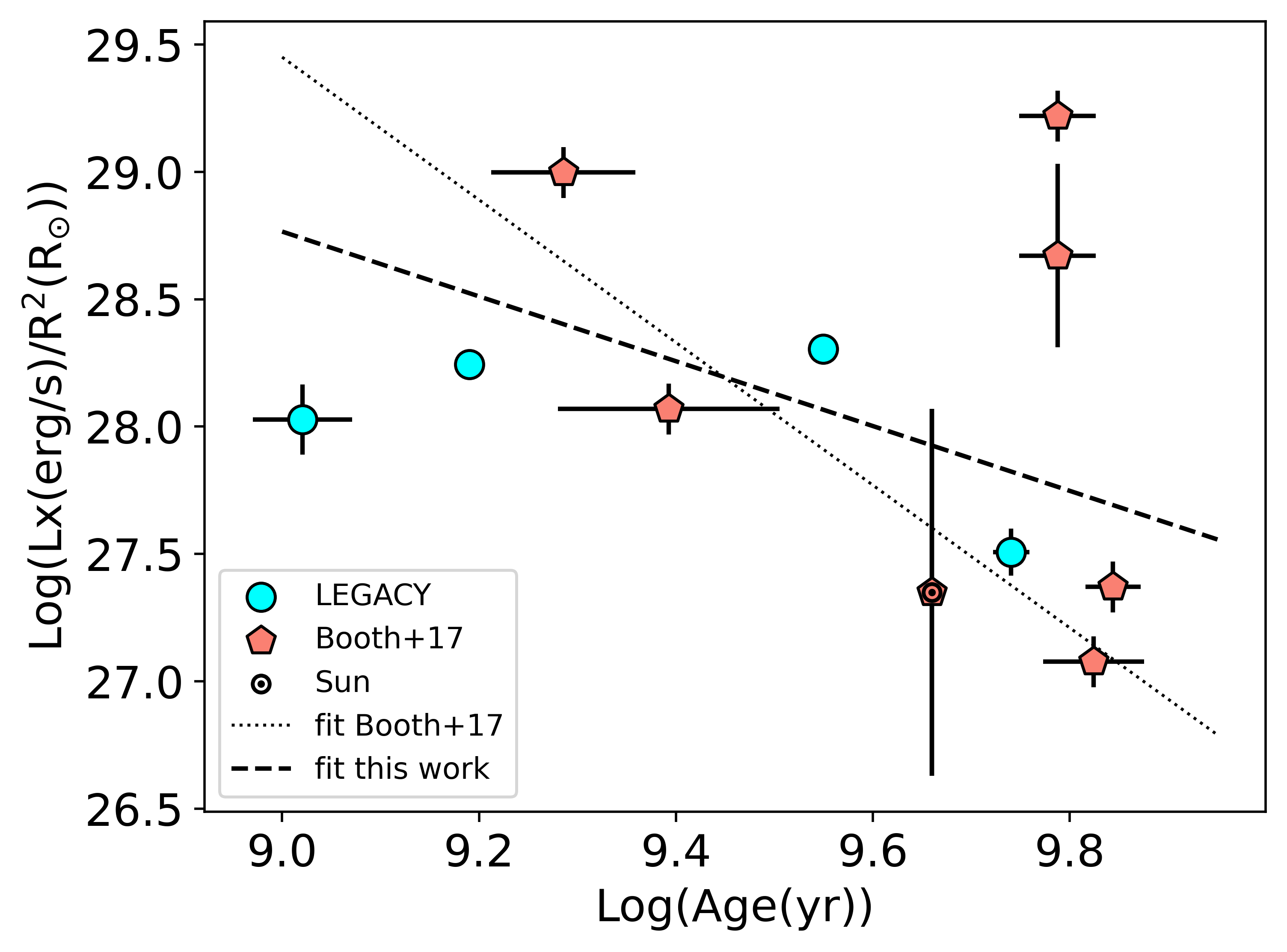}
    \caption{Distribution of stars from the \textit{Kepler} LEGACY (cyan dots) with $\rm T_{eff} \leq 6250$ and stars from B17 (orange pentagons) including the Sun (indicated by the usual symbol) in the $\rm Log(L_x/R^2)$-$\rm Log(Age)$ plane.}
    \label{fig:LogLxR2_LogAge_Teff_6250}
\end{figure}

In Fig.~\ref{fig:LogLxR2_LogAge_Teff_6250}, the distribution of stars in the Log(L$_x$/R$^2$) vs Log(Age) plane and the fit are showed.

\section{Log(R$\rm _x$)-Log(Age) fit dependence on $\rm (B-V)_0$}
\label{App:fig_BV}

\begin{figure}
    \centering
    \includegraphics[width=0.8\linewidth]{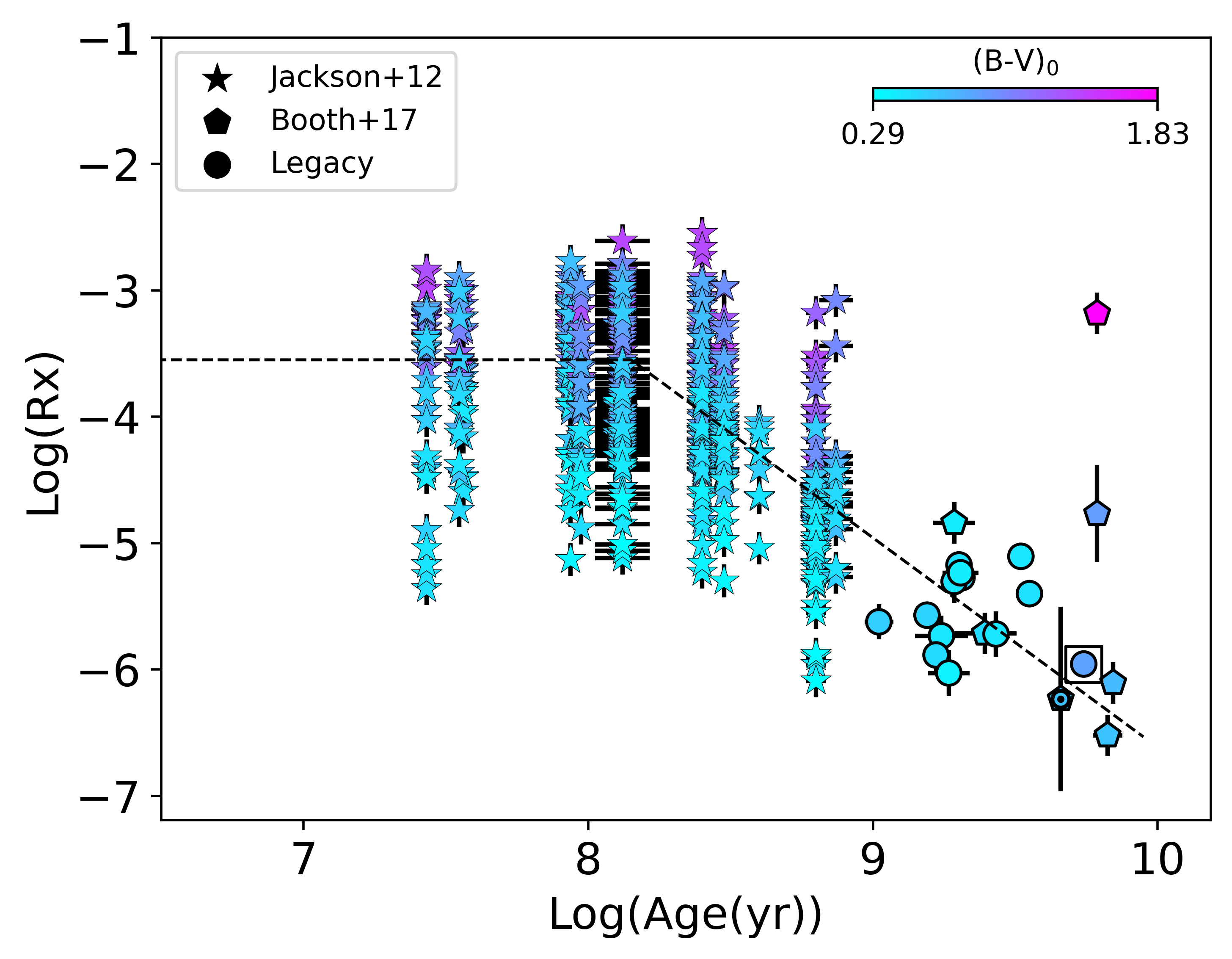}  
    \caption{Distribution of stars in the $\rm Log(R_x)$-$\rm Log(Age)$ plane, where the star, dot and pentagon markers indicate targets from J12, \textit{Kepler} LEGACY and B17, respectively. The empty square highlights the position of KIC8006161. The black dashed line indicates the fit of Eq.~\ref{Eq:Rx_Age}. The data is colour coded according to $\rm (B-V)_0$.}
    \label{fig:LogRx_LogAge}
\end{figure}

\begin{figure*}
    \centering

    \begin{subfigure}{
        \includegraphics[width=0.3\textwidth]{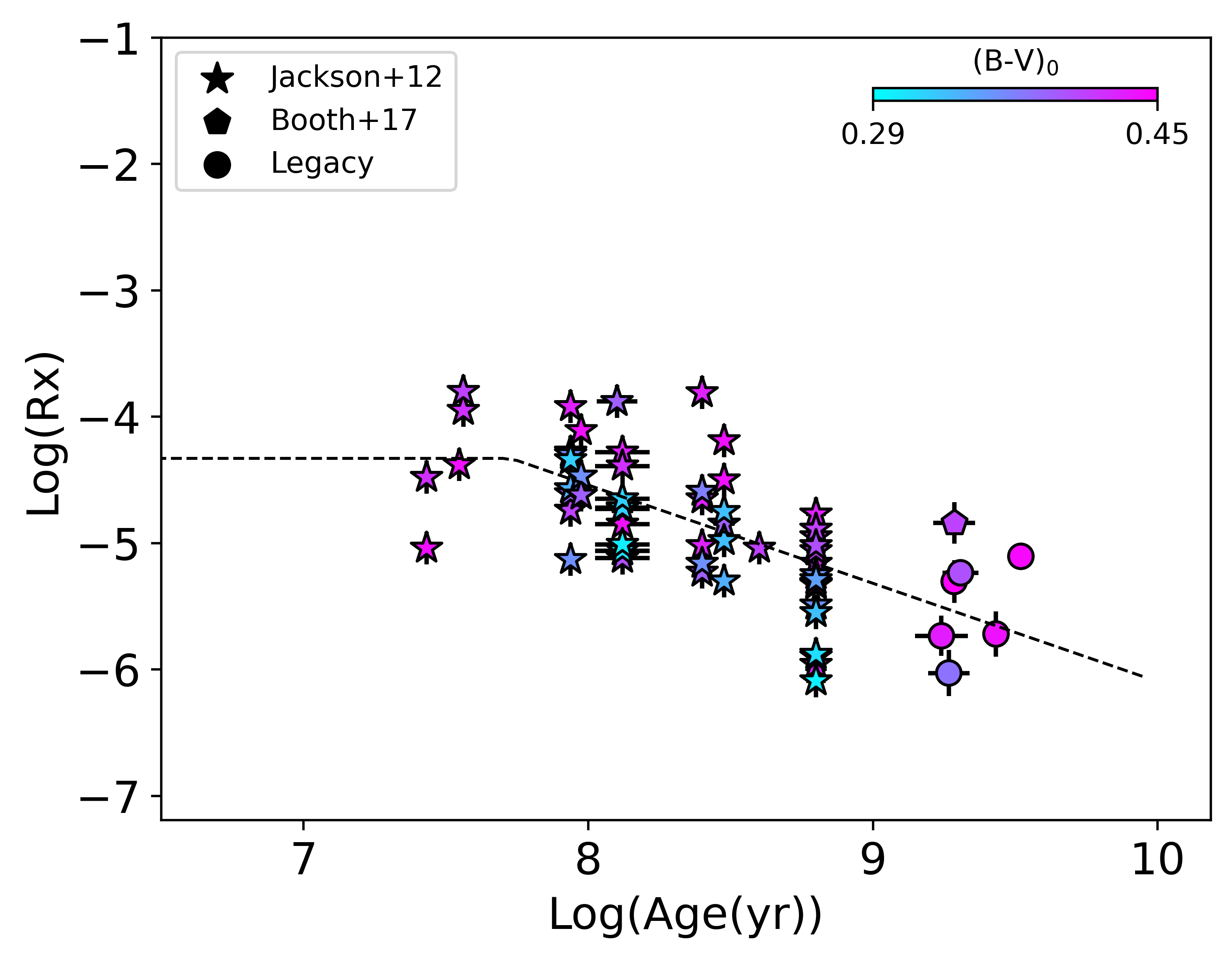}
        }
    \end{subfigure}
    \hfill
    \begin{subfigure}{
        \includegraphics[width=0.3\textwidth]{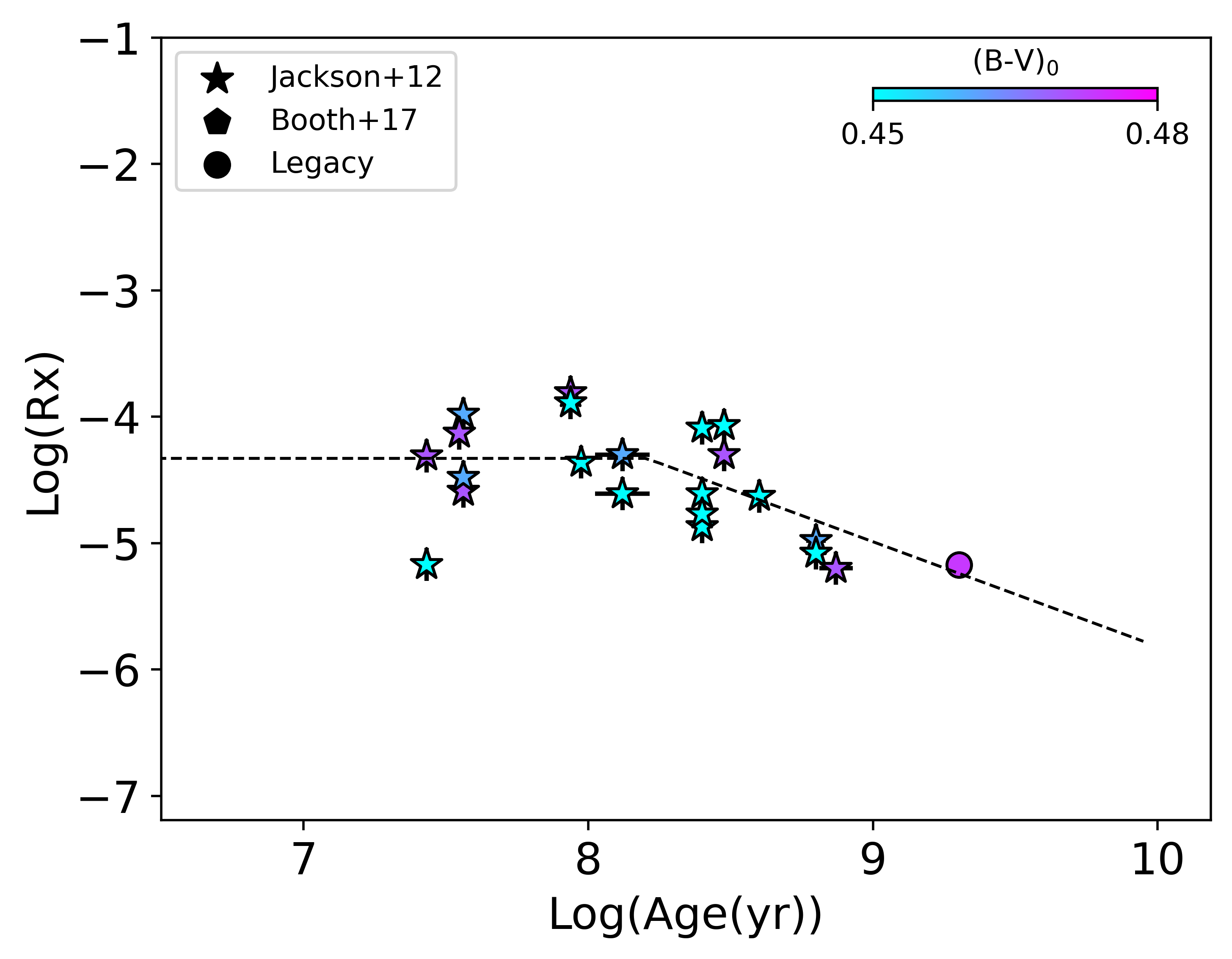}
        }
    \end{subfigure}
    \hfill
    \begin{subfigure}{
        \includegraphics[width=0.3\textwidth]{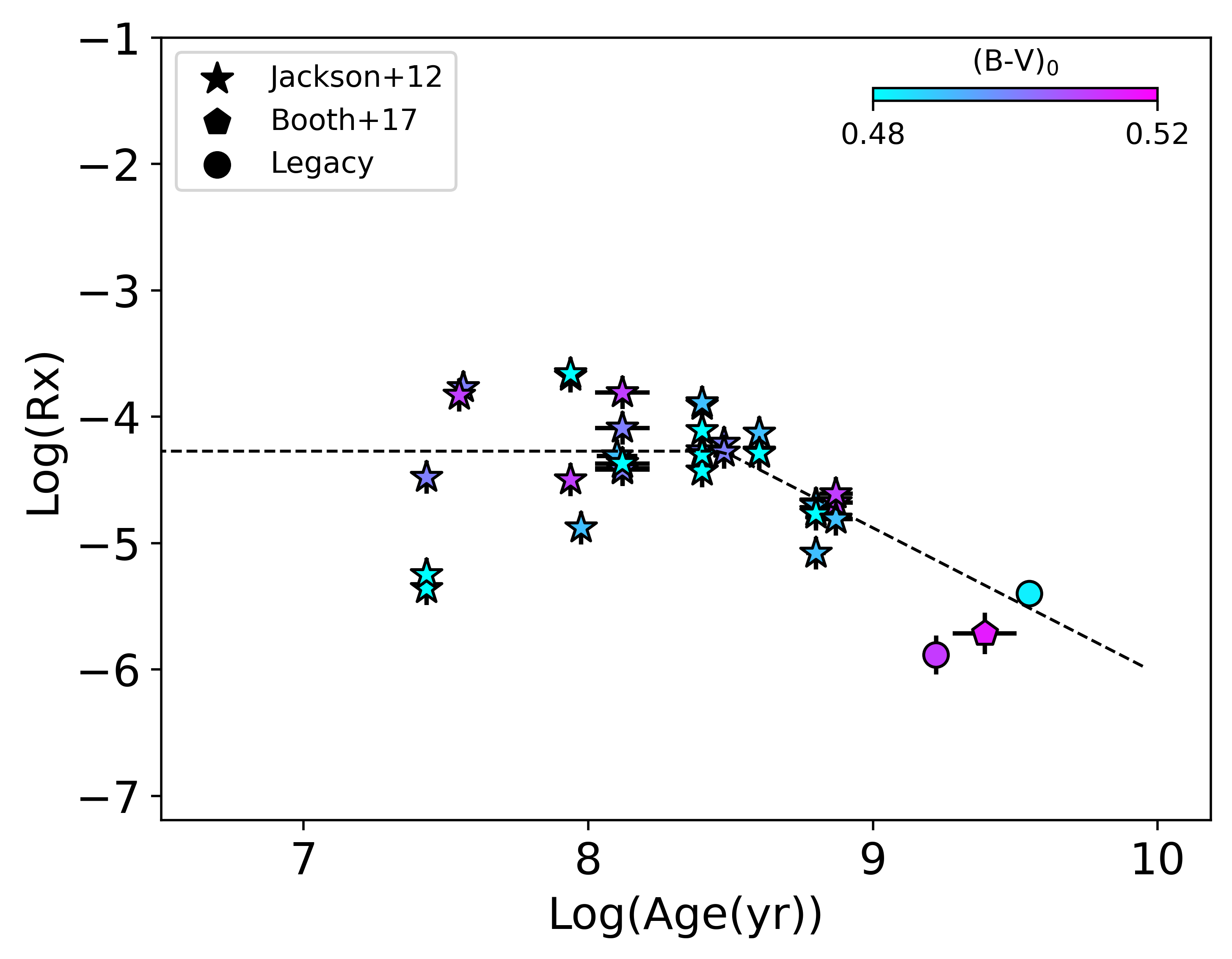}
        }
    \end{subfigure}

    \par\vspace{0.5em} 

    \begin{subfigure}{
        \includegraphics[width=0.3\textwidth]{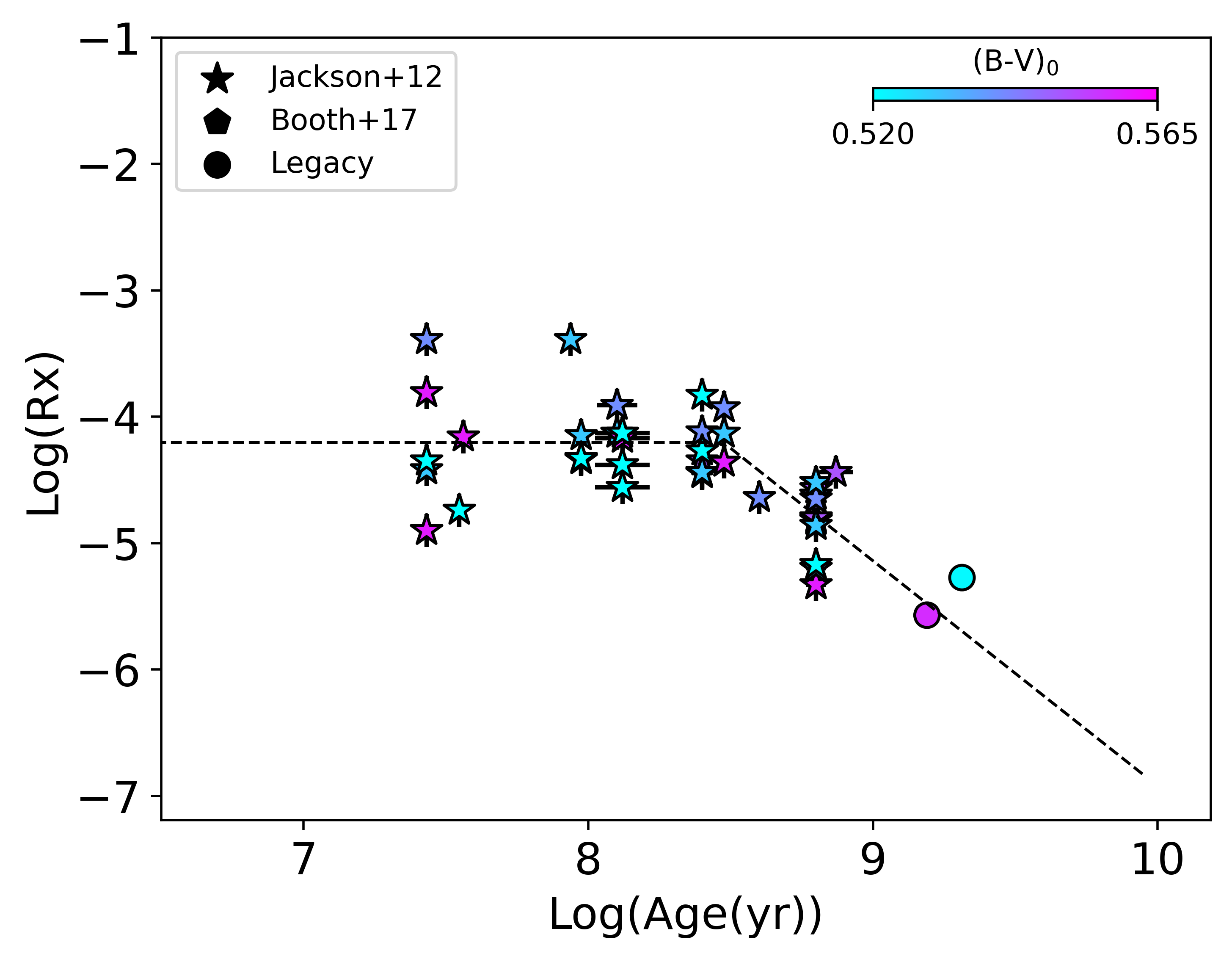}
        }
    \end{subfigure}
    \hfill
    \begin{subfigure}{
        \includegraphics[width=0.3\textwidth]{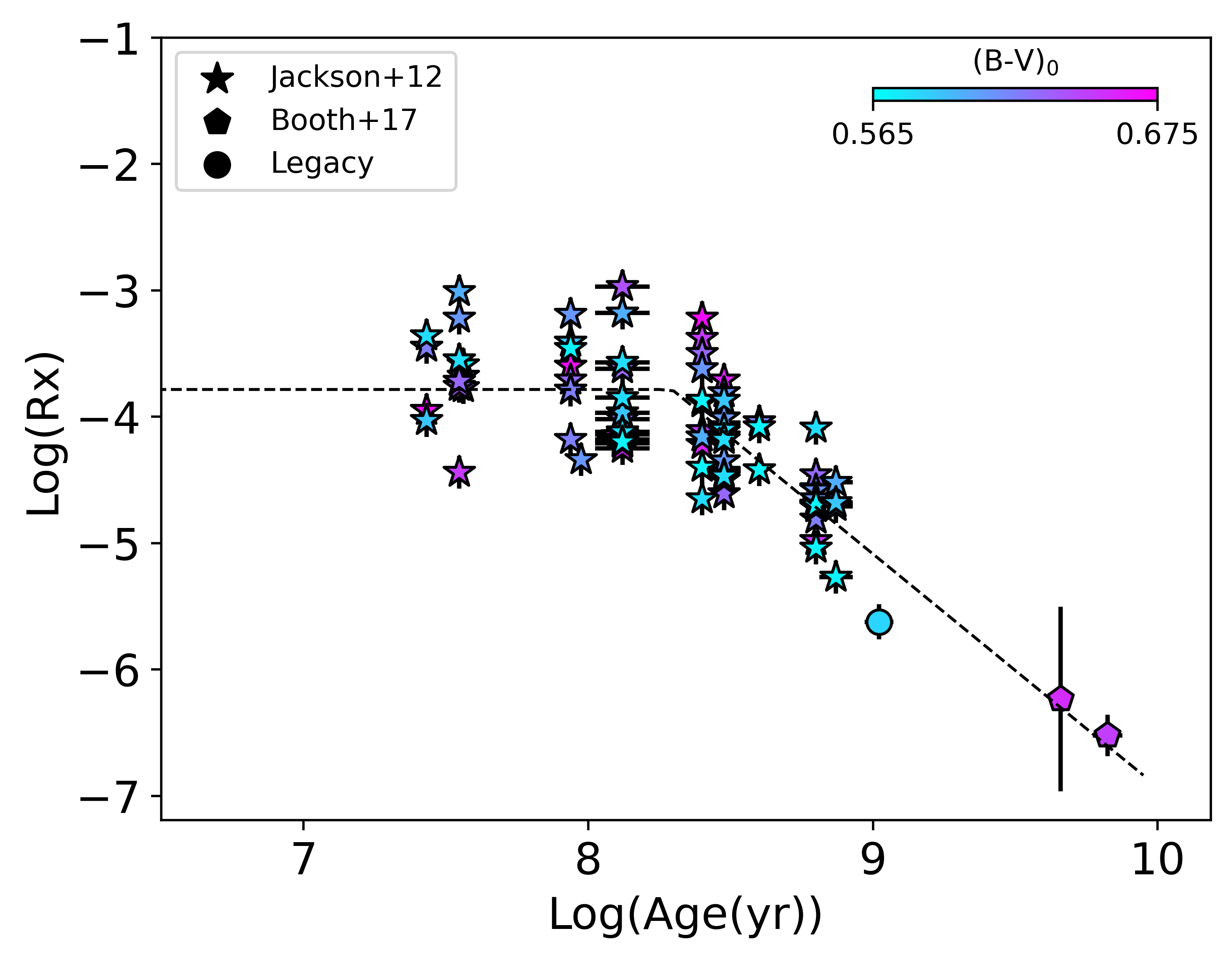}
        }
    \end{subfigure}
    \hfill
    \begin{subfigure}{
        \includegraphics[width=0.3\textwidth]{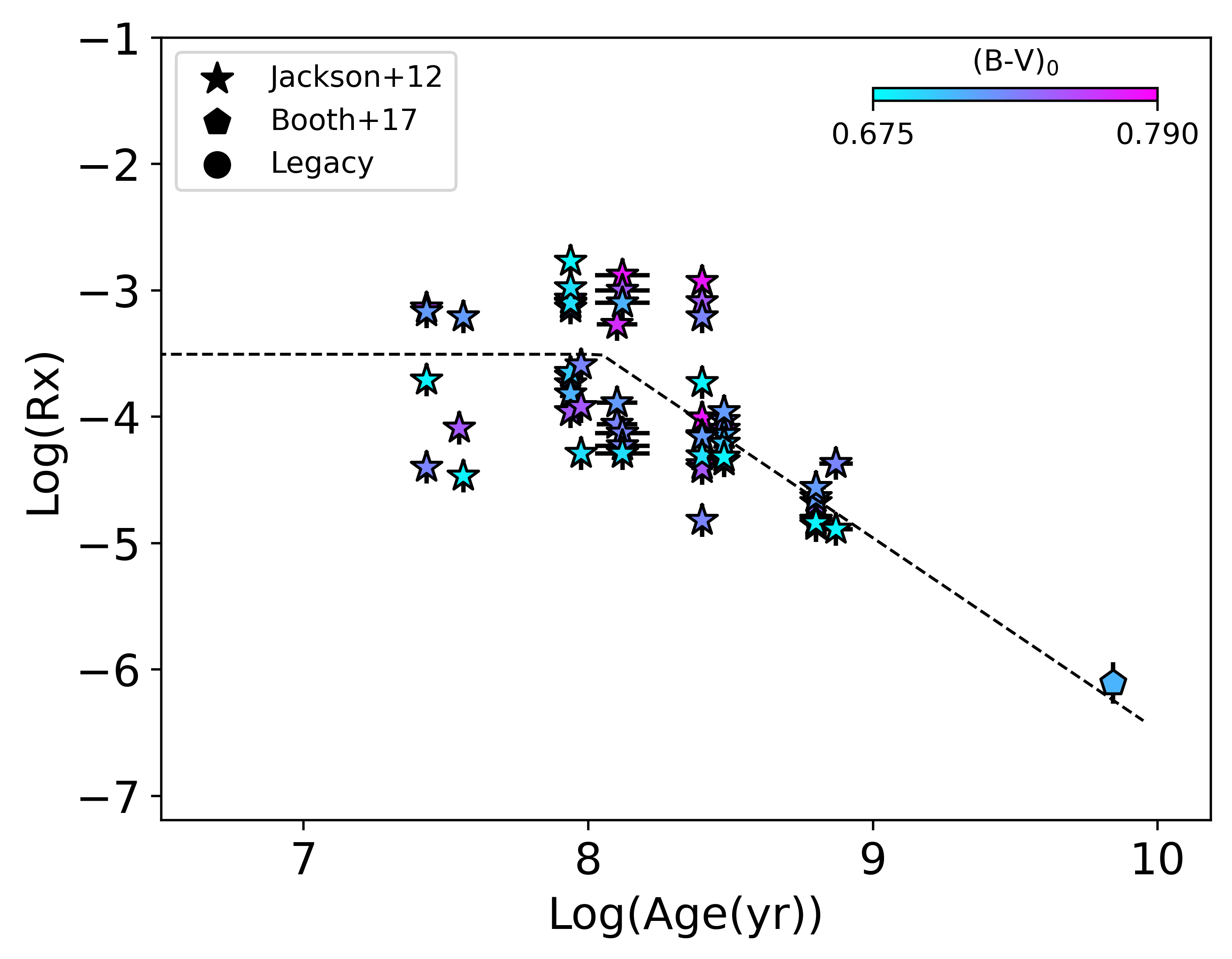}
        }
    \end{subfigure}

    \par\vspace{0.5em}

    \begin{subfigure}{
        \includegraphics[width=0.3\textwidth]{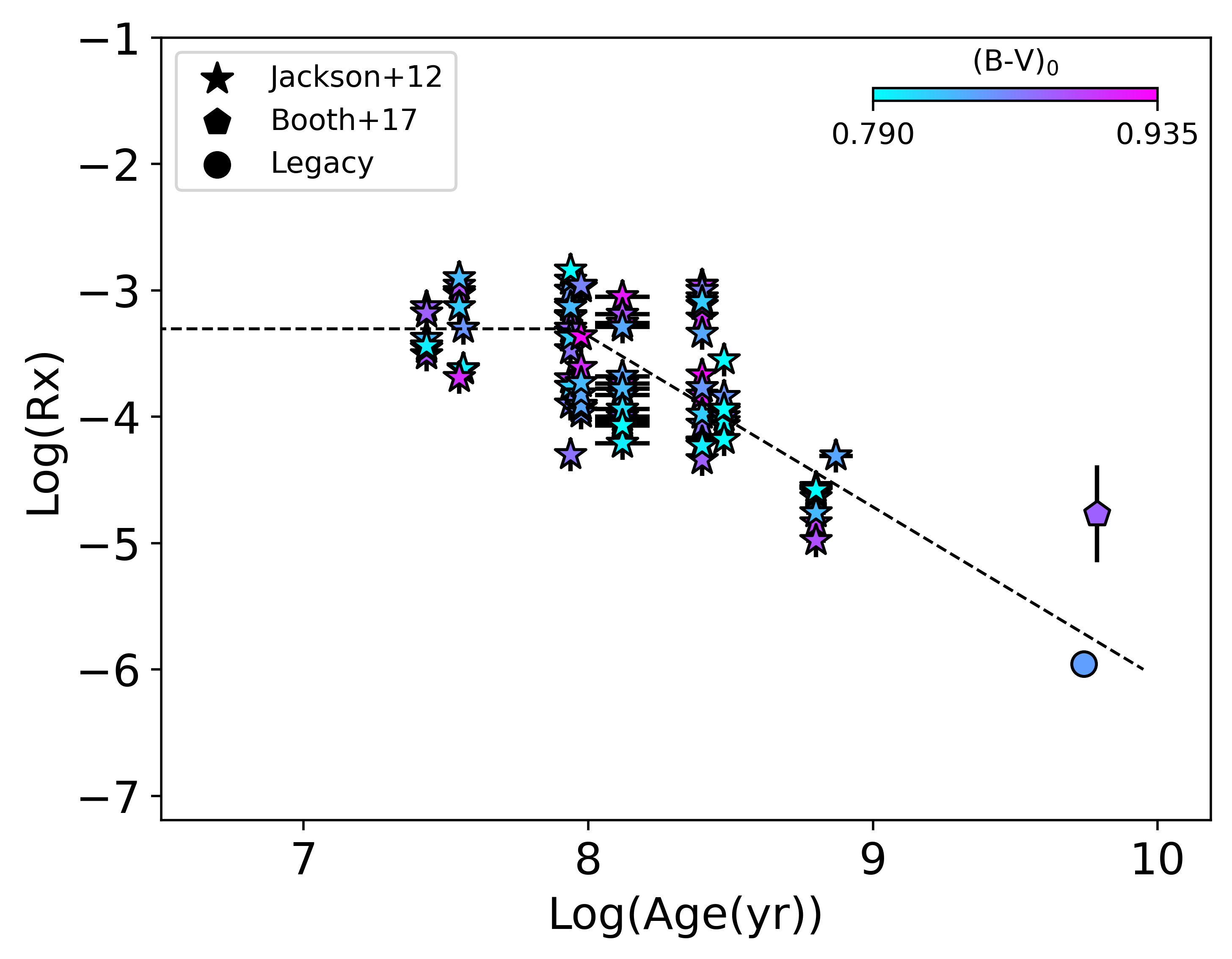}
        }
    \end{subfigure}
    \hfill
    \begin{subfigure}{
        \includegraphics[width=0.3\textwidth]{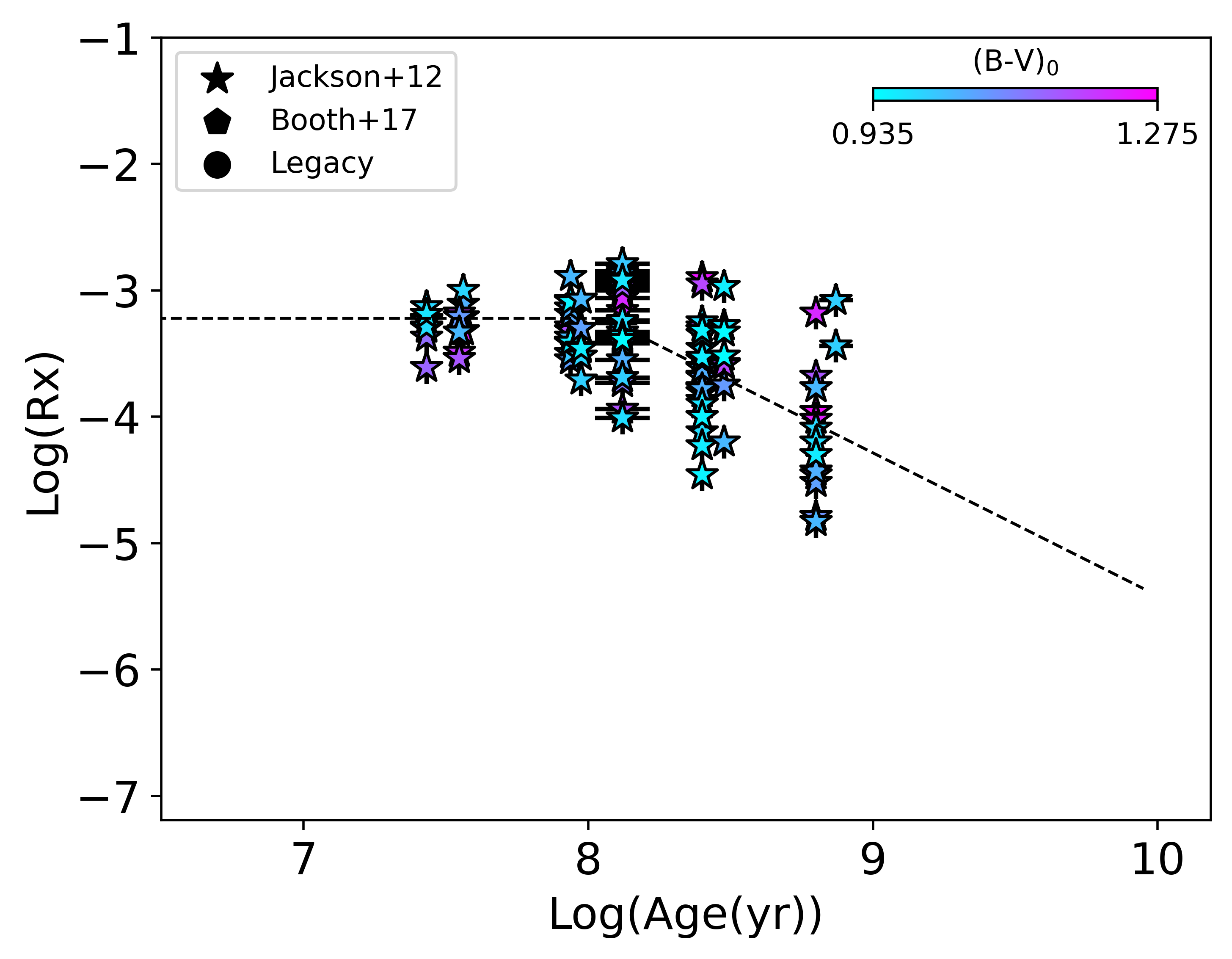}
        }
    \end{subfigure}
    \hfill
    \begin{subfigure}{
        \includegraphics[width=0.3\textwidth]{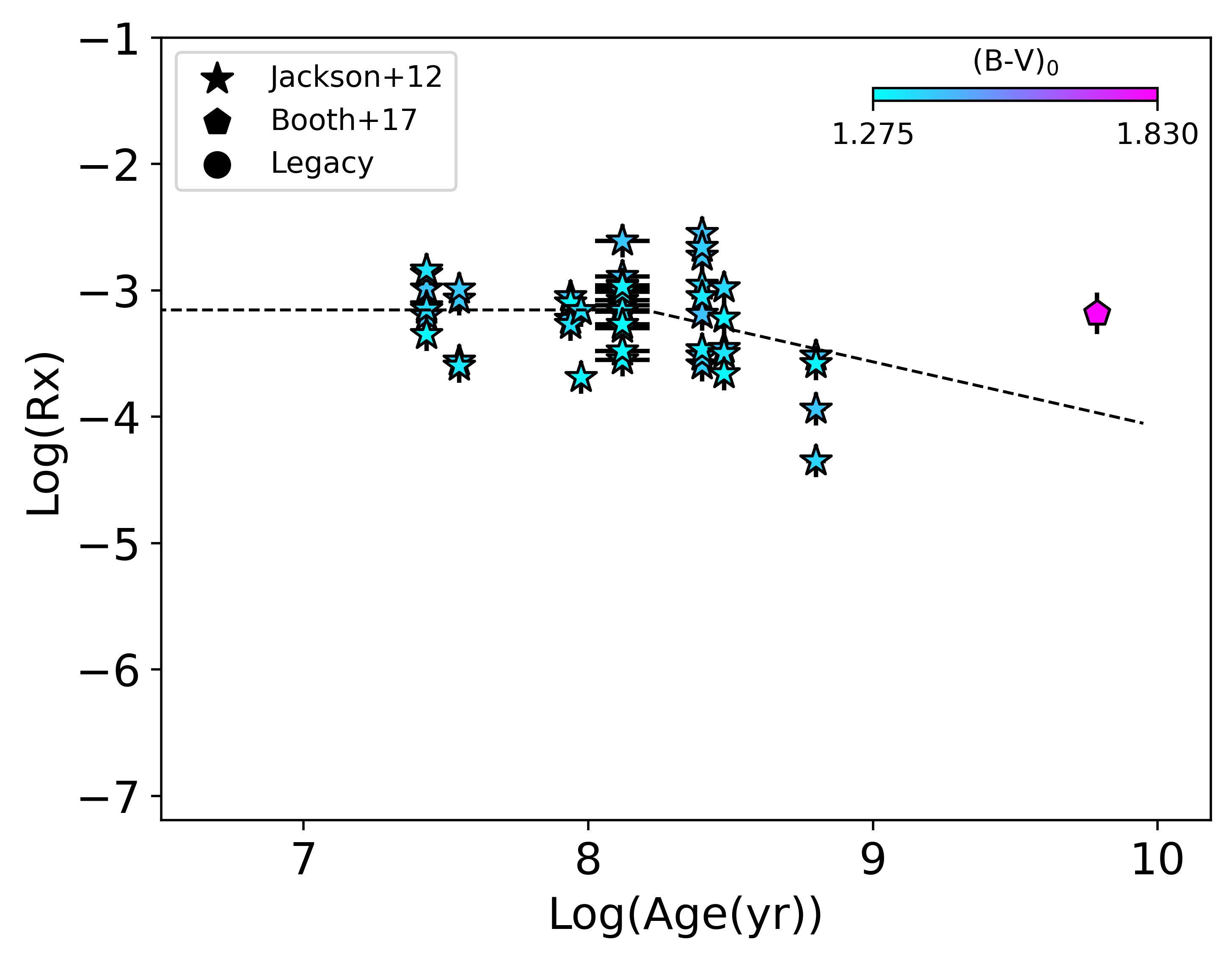}
        }
    \end{subfigure}

    \caption{Fit of the $\rm Log(R_x)$-$\rm Log(Age)$ relationship divided by $\rm (B-V)_0$ colour bin. The meaning of the symbols is the same one as in Fig.~\ref{fig:LogRx_LogAge}.}
    \label{Fig:Rx_Age_BV}
\end{figure*}

\begin{figure*}
    \centering
    \includegraphics[width=0.3\textwidth]{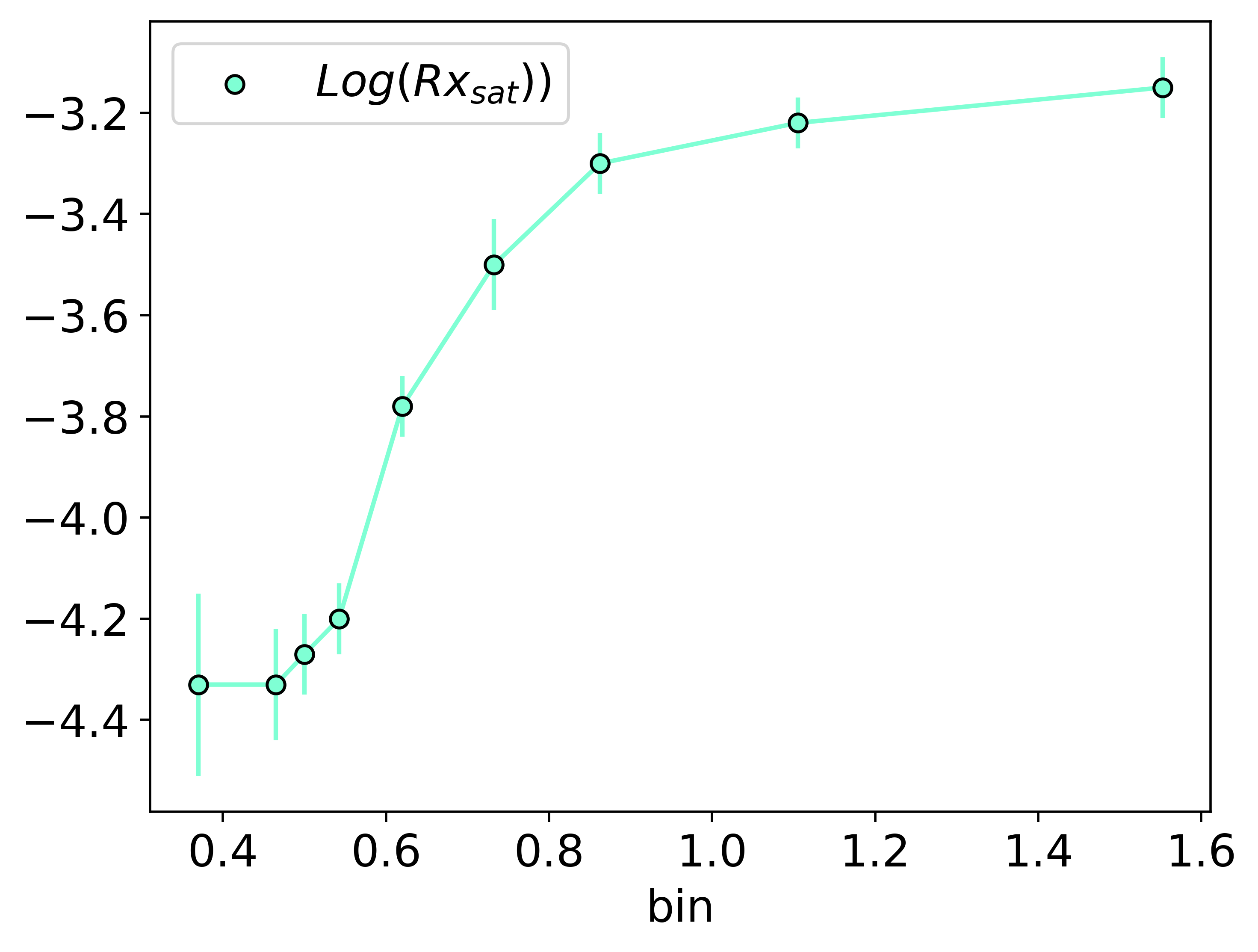}
    \includegraphics[width=0.3\textwidth]{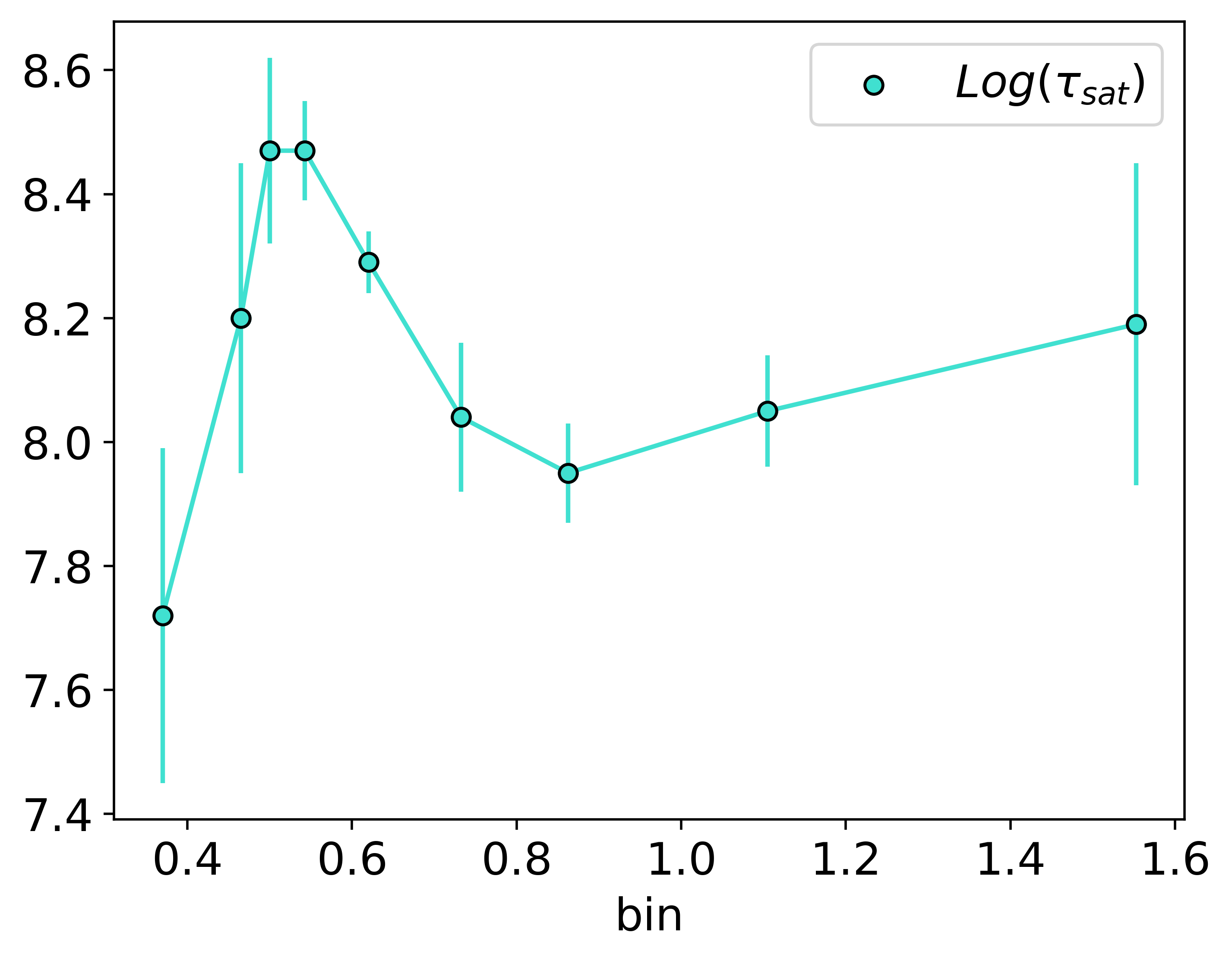}
    \includegraphics[width=0.3\textwidth]{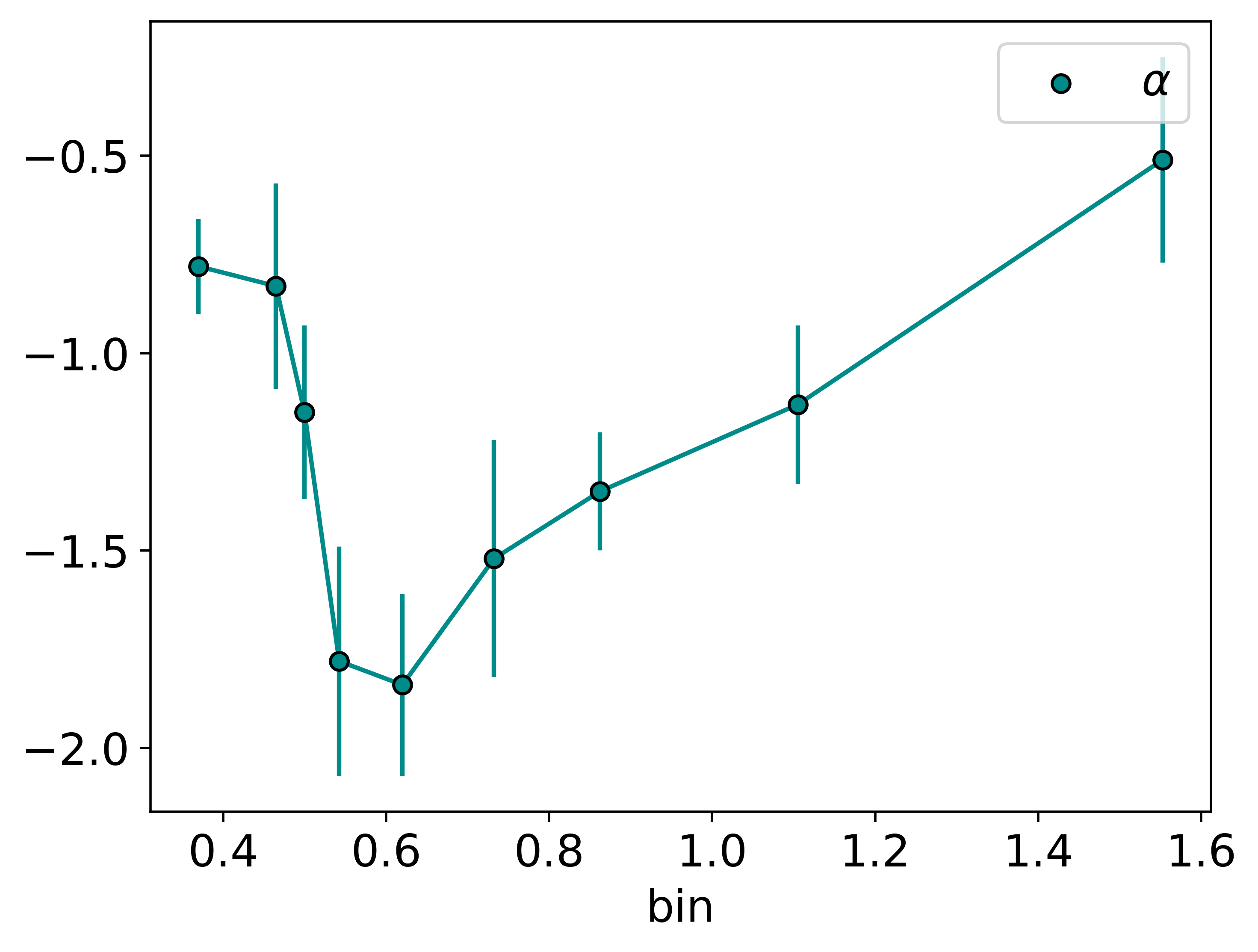}
    \caption{From left to right: trends of the fitting parameters listed in Table~\ref{Tab:Rx_Age}, $\rm Log(R_{x,sat})$, $\rm Log(\tau_{sat})$ and $\alpha$ as function of the $\rm (B-V)_0$ colour bin.}
    \label{fig:params_trends}
\end{figure*}

In Fig.~\ref{fig:LogRx_LogAge} and \ref{Fig:Rx_Age_BV}, the fit of the $\rm Log(R_x)$-$\rm Log(Age)$ relationship is represented for the global and split ranges of the stellar $\rm (B-V)_0$ colour, respectively. In Fig.~\ref{fig:params_trends}, the trends of the fitting parameters are indicated, as a function of the $\rm (B-V)_0$ colour bins. We also studied the impact of removing stars with $\rm T_{eff}(K) > 6250$ from the fitting procedure, finding a negligible impact.

\section{Log(R$\rm _x$)-Log($\rm Ro$) relationship for \textit{Kepler} LEGACY stars with $\rm T_{eff}(K) \leq 6250$}
\label{App:Rx_Ro}

In Table~\ref{Tab:Rx_Ro_reduced} we present the results obtained in the fitting procedure on the Log(R$\rm _x$)-Log($\rm Ro$) relationship when excluding stars with $\rm T_{eff}(K) > 6250$.

\begin{table}
\noindent
\captionof{table}{Results of the fitting procedure determined with an analogous procedure to Table~\ref{Tab:Rx_Ro}, but for \textit{Kepler} LEGACY stars with $\rm T_{eff}(K) \leq 6250$.}
\centering
\resizebox{\columnwidth}{!}{
\begin{tabular}{|l|c|c|c|c|}
 \cline{2-5}
 \multicolumn{1}{c|}{} & a & b & c & $\rm Log(Ro_{sat})$ \\
 \hline
 $\rm Ro_{(V-Ks)_0}$, all & $-0.26 \pm 0.08$  & $-3.64 \pm 0.11$ & $-1.86 \pm 0.14$ & $-0.68 \pm 0.04$ \\
 $\rm Ro_{(V-Ks)_0}$, K-F & $-0.29 \pm 0.10$  & $-3.69 \pm 0.14$ & $-1.84 \pm 0.15$ & $-0.68 \pm 0.05$ \\
 \hline
 $\rm Ro_{T_{eff}}$, all & $-0.13 \pm 0.09$  & $-3.41 \pm 0.15$ & $-1.58 \pm 0.12$ & $-0.98 \pm 0.05$ \\
 $\rm Ro_{T_{eff}}$, K-F & $-0.012 \pm 0.118$  & $-3.27 \pm 0.18$ & $-1.89 \pm 0.14$ & $-0.98 \pm 0.05$ \\
 \hline
\end{tabular}}
\label{Tab:Rx_Ro_reduced}
\end{table}

\section{Log(R$\rm _x$)-Log($\rm Ro$) fit dependence on $\rm [Fe/H]$}
\label{App:fig_Rx_Ro_FeH} 

\begin{figure*}
    \centering
    \begin{subfigure}{
        \includegraphics[width=0.4\textwidth]{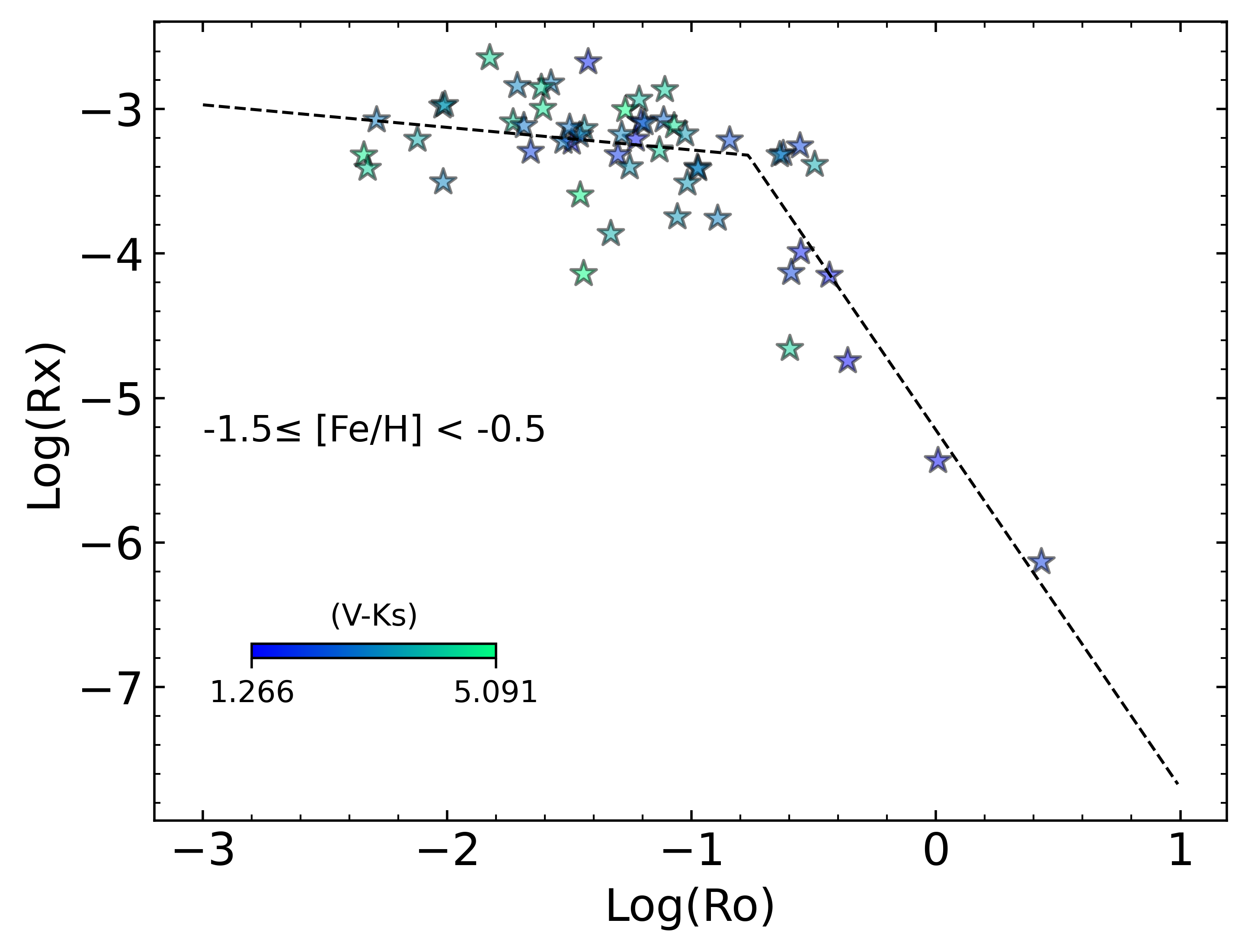}
        }
    \end{subfigure}
    \hspace{0.02\textwidth}
    \vspace{-3.4\baselineskip} 
    \begin{subfigure}{
        \includegraphics[width=0.4\textwidth]{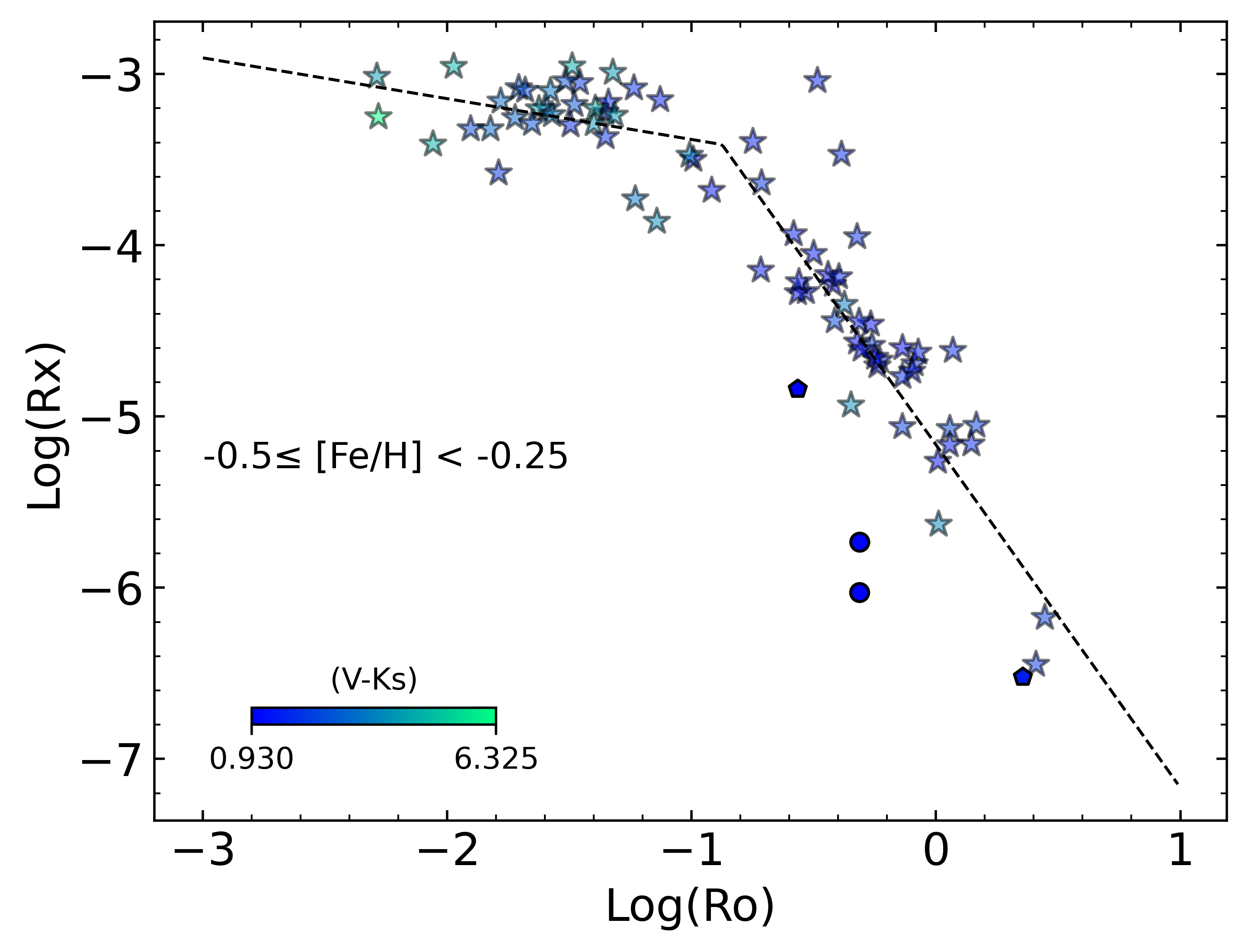}
        }
    \end{subfigure}

    \par\vspace{0.5em} 

    \begin{subfigure}{
        \includegraphics[width=0.4\textwidth]{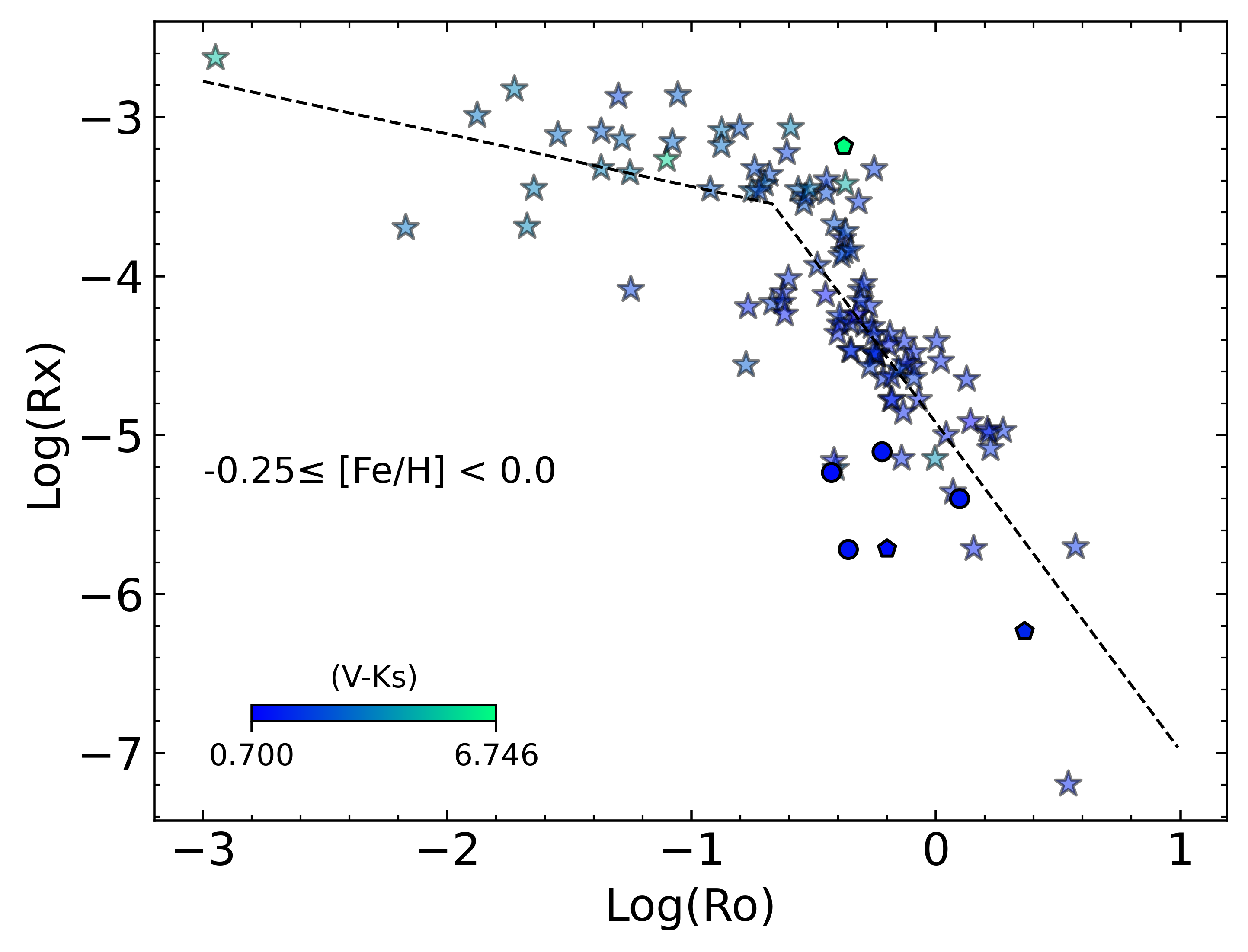}
        }
    \end{subfigure}
    \hspace{0.02\textwidth}
    \vspace{-1.4\baselineskip} 
    \begin{subfigure}{
        \includegraphics[width=0.4\textwidth]{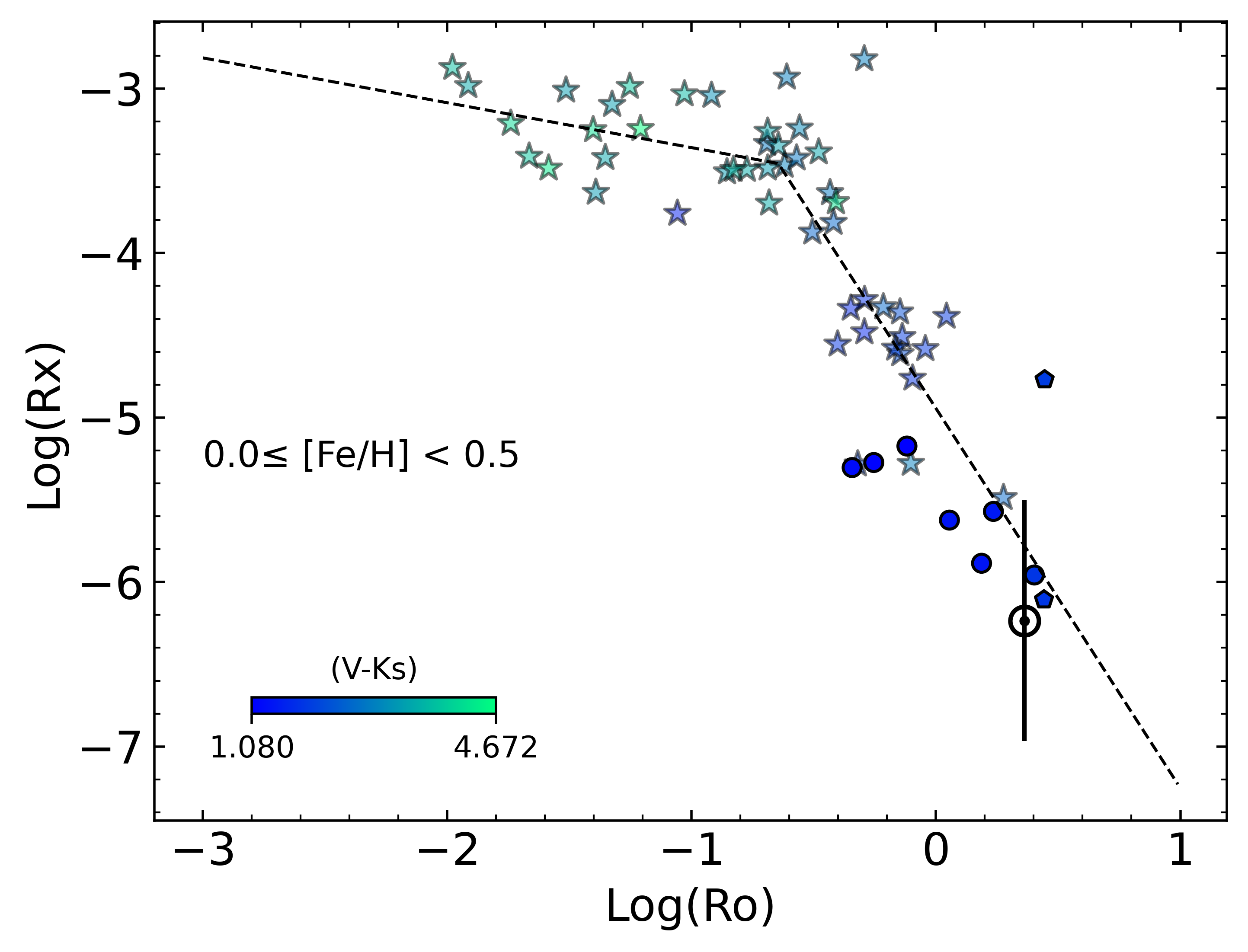}
        }
    \end{subfigure}
    \caption{Distributions of stars from W11 (star symbols), \textit{Kepler} LEGACY (dots) and B17 (pentagons) in the $\rm Log(R_x)$–$\rm Log(Ro)$ plane, split in four metallicity bins, with $\rm [Fe/H]$ increasing from left to right, top to bottom. $\rm Ro$ is computed as in W18, and the stars are colour coded according to $\rm (V-Ks)_0$.}
    \label{Fig:Rx_Ro_VKs_FeH}
\end{figure*}

\begin{figure*}
    \centering
    \begin{subfigure}{
        \includegraphics[width=0.4\textwidth]{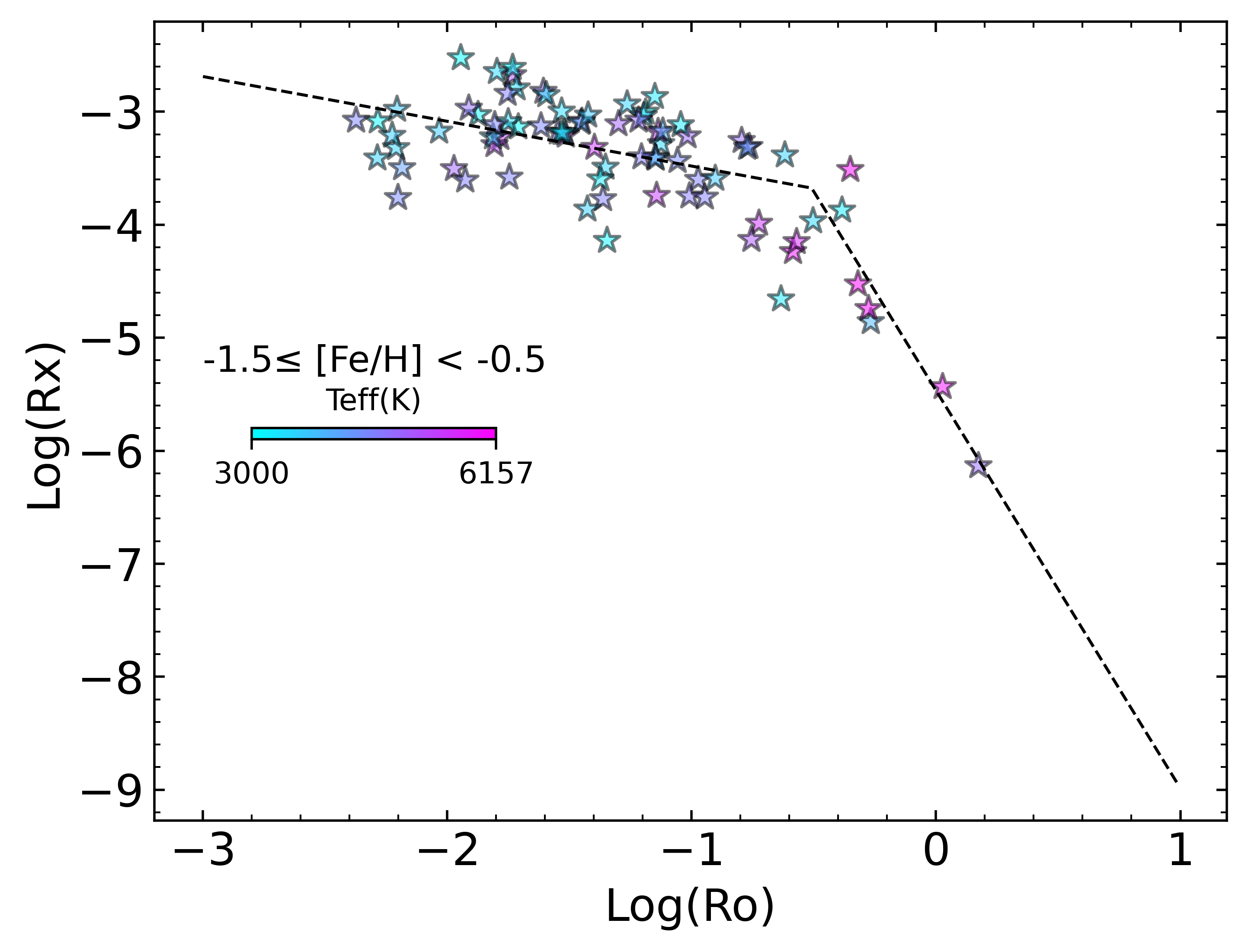}
        }
    \end{subfigure}
    \hspace{0.02\textwidth}
    \vspace{-3.0\baselineskip} 
    \begin{subfigure}{
        \includegraphics[width=0.4\textwidth]{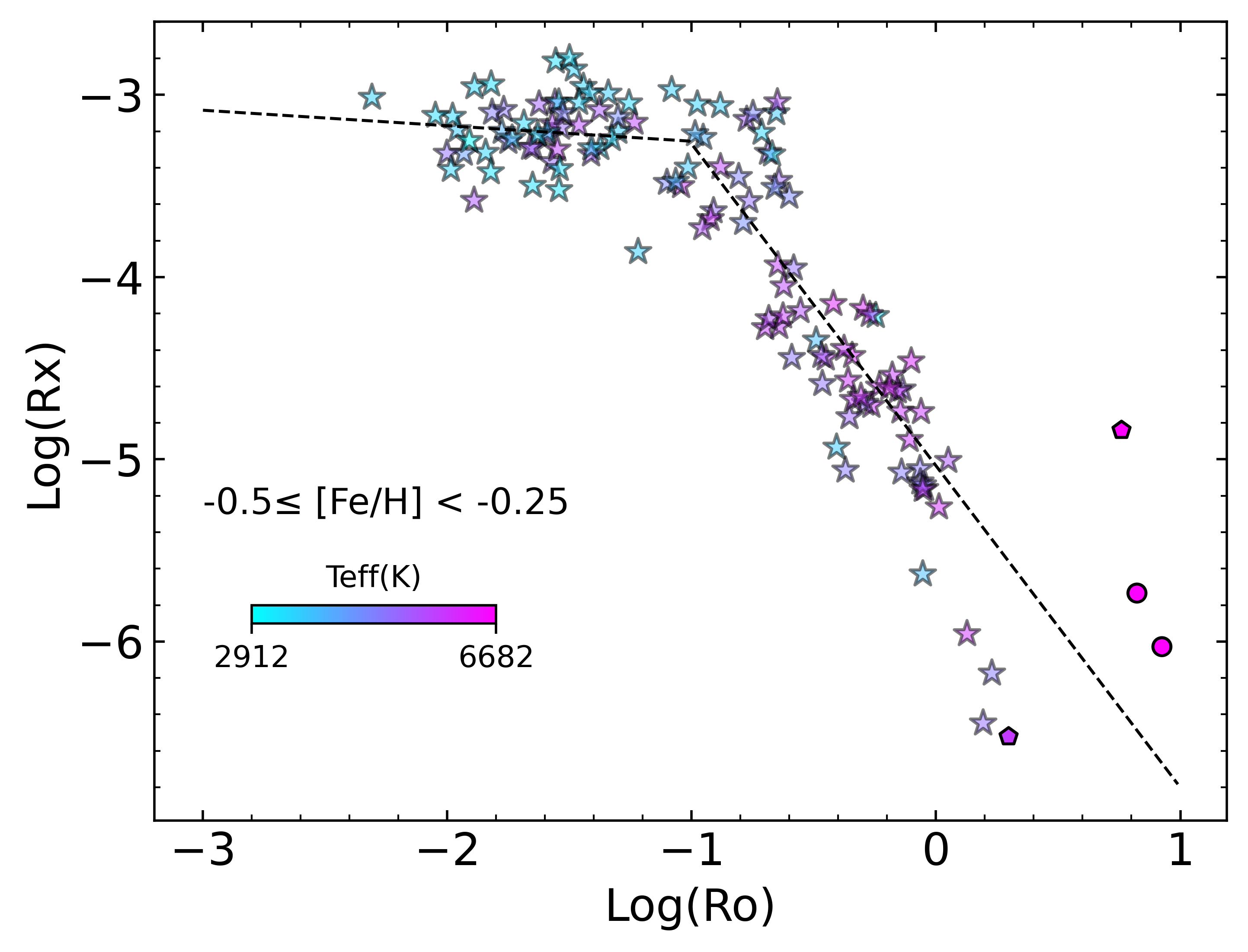}
        }
    \end{subfigure}

    \par\vspace{0.0em} 

    \begin{subfigure}{
        \includegraphics[width=0.4\textwidth]{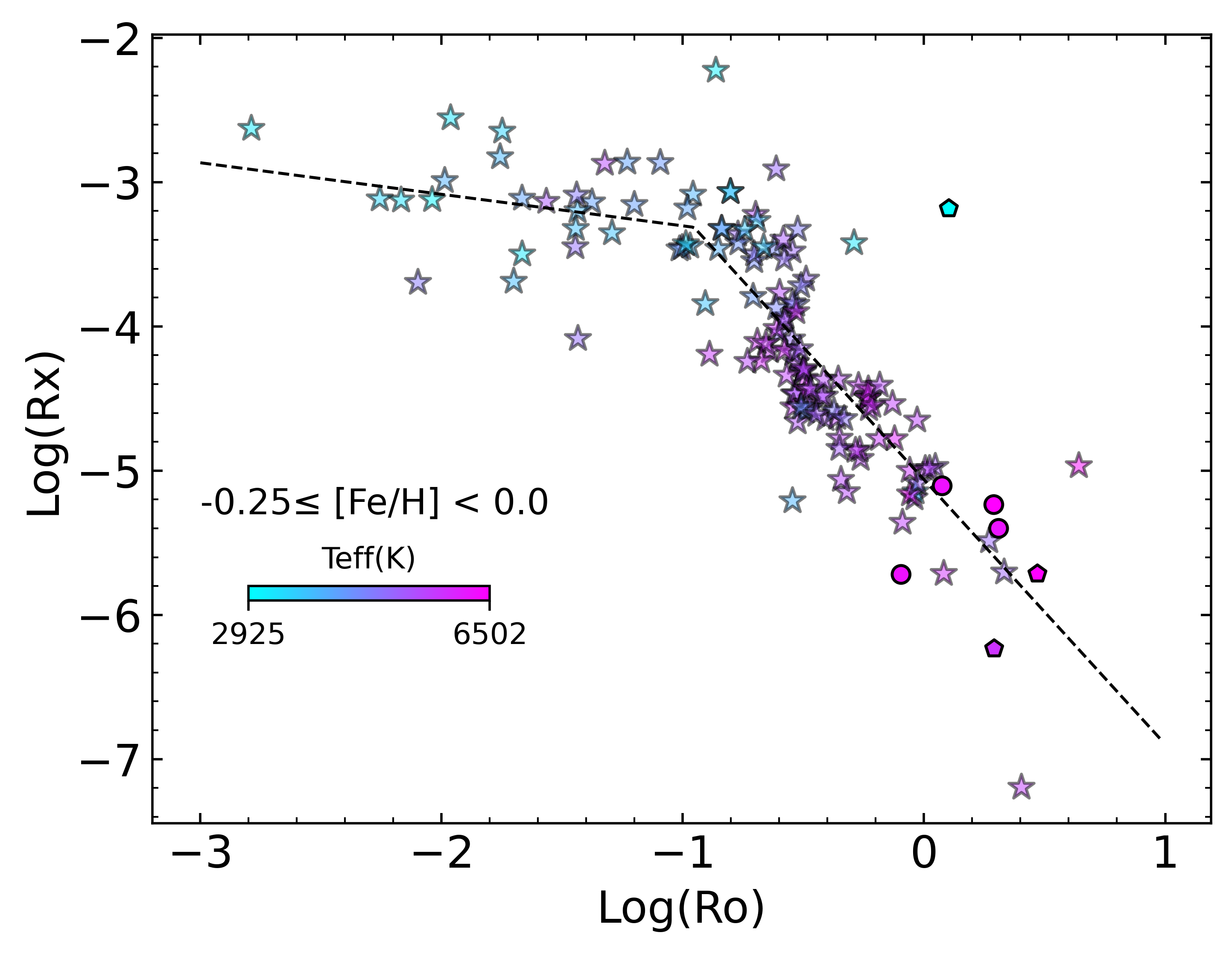}
        }
    \end{subfigure}
    \hspace{0.02\textwidth}
    \vspace{-1.4\baselineskip} 
    \begin{subfigure}{
        \includegraphics[width=0.4\textwidth]{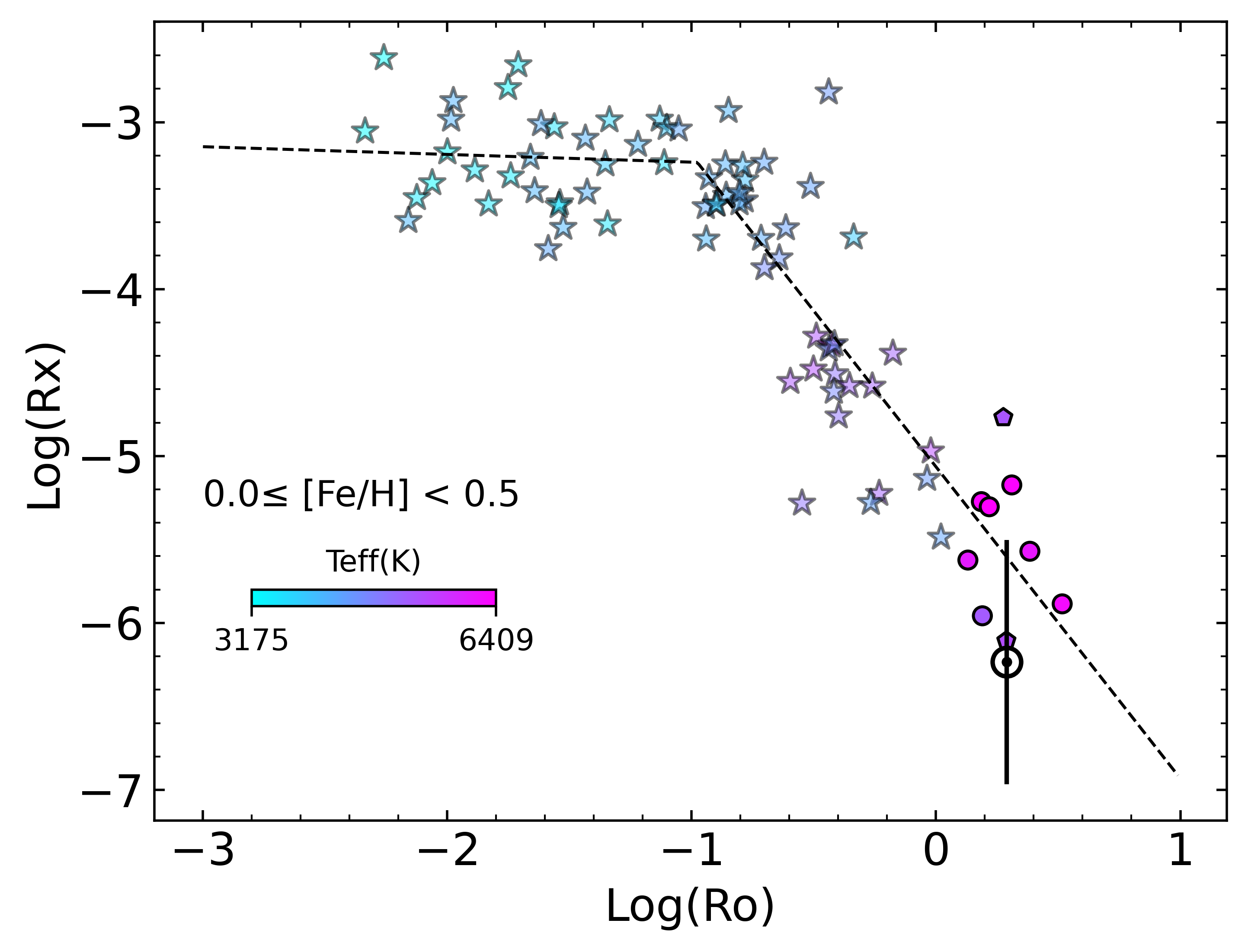}
        }
    \end{subfigure}
    \caption{Distributions of stars from W11 (star symbol), \textit{Kepler} LEGACY (dots) and B17 (pentagons) in the $\rm Log(R_x)$–$\rm Log(Ro)$ plane, split in four metallicity bins, with $\rm [Fe/H]$ increasing from left to right, top to bottom. $\rm Ro$ is computed as in \citet{Cranmer2011}, and the stars are coloured coded according to $\rm T_{eff}$.}
    \label{Fig:Rx_Ro_Teff_FeH}
\end{figure*}

\begin{figure*}
    \centering
    \includegraphics[width=0.24\textwidth]{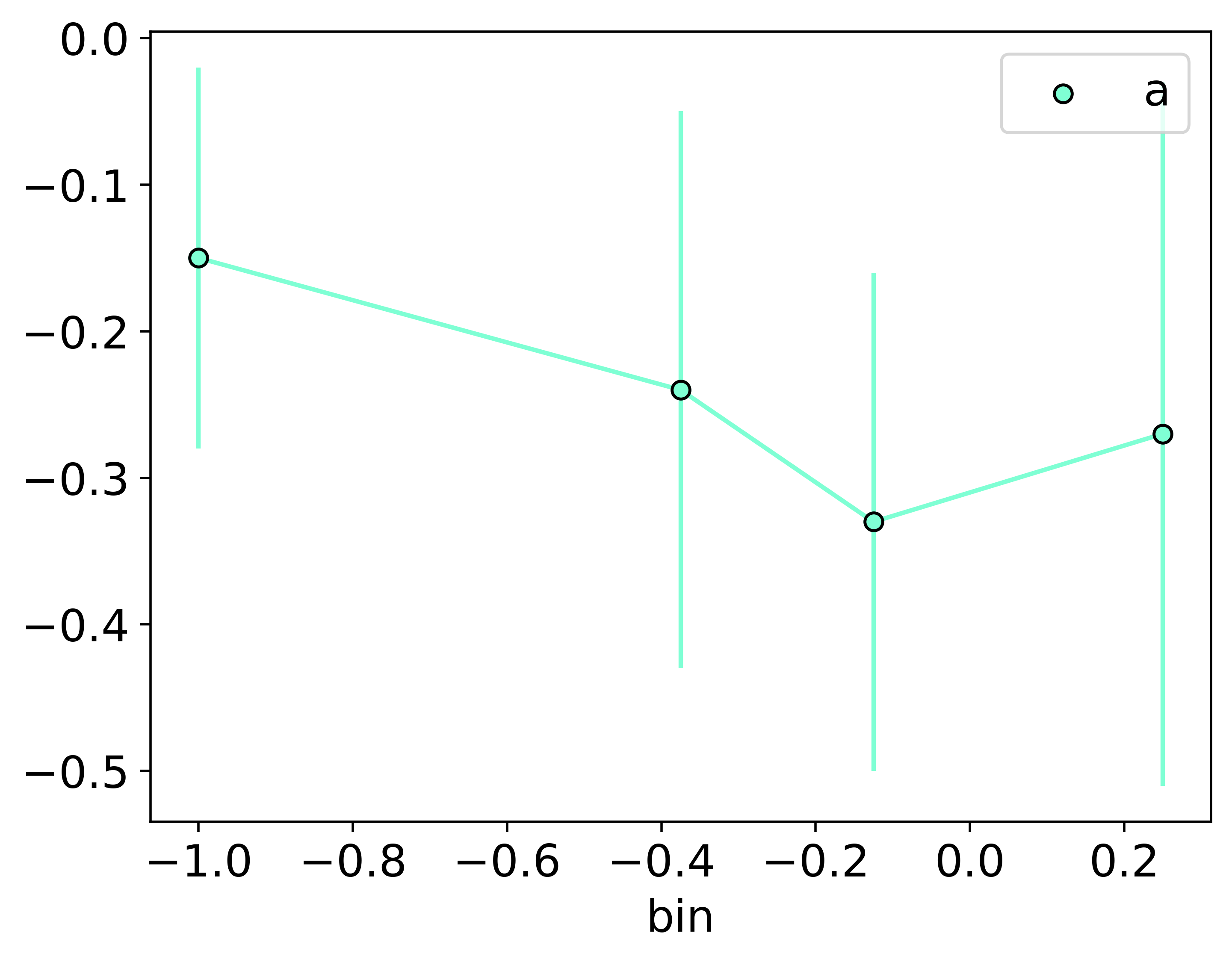}
    \includegraphics[width=0.24\textwidth]{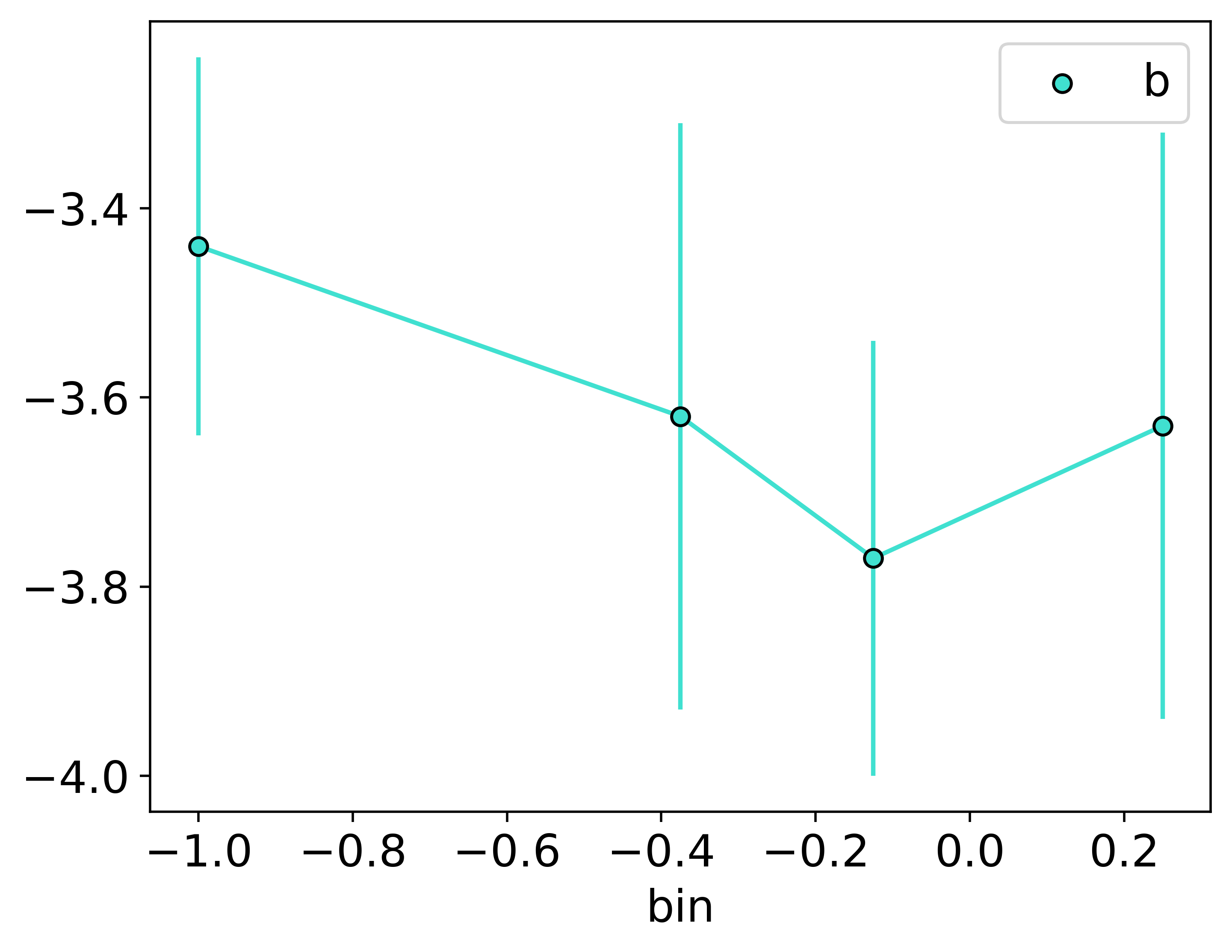}
    \includegraphics[width=0.24\textwidth]{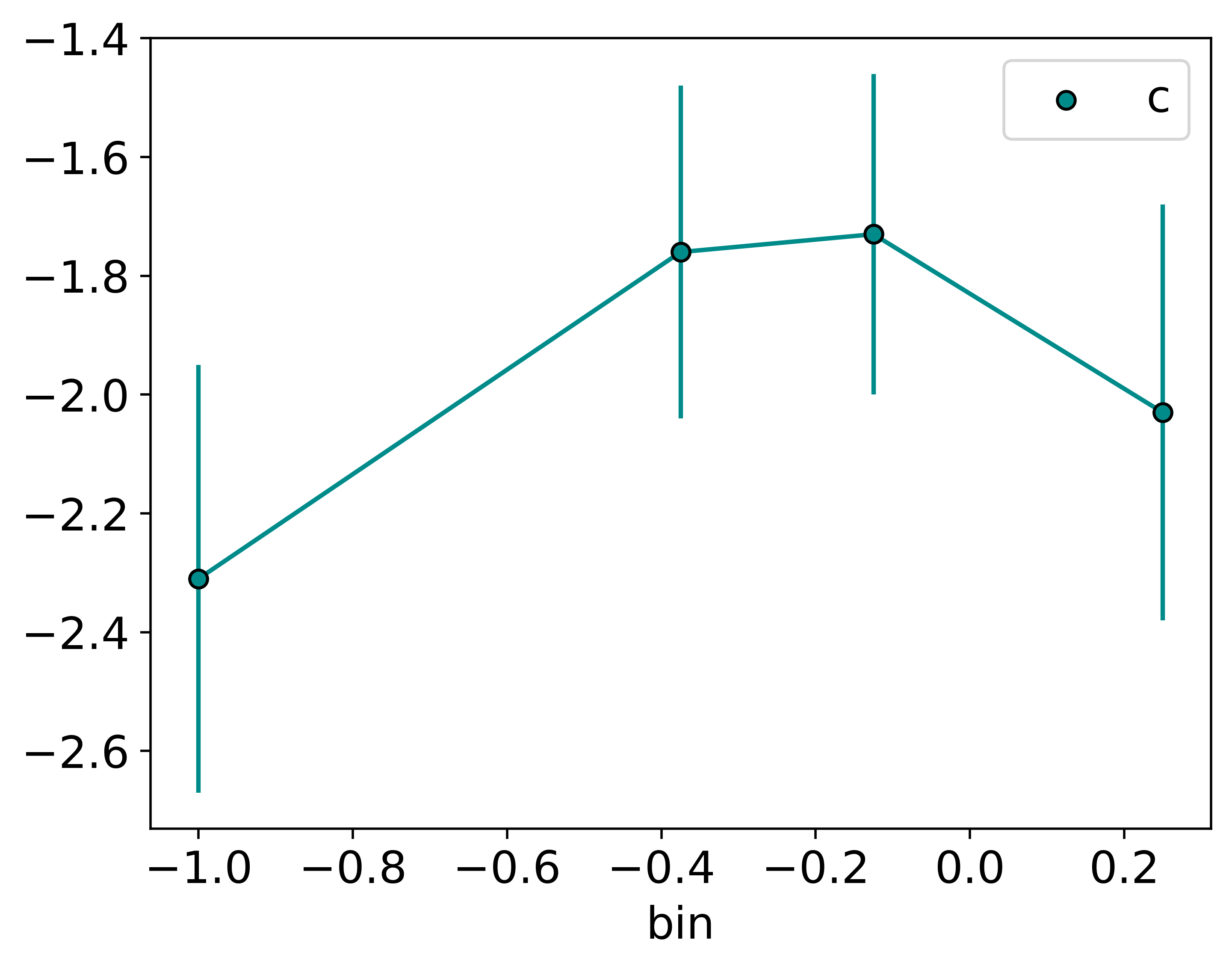}
    \includegraphics[width=0.24\textwidth]{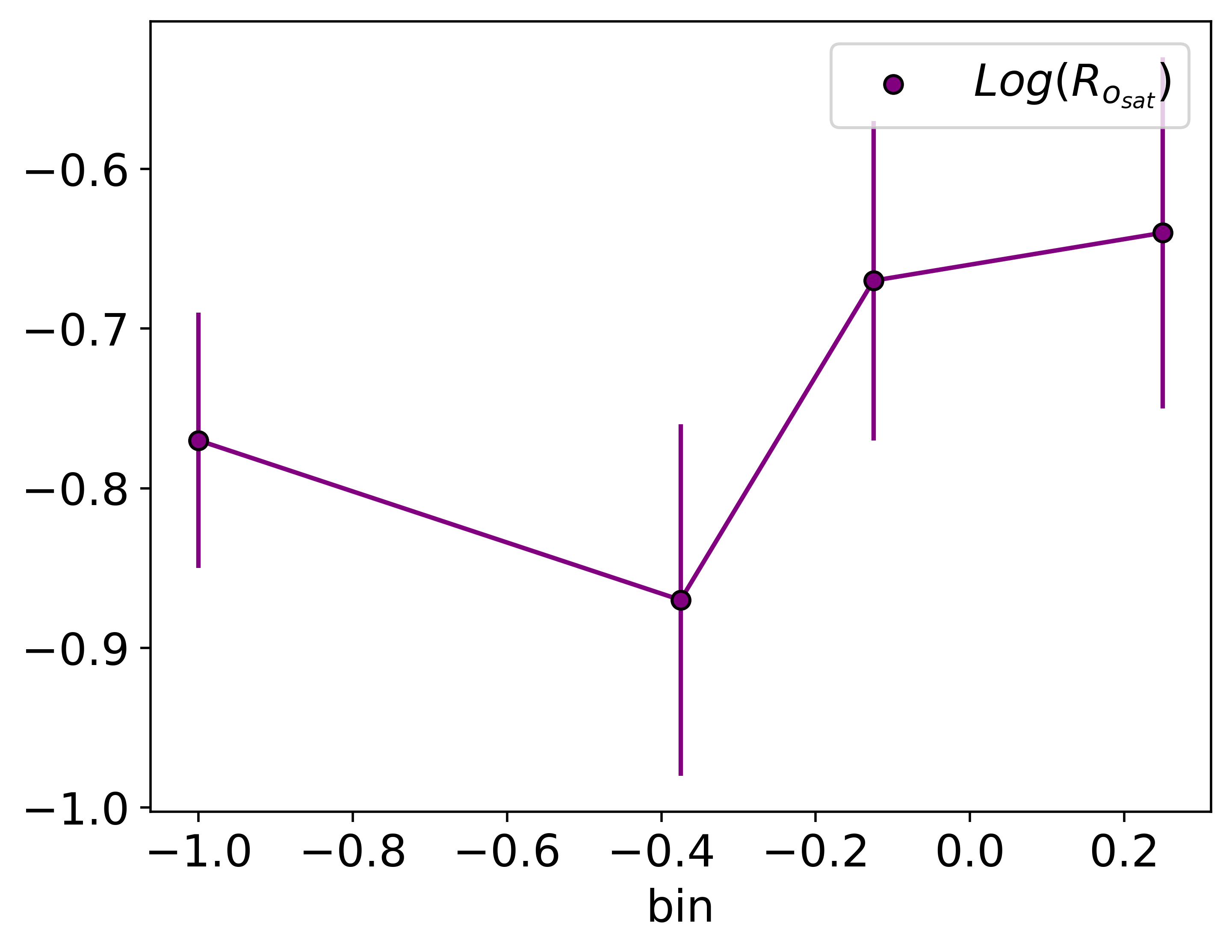}
    \caption{From left to right: trends of the fitting parameters listed in Table~\ref{Tab:Rx_Ro_FeH}, ``a'', ``b'', ``c'' and $\rm Log(Ro_{{sat}})$ as a function of the metallicity bin, for $\rm Ro$ computed as in W18.}
    \label{fig:params_trends_metallicity_bins_VKs}
\end{figure*}

\begin{figure*}
    \centering
    \includegraphics[width=0.24\textwidth]{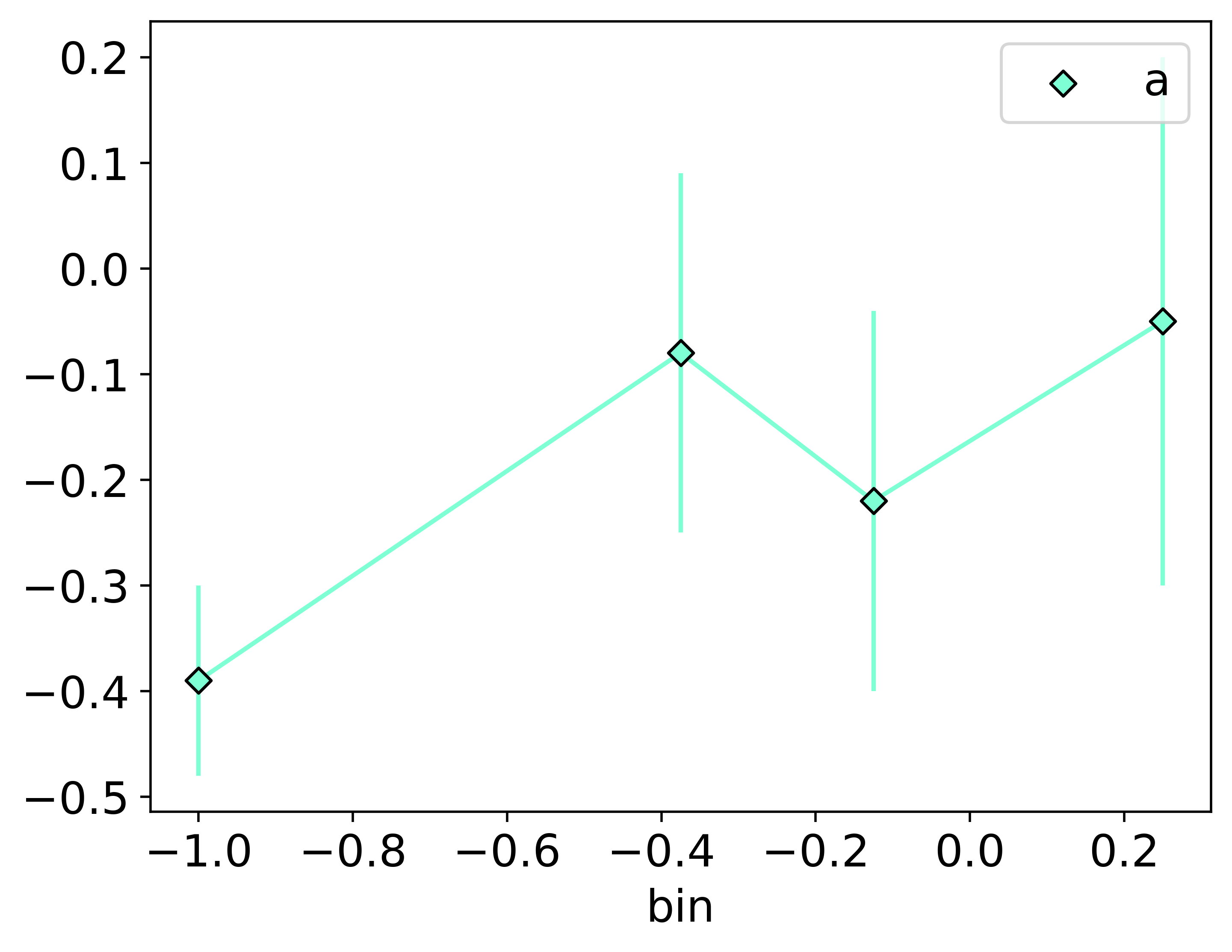}
    \includegraphics[width=0.24\textwidth]{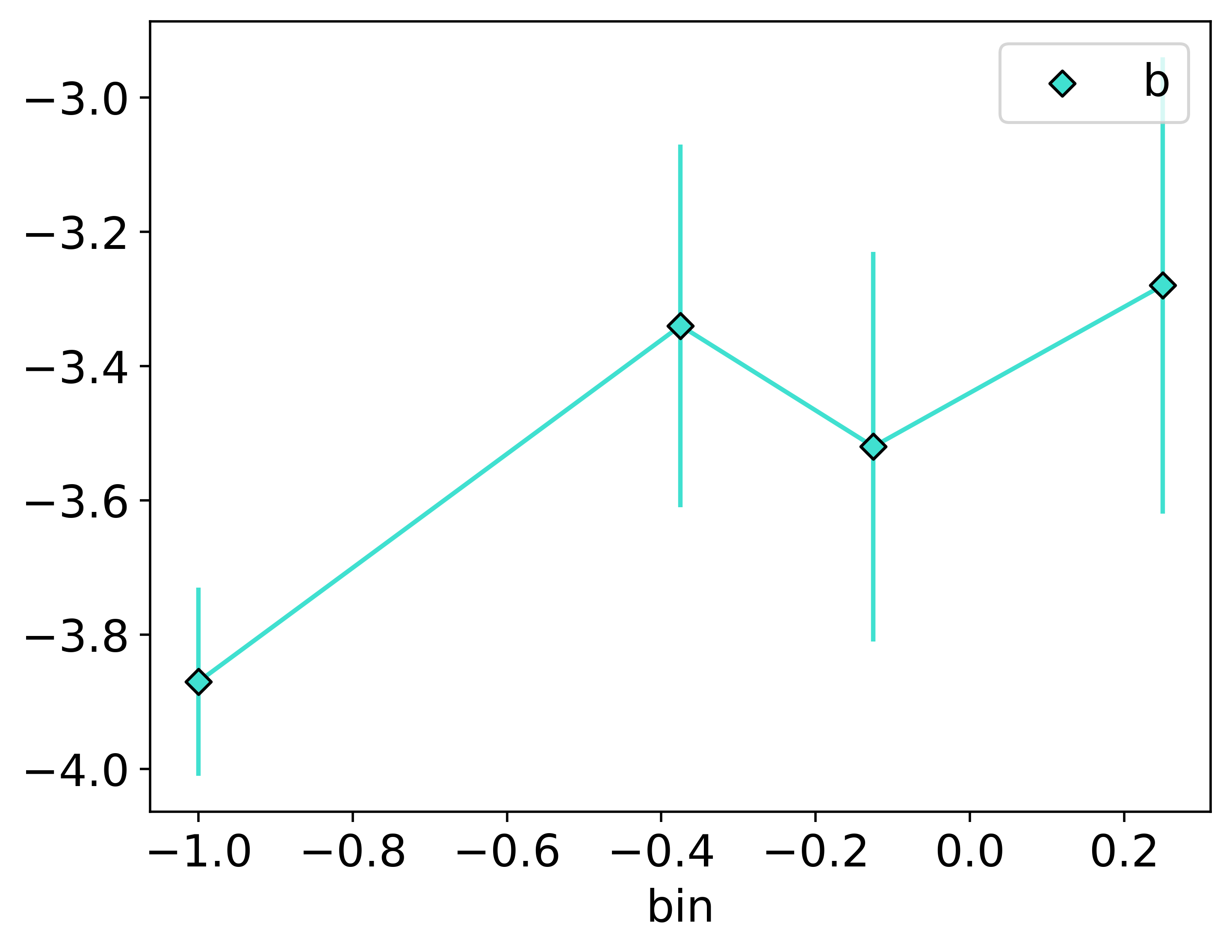}
    \includegraphics[width=0.24\textwidth]{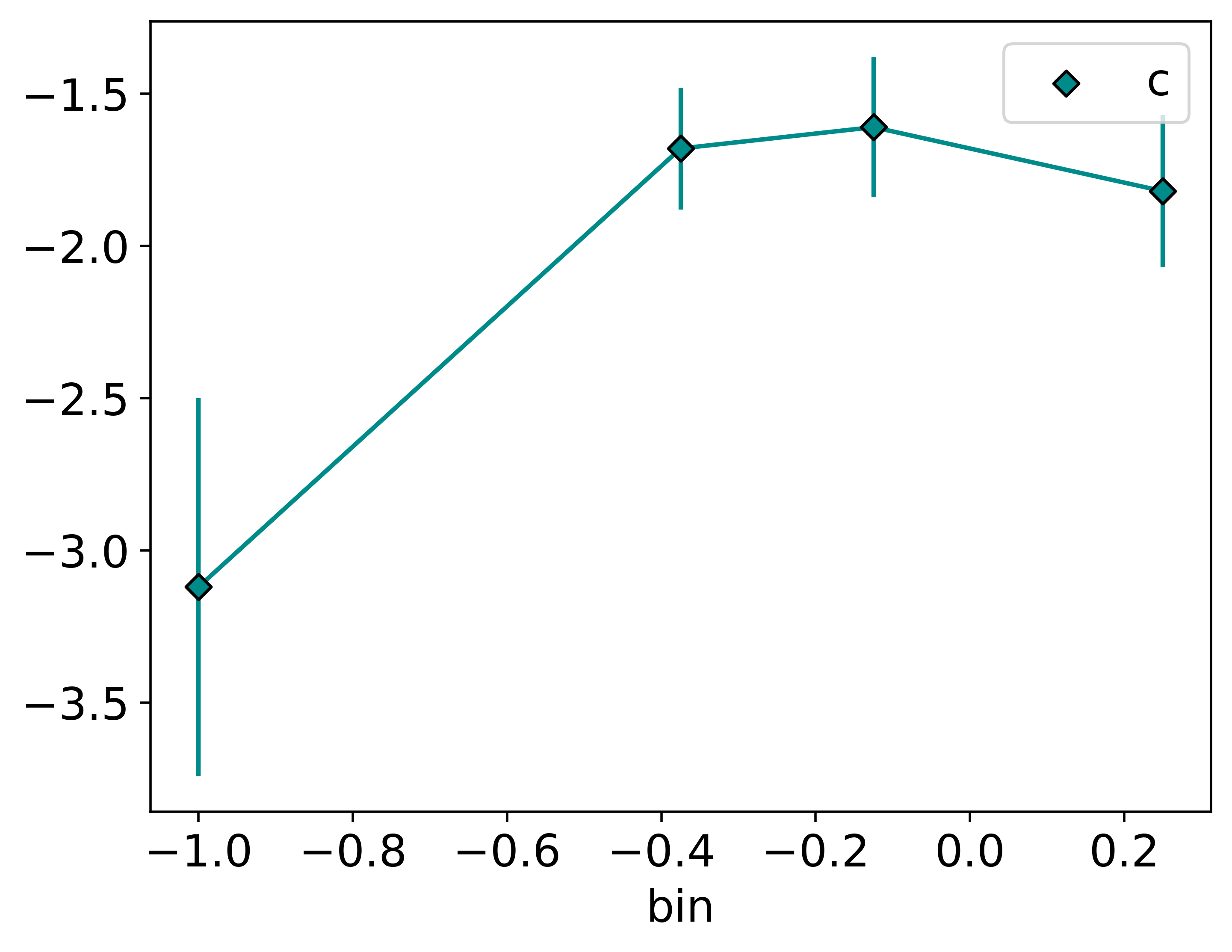}
    \includegraphics[width=0.24\textwidth]{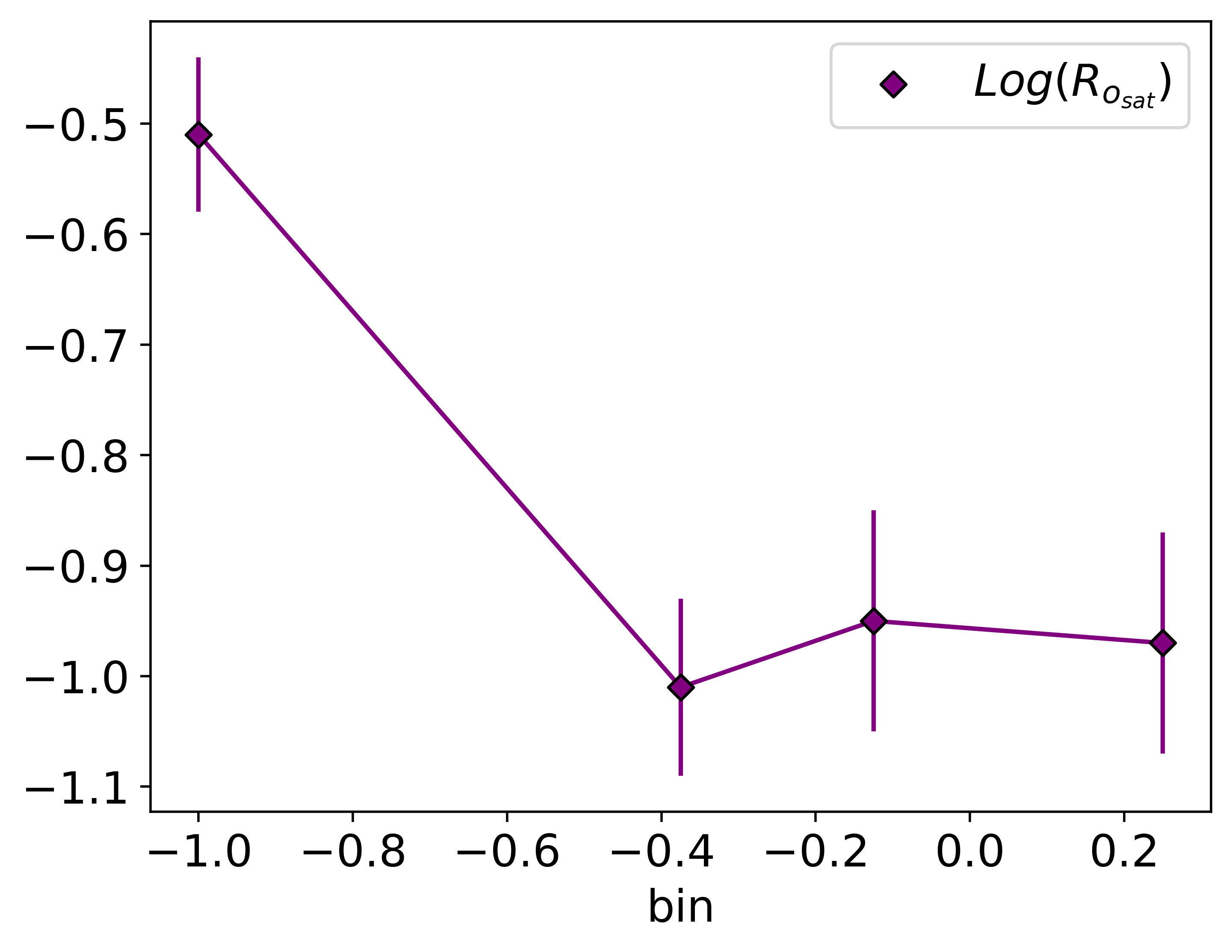}
    \caption{From left to right: trends of the fitting parameters listed in Table~\ref{Tab:Rx_Ro_FeH}, ``a'', ``b'', ``c'' and $\rm Log(Ro_{{sat}})$ as a function of the metallicity bin, for $\rm Ro$ computed as in \citet{Cranmer2011}.}
    \label{fig:params_trends_metallicity_bins_Teff}
\end{figure*}

In Figs.~\ref{Fig:Rx_Ro_VKs_FeH} and \ref{Fig:Rx_Ro_Teff_FeH}, we show the fit of the $\rm Log(R_x)$-$\rm Log(Ro)$ relationship for the global sample of stars divided in four metallicity bins, for $\rm \tau_{conv}$ computed as in \citet{Wright2018} and \citet{Cranmer2011}, respectively. In Figs.~\ref{fig:params_trends_metallicity_bins_VKs} and \ref{fig:params_trends_metallicity_bins_Teff}, the relative trend of the characteristic parameters of the fits are showed.

\section{Impact of recalibrated prescriptions on planetary mass loss}
\label{App:Radius_valley} 

\begin{figure*}
    \centering
    \begin{subfigure}{
        \includegraphics[width=0.45\textwidth]{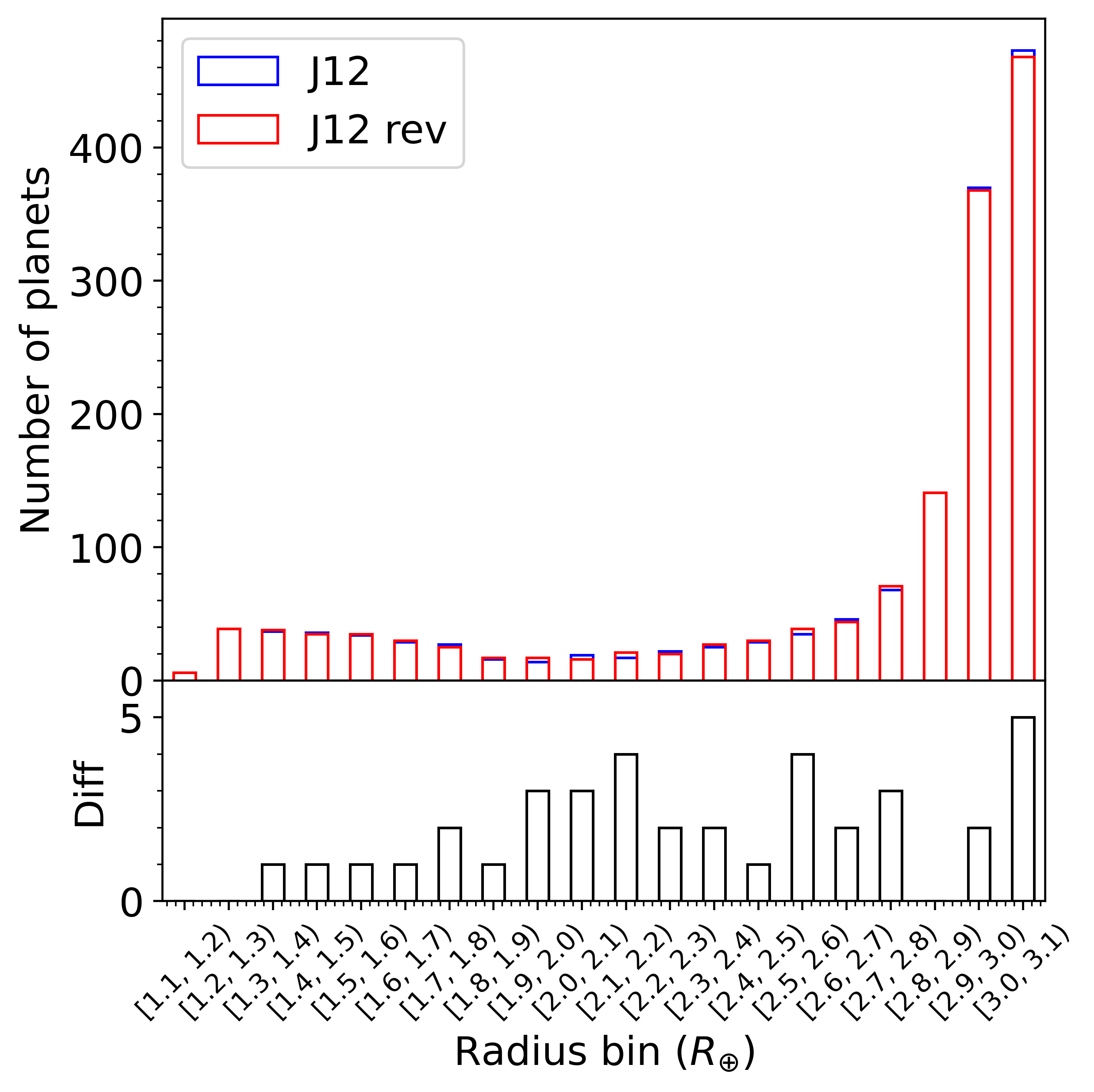}
        }
    \end{subfigure}
    \hfill
    \begin{subfigure}{
        \includegraphics[width=0.45\textwidth]{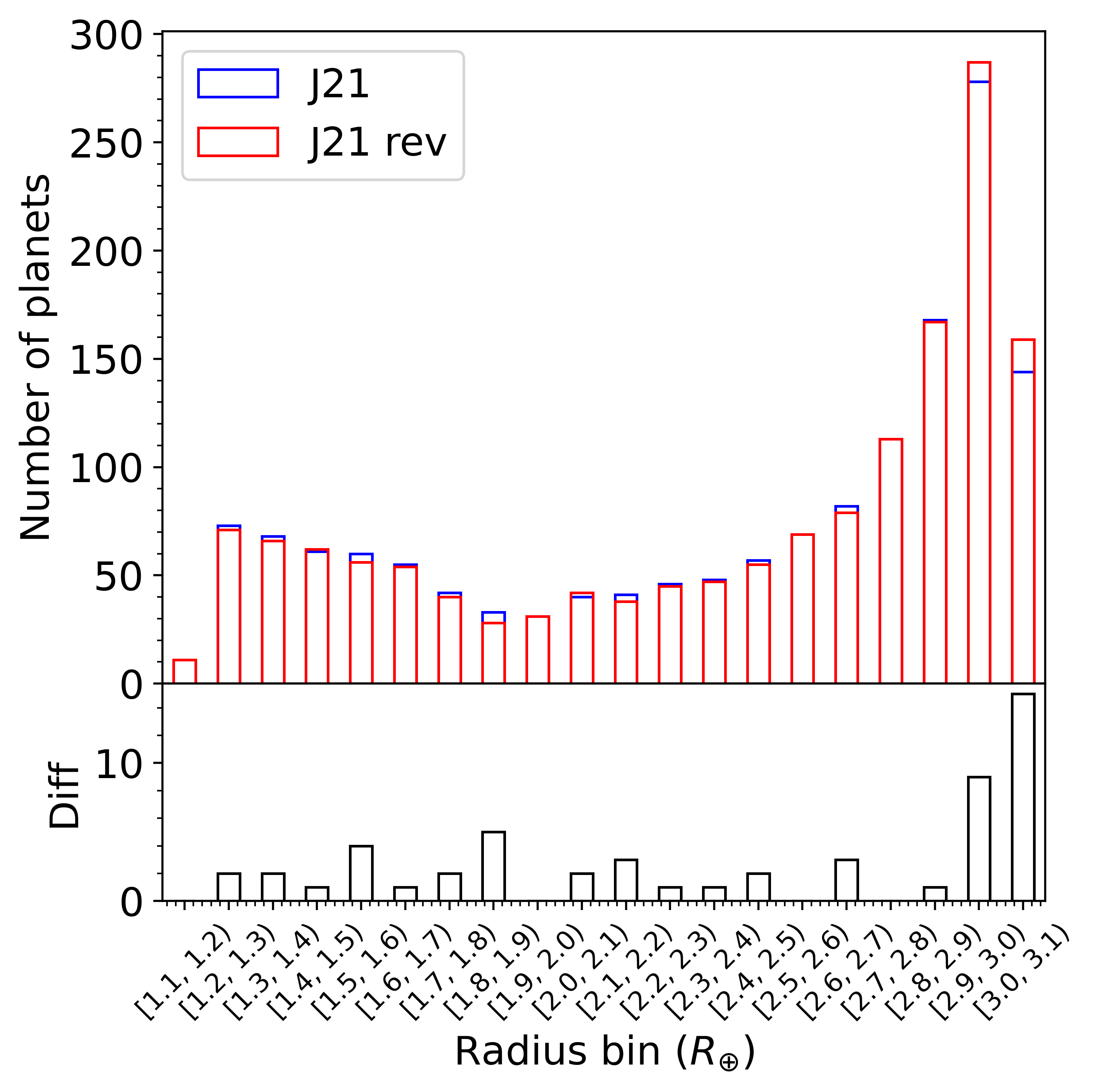}
        }
    \end{subfigure}
    \caption{Left panel: comparison between the number of planets per radius bin at 4.57 Gyr, obtained by simulating the atmospheric mass loss due to the X-ray luminosity computed with the original prescription of \citet{Jackson2012} (blue bars) and the revisited one (this work, red bars) and EUV irradiation as in \citet{SanzForcada2011}, for a $\rm 1~M_{\odot}$ parent star at solar metallicity. In the bottom panel, the numerical difference (in modulus) of planets per bin for the two considered prescriptions is shown. Right panel: the meaning of the plot is the same as for the left panel, but for the recalibrated prescription of \citet{Johnstone2021} as in \citet{Pezzotti2021} (blue bars) and the revisited one (this work, red bars).}
    \label{Fig:planets_radius_bin}
\end{figure*}

In Fig.~\ref{Fig:planets_radius_bin} we show the comparison between the number of planets per radius bin obtained when using J12 and J12 rev, and analogously J21 and J21 rev prescriptions in our SPI code for the computation of the X-ray luminosity tracks.

\end{appendix}

\end{document}